\documentclass[twocolumn]{aastex63}

\usepackage{gensymb}
\usepackage{booktabs}

\newcommand{\gppmfit}{\texttt{gppm\_fit}}

\newcommand{\Rjup}{{$\mathrm{R_{J}}$}}

\newcommand{\Mjup}{{$\mathrm{M_{J}}$}}
\newcommand{\Magellan}{\textit{Magellan}}

\shorttitle{An ACCESS \& LRG-BEASTS transmission spectrum of WASP-103b}
\shortauthors{Kirk et al.}

\begin{document}

\title{ACCESS \& LRG-BEASTS: a precise new optical transmission spectrum of the ultrahot Jupiter WASP-103b}

\correspondingauthor{James Kirk}
\email{james.kirk@cfa.harvard.edu}

\author[0000-0002-4207-6615]{James Kirk}
\affil{Center for Astrophysics $\vert$ Harvard \& Smithsonian, 60 Garden Street, Cambridge, MA 02138, USA}

\author[0000-0002-3627-1676]{Benjamin V.\ Rackham}
\altaffiliation{51 Pegasi b Fellow}
\affiliation{Department of Earth, Atmospheric and Planetary Sciences, and Kavli Institute for Astrophysics and Space Research, Massachusetts Institute of Technology, Cambridge, MA 02139, USA}

\author[0000-0003-4816-3469]{Ryan J.\ MacDonald}
\affil{Department of Astronomy and Carl Sagan Institute, Cornell University, 122 Sciences Drive, 14853, Ithaca, NY, USA}

\author[0000-0003-3204-8183]{Mercedes L\'{o}pez-Morales}
\affil{Center for Astrophysics $\vert$ Harvard \& Smithsonian, 60 Garden Street, Cambridge, MA 02138, USA}

\author[0000-0001-9513-1449]{N\'{e}stor Espinoza}
\affil{Space Telescope Science Institute, 3700 San Martin Drive, Baltimore, MD 21218, USA}

\author[0000-0001-9699-1459]{Monika Lendl}
\affil{Observatoire de Gen\'{e}ve, University of Geneva, Chemin Pegasi 51, 1290 Sauverny, Switzerland}

\author{Jamie Wilson}
\affil{Astrophysics Research Centre, School of Mathematics and Physics, Queens University Belfast, Belfast BT7 1NN, UK}

\author{David J.\ Osip}
\affil{Las Campanas Observatory, Carnegie Institution of Washington, Colina el Pino, Casilla 601 La Serena, Chile}

\author[0000-0003-1452-2240]{Peter J.\ Wheatley}
\affiliation{Department of Physics, University of Warwick, Gibbet Hill Road, Coventry, CV4 7AL, UK}
\affiliation{Centre for Exoplanets and Habitability, University of Warwick, Gibbet Hill Road, Coventry, CV4 7AL, UK}

\author[0000-0002-3849-8276]{Ian Skillen}
\affiliation{Isaac Newton Group of Telescopes, Apartado de Correos 321, E-38700 Santa Cruz de La Palma, Spain}

\author[0000-0003-3714-5855]{D\'{a}niel Apai}
\affil{Department of Astronomy/Steward Observatory, The University of Arizona, 933 N.\ Cherry Avenue, Tucson, AZ 85721, USA}
\affil{Lunar and Planetary Laboratory, The University of Arizona, 1629 E.\ Univ.\ Blvd, Tucson, AZ 85721}
\affil{Earths in Other Solar Systems Team, NASA Nexus for Exoplanet System Science}

\author[0000-0003-2831-1890]{Alex Bixel}
\affil{Department of Astronomy/Steward Observatory, The University of Arizona, 933 N.\ Cherry Avenue, Tucson, AZ 85721, USA}
\affil{Earths in Other Solar Systems Team, NASA Nexus for Exoplanet System Science}

\author[0000-0002-9308-2353]{Neale P.\ Gibson}
\affil{School of Physics, Trinity College Dublin, The University of Dublin, Dublin 2, Ireland}

\author[0000-0002-5389-3944]{Andr\'{e}s Jord\'{a}n}
\affil{Facultad de Ingenier\'{i}a y Ciencias, Universidad Adolfo Ib\'{a}\~{n}ez, Av.\ Diagonal las Torres 2640, Pe\~{n}alol\'{e}n, Santiago, Chile}
\affil{Millennium Institute for Astrophysics, Santiago, Chile}

\author[0000-0002-8507-1304]{Nikole K.\ Lewis}
\affil{Department of Astronomy and Carl Sagan Institute, Cornell University, 122 Sciences Drive, 14853, Ithaca, NY, USA}

\author[0000-0002-2574-356X]{Tom Louden}
\affil{Department of Physics, University of Warwick, Gibbet Hill Road, Coventry, CV4 7AL, UK}
\affiliation{Centre for Exoplanets and Habitability, University of Warwick, Gibbet Hill Road, Coventry, CV4 7AL, UK}

\author[0000-0002-6167-3159]{Chima D.\ McGruder}
\altaffiliation{NSF Fellow}
\affil{Center for Astrophysics $\vert$ Harvard \& Smithsonian, 60 Garden Street, Cambridge, MA 02138, USA}

\author[0000-0002-6500-3574]{Nikolay Nikolov}
\affil{Space Telescope Science Institute, 3700 San Martin Drive, Baltimore, MD 21218, USA}

\author[0000-0003-0650-5723]{Florian Rodler}
\affil{European Southern Observatory, Alonso de Cordova 3107, Vitacura, Santiago de Chile, Chile}

\author[0000-0001-6205-6315]{Ian C.\ Weaver}
\affil{Center for Astrophysics $\vert$ Harvard \& Smithsonian, 60 Garden Street, Cambridge, MA 02138, USA}



\begin{abstract}

We present a new ground-based optical transmission spectrum of the ultrahot Jupiter WASP-103b ($T_{eq} = 2484$\,K). Our transmission spectrum is the result of combining five new transits from the ACCESS survey and two new transits from the LRG-BEASTS survey with a reanalysis of three archival Gemini/GMOS transits and one VLT/FORS2 transit. Our combined 11-transit transmission spectrum covers a wavelength range of 3900--9450\,\AA\ with a median uncertainty in the transit depth of 148 parts-per-million, which is less than one atmospheric scale height of the planet. In our retrieval analysis of WASP-103b's combined optical and infrared transmission spectrum, we find strong evidence for unocculted bright regions ($4.3\sigma$) and weak evidence for H$_2$O ($1.9\sigma$), HCN ($1.7\sigma$), and TiO ($2.1\sigma$), which could be responsible for WASP-103b's observed temperature inversion. Our optical transmission spectrum shows significant structure that is in excellent agreement with the extensively studied ultrahot Jupiter WASP-121b, for which the presence of VO has been inferred. For WASP-103b, we find that VO can only provide a reasonable fit to the data if its abundance is implausibly high and we do not account for stellar activity. Our results highlight the precision that can be achieved by ground-based observations and the impacts that stellar activity from F-type stars can have on the interpretation of exoplanet transmission spectra.

\end{abstract}



\section{Introduction} \label{sec:intro}

Ultrahot Jupiters ($T_{eq} \gtrsim 2000$\,K) are becoming increasingly targeted for studies of their atmospheres \citep[e.g.,][]{Haynes2015,Evans2016,Evans2017,Evans2018,Nugroho2017,Sheppard2017,Arcangeli2018,Hoeijmakers2018,Lothringer2018,Lothringer2019,Hoeijmakers2019,Fu2020,Gibson2020,Changeat2021}. In part this is due to their high temperatures leading to large atmospheric scale heights and relatively cloud-free atmospheres, as condensation is suppressed. This latter point makes hotter planets particularly favorable for atmospheric studies, as the majority of studied exoplanets show muted features due to clouds and hazes \citep[e.g.,][]{Sing2016,Wakeford2019}. However, the amplitudes of water features decrease as equilibrium temperatures rise above 2500\,K due to the thermal dissociation of water and impact of H$^-$ opacity \citep[e.g.,][]{Arcangeli2018,Lothringer2018,Parmentier2018,Gao2020}.

Recently, several ultrahot Jupiters have revealed a plethora of atomic absorption in high-resolution transmission spectra \citep[e.g.,][]{Hoeijmakers2018,Hoeijmakers2019,Hoeijmakers2020,Sing2019,Yan2019,Cabot2020,Ehrenreich2020,Gibson2020,Nugroho2020,Borsa2021,Yan2021}. These planets also show the strongest evidence for temperature inversions among exoplanet atmospheres, such as in WASP-121b \citep{Evans2017,Evans2020}, WASP-33b \citep{Haynes2015,Nugroho2017}, WASP-18b \citep{Sheppard2017,Arcangeli2018}, WASP-76b \citep{Fu2020}, and WASP-103b \citep{Kreidberg2018}. However, of these exoplanets only WASP-121b, WASP-33b, and WASP-76b show evidence for TiO or VO (albeit in some cases this evidence is disputed, \citealp[e.g.,][]{Merritt2020,Borsa2021}), which have long been predicted to be responsible for hot Jupiter temperature inversions \citep[e.g.,][]{Hubeny2003,Fortney2008,Spiegel2009}. 

In this paper we present a new optical transmission spectrum of WASP-103b that reveals tentative evidence ($2.1\sigma$) for TiO in its atmosphere, which may be responsible for the planet's observed temperature inversion \citep{Kreidberg2018}.

\subsection{WASP-103b}

\object{WASP-103b}, discovered by \cite{Gillon2014}, is an ultrahot Jupiter ($\mathrm{M} = 1.51 \pm 0.11$\,\Mjup, $\mathrm{R} = 1.623^{+0.051}_{-0.053}$\,\Rjup, $\mathrm{T_{eq}} = 2484 \pm 67$\,K; \citealt{Delrez2018}). Its high equilibrium temperature owes to the fact that it orbits its F8V host star in 0.93 days \citep{Gillon2014,Delrez2018}. Because of its very high temperature, WASP-103b has a large scale height of 600\,km, which contributes 152\,ppm to the transit depth (assuming the parameters from \citealt{Delrez2018} and a mean molecular weight of 2.3\,amu). 

WASP-103b has been the target of atmospheric studies on several previous occasions. \cite{Southworth2015} presented a transmission photometry study of the planet using transits acquired in seven different photometric bands, from $g'$ to $z'$. They found a steep slope (gradient, $\alpha(\lambda) = 19.0 \pm 1.5$) in their transmission spectrum rising towards their bluest wavelength bins and extending over ${\sim}10$ atmospheric scale heights. \cite{Southworth2015} concluded unocculted starspots were unlikely to cause the observed slope due to the lack of the host star's photometric variability \citep{Gillon2014} and the temperature of the host ($T_{\mathrm{eff}} = 6110$\,K), which they suggested was too hot for significant spot-induced activity.

\cite{Southworth2016} subsequently revisited the earlier transmission spectrum of \cite{Southworth2015} after new high-resolution photometry revealed the presence of a blended background star \citep{Wollert2015,Ngo2016}. While correcting for this third light source did reduce the gradient of the slope in their data to $\alpha(\lambda) = 11.2 \pm 0.9$, it remained significantly steeper than expected for Rayleigh scattering ($\alpha(\lambda) = 4$). 

\cite{Turner2017} extended the observations of WASP-103b to the near-UV with ground-based observations of the planet in the U band. The transit depth they derived was consistent with the bluest bin of \cite{Southworth2016}, indicating that the optical slope may not continue to rise into the near-UV.

The presence of a slope in the optical transmission spectrum of WASP-103b was challenged by \cite{Lendl2017}, who presented results using the multi-object spectrograph GMOS on the Gemini-North telescope. They found no signs of the strong slope reported by \cite{Southworth2015} and instead found the atmosphere to be cloud-free with absorption from sodium and potassium. 

Following these studies in transmission, \cite{Cartier2017} presented HST/WFC3 secondary eclipse observations of WASP-103b. Their near-IR emission spectrum was featureless to a precision of 175\,ppm, finding the planet's atmosphere to be approximately isothermal over the wavelength range probed (1.1--1.7\,$\mu$m). However, they also noted that a thermally inverted atmosphere and a monotonically decreasing atmosphere with C/O $>1$ were consistent with their data.

\cite{Delrez2018} presented 16 new ground-based secondary eclipses of WASP-103b in the $z'$ and $K_S$ bands, and combined these with the HST/WFC3 results of \cite{Cartier2017}. They found the $z'$ and HST/WFC3 observations to be best fitted with a vertically isothermal atmosphere. However, they detected an anomalously deep eclipse in the $K_S$ band, which suggested the presence of a thermal inversion, but they noted that the $K_S$ band is not expected to contain strong spectral features and so could not explain this feature. They additionally reanalyzed the transit data from \cite{Southworth2015}, finding consistent results, indicating that the slope seen by \cite{Southworth2016} is inherent to the data and not dependent on the data reduction method. 

\cite{Kreidberg2018} presented phase curve measurements of WASP-103b observed with HST/WFC3 and Spitzer/IRAC. They found the phase curves to show large amplitudes and negligible hot spot offsets, indicating poor circulation of heat from the day side to the nightside of the planet. In the WFC3/IR/G141 bandpass (1.1--1.7\,$\mu$m), \cite{Kreidberg2018} found the phase-resolved spectra to be consistent with blackbodies with brightness temperatures ranging from 1880\,K on the nightside to 2930\,K on the day side. However, they observed a significantly higher brightness temperature in the Spitzer 4.5\,$\mu$m band that is likely due to CO in emission, indicating a thermal inversion in the planet's atmosphere. As a result, \cite{Kreidberg2018} speculated that either TiO, VO, or FeH may be present in the planet's upper atmosphere. By running atmospheric retrievals on their infrared data, they found WASP-103b to have a  $23^{+29}_{-13} \times$ solar metallicity atmosphere but did not detect water, which they attributed to its dissociation.

Most recently, \cite{Wilson2020} presented new VLT/FORS2 data of WASP-103b covering the wavelength region of 4000--6000\,\AA. By running retrievals on the combined VLT, Gemini, and HST transmission spectrum, they found the optical transmission spectrum to be featureless, with evidence for Na at $2\sigma$, but they did detect H$_2$O at $4\sigma$ confidence.

Given the contrasting results regarding WASP-103b's optical transmission spectrum and the potential for a previously undetected species at optical wavelengths giving rise to the planet's temperature inversion, we present here a new optical transmission spectrum of WASP-103b. Our new transmission spectrum combines five transits from the ACCESS survey on Magellan/Baade \citep{Jordan2013,Rackham2017,Bixel2019,Espinoza2019,McGruder2020, Weaver2020,Weaver2021} and two transits from the LRG-BEASTS survey on the William Herschel Telescope \citep{Kirk2017,Kirk2018,Kirk2019,Louden2017,Alderson2020}. We combine these new transit observations with a re-analysis of the published Gemini/GMOS data \citep{Lendl2017} and VLT/FORS2 data \citep{Wilson2020}, and include these with the published HST/WFC3/IR/G141 and Spitzer/IRAC transmission spectrum of \cite{Kreidberg2018} for our retrieval analyses. 

This paper is organized as follows.
We describe the observations in \autoref{sec:observations}, the data reduction in \autoref{sec:data_reduction}, our light curve fitting procedures in \autoref{sec:data_analysis}, and our corrections for third-light and night-side contamination in \autoref{sec:3rd_light_corr}.
We present the transmission spectrum of WASP-103b in \autoref{sec:results} and interpret it using the petitRADTRANS and POSEIDON retrieval frameworks in \autoref{sec:pRT} and \autoref{sec:POSEIDON}, respectively.
We discuss our results in \autoref{sec:discussion} and, finally, summarize our conclusions in \autoref{sec:conclusions}.

\section{Observations}
\label{sec:observations}

\subsection{ACCESS}
\label{sec:obs_ACCESS}

We observed five transits of WASP-103b in the 2015--2017 observing seasons with the Inamori-\Magellan{} Areal Camera and Spectrograph (IMACS) \citep{Dressler2011} on the 6.5-m \Magellan{} Baade Telescope at Las Campanas Observatory.
Our observation design largely mirrored that of previous ACCESS observations \citep{Jordan2013, Rackham2017, Bixel2019, Espinoza2019, McGruder2020, Weaver2020,Weaver2021}.
For each observation, we used the $f/2$ camera of IMACS in multi-object spectroscopy mode with 2$\times$2 binning (0.4\,\arcsec\,pixel$^{-1}$).
Conditions were clear for all observations except the first two nights (`ACCESS n1' and `ACCESS n2'), for which passing clouds impacted the data quality during the 2.6-hr transit for $\sim$1\,hr and ${\sim}$0.25\,hr, respectively.
Further details of these observations are provided in \autoref{tab:magellan_obs_log}.

We used a custom multislit mask with large slits ($10\,\arcsec$ wide by $21\,\arcsec$ long) for collecting spectra of \object{WASP-103} and six comparison stars simultaneously.
After the first three observations, we redesigned the mask so that each spectrum would only be dispersed across a single chip.
The second mask includes six comparison stars as well, four of which are repeated from the first mask (\autoref{tab:magellan_comps}).
We designed the two science masks for use with the 300 line/mm and 150 line/mm grisms, respectively, though the second is also compatible with the 300 line/mm grism over a narrower wavelength range.
For each science mask, we also made a mask with narrow (0.5\,\arcsec) slits for wavelength calibration.
After the first two unfiltered observations, we began using
the WB5600--9200 filter to block any potential contribution from second-order light.
For our final \Magellan{}/IMACS observation, we used the GG495 filter for increased sensitivity at bluer optical wavelengths.

We selected exposure times to provide a maximum count rate of $\sim$35,000 analog-to-digital units (ADU) in order to remain within the linear regime of the IMACS CCDs \citep[see Section~3.8 of][]{Bixel2019}.
We used the ``turbo'' readout mode for the first three transits and opted for the slightly slower and less noisy ``fast'' readout mode for latter transits.

For each observation, we collected a series of quartz lamp flats with the same mask, disperser, and binning as the science frames. Immediately before or after the science observations, we also collected a series of exposures of HeArNe lamps through the narrow-slit calibration mask to aid with wavelength calibration.

\begin{deluxetable*}{lrccccccccr}
    \tabletypesize{\scriptsize}
    \tablewidth{\linewidth}
    \tablecaption{Log of Magellan/IMACS observations. \label{tab:magellan_obs_log}}
    \tablehead{\colhead{Transit}      &
               \colhead{Date}         &
               \colhead{Time}         &
               \colhead{Airmass}      &
               \colhead{Seeing}       &
               \colhead{Grism}    &
               \colhead{Mask}         &
               \colhead{Filter}         &
               \colhead{Frames}       & 
               \colhead{Exposure}     & 
               \colhead{Readout +}    \\
               \colhead{}             &
               \colhead{(UTC)}        &
               \colhead{(UTC)}        &
               \colhead{}             &
               \colhead{}             &
               \colhead{}             &
               \colhead{}             &
               \colhead{}             &
               \colhead{}             & 
               \colhead{Times (s)}    & 
               \colhead{Overhead (s)} }
\startdata
ACCESS n1 & 5 Jun 2015  & 03:31--08:00 & $1.28 \rightarrow 1.24 \rightarrow 2.19$ & 0.7--1.2 & 300 line/mm (17.5\degree) & 3215 & Spectroscopic & 261 & 33     & 29 \\
ACCESS n2 & 6 Jun 2015  & 01:30--06:28 & $1.74 \rightarrow 1.24 \rightarrow 1.47$ & 0.7--1.4 & 300 line/mm (17.5\degree) & 3215 & Spectroscopic & 289 & 33     & 29 \\
ACCESS n3 & 2 Jul 2015  & 00:30--04:05 & $1.49 \rightarrow 1.24 \rightarrow 1.34$ & 0.7--1.0 & 300 line/mm (17.5\degree) & 3215 & WB5600--9200  & 210 & 33     & 29 \\
ACCESS n4 & 5 Apr 2017  & 05:45--10:13 & $1.65 \rightarrow 1.24 \rightarrow 1.40$ & 0.7--1.7 & 150 line/mm (18.8\degree) & 3397 & WB5600--9200  & 348 & 10--25 & 31 \\
ACCESS n5 & 1 May 2017 & 03:34--08:43 & $1.89 \rightarrow 1.24 \rightarrow 1.45$ & 1.0--1.8 & 300 line/mm (17.5\degree) & 3397 & GG495         & 262 & 40     & 31 \\
\enddata
\tablecomments{
Mask numbers may be used to query information about the masks via the IMACS Slit Mask Manager (\url{http://masks.lco.cl/search/}).
}
\end{deluxetable*}

\begin{deluxetable}{lllll}
    \tabletypesize{\scriptsize}
    \tablewidth{\linewidth}
    \tablecaption{
        Comparison stars for Magellan/IMACS observations.
        $Gaia$ Data Release 2 \citep{GaiaDR2} identifier, R.\,A., Decl., $Gaia$ G-band magnitude, and relevant IMACS mask(s) are listed.
        \label{tab:magellan_comps}
    }
    \tablehead{\colhead{$Gaia$ DR2 ID}   &
               \colhead{R.\,A. (J2000.0)}   &
               \colhead{Decl. (J2000.0)}  &
               \colhead{$G$}           &
               \colhead{Mask(s)}}
\startdata
4439093346050733440 & 16:36:41.654 & +07:13:52.99 & 11.93 & 3215, 3397 \\
4439098839311731968 & 16:37:13.193 & +07:16:16.58 & 12.28 & 3215, 3397 \\
4439102382661930496 & 16:36:39.917 & +07:16:53.63 & 12.20 & 3215, 3397 \\
4439089875717149696 & 16:36:34.512 & +07:07:02.69 & 12.70 & 3215, 3397 \\
4439092139162752128 & 16:36:48.826 & +07:11:18.59 & 12.82 & 3215 \\
4439076715937339648 & 16:36:47.243 & +07:01:06.98 & 12.98 & 3215 \\
4439092418337881088 & 16:36:32.904 & +07:09:20.49 & 12.53 & 3397 \\
4439078777521638144 & 16:37:17.850 & +07:05:51.90 & 13.28 & 3397 \\
\enddata
\end{deluxetable}

\subsection{LRG-BEASTS}
\label{sec:obs_LRG-BEASTS}

The LRG-BEASTS transits of WASP-103b were observed on the nights of 1 Jun 2016 (referred to as `LRG-BEASTS n1' hereafter) and 26 Jun 2016 (referred to `LRG-BEASTS n2'). Both transits were observed with the low-resolution ($R \approx 300$) grism spectrograph ACAM \citep{Benn2008} on the 4.2-m William Herschel Telescope (WHT) in La Palma. This is the same instrument as has been used in the previous five LRG-BEASTS studies \citep{Kirk2017,Kirk2018,Kirk2019,Louden2017,Alderson2020}. 

Unlike IMACS, ACAM is a long-slit spectrograph. For both of the LRG-BEASTS nights, we used the wide 7.6\,\arcmin $\times$ 40\,\arcsec\ slit, allowing us to simultaneously observe the spectrum of WASP-103 and a comparison star through the same slit. While a wide slit is useful to avoid differential slit losses between the target and the comparison, it does mean that other stars can fall within the slit and whose light can contaminate both the target and comparison\footnote{We note that the blending of the background star detected by \cite{Wollert2015} and \cite{Ngo2016} is unavoidable at the spatial resolution of the spectrographs used here.}. We therefore had to balance our desire to have a comparison star with a similar magnitude and color to the target while avoiding both possible contaminating stars and the edges of the detector. For our LRG-BEASTS observations of WASP-103, we selected the bright HD\,149891 as our comparison star, which has a V magnitude of 8.7 and $B-V$ color of 1.1. By comparison, WASP-103 has a V magnitude of 12.1 and $B-V$ color of 0.6. 

Given the brightness of the comparison, we had to use short, 9 second exposures to avoid saturation. In order to reduce the overhead, we used a fast readout mode and we windowed the CCD into two windows ([530:800, 1100:3100] for the target and [1302:1572, 1100:3100] for the comparison), which reduced the overhead to 5\,s.

On the first night, we acquired 1400 spectra of the target with an airmass varying between 1.40 $\rightarrow$ 1.08 $\rightarrow$ 1.43. The moon did not rise during our observations on this night. On the second night we took 1068 spectra of the target with an airmass varying between 1.09 $\rightarrow$ 1.08 $\rightarrow$ 1.70, with a moon illumination of 58\% at a distance of 108$^\circ$ from the target. Biases, tungsten lamp flats, sky flats, and CuAr plus CuNe arc spectra were taken at the start and end of both nights.

\subsection{Gemini/GMOS and VLT/FORS2 transits}
\label{sec:obs_GMOS}

Three transits of WASP-103b were observed with the Gemini North GMOS multi-object spectrograph on 27 Jun 2015 (`GMOS n1'), 10 Jul 2015 (`GMOS n2'), and 3 May 2016 (`GMOS n3'). A single transit of WASP-103b was observed using VLT/FORS2 on 1 May 2017. This was the same transit as observed by ACCESS (`n5'). For all GMOS and FORS2 transits, two comparison stars were observed in addition to WASP-103. We refer the reader to \cite{Lendl2017} and \cite{Wilson2020} for a description of these observations.

\section{Data Reduction}
\label{sec:data_reduction}

\subsection{ACCESS}
\label{sec:dr_ACCESS}

We reduced the \Magellan/IMACS data using our custom Python-based pipeline described previously \citep{Jordan2013, Rackham2017, Bixel2019, Espinoza2019, McGruder2020, Weaver2020, Weaver2021}.
The pipeline performs standard procedures for multi-object spectroscopy, including bias and flat calibration, bad-pixel and cosmic-ray correction, sky subtraction, spectral tracing and extraction, and wavelength calibration. It outputs for each observation sets of wavelength-calibrated 1D spectra for the target and comparison stars and arrays of state parameters useful for detrending systematics.
These include time, airmass, $x$ and $y$ detector positions of the spectra, sky background level, and the full-width half-maximum (FWHM) of the spectra.
As with previous ACCESS studies \citep{Rackham2017, Bixel2019, Espinoza2019, McGruder2020, Weaver2020,Weaver2021}, we found that flat fielding introduced additional noise into our high-SNR science images, and so we ultimately chose not to apply a flat-field correction. To extract the 1D spectra, the ACCESS pipeline uses standard aperture photometry. For all nights, we used a 15-pixel-wide aperture, corresponding to 6\,\arcsec, which we selected to limit the impact of seeing variations while also leaving adequate space in the spatial direction for estimating the sky background.

The extracted, wavelength-calibrated mean-averaged spectra for each ACCESS observation of WASP-103 are shown in \autoref{fig:stellar_spectra}.

\subsection{LRG-BEASTS}
\label{sec:dr_LRG-BEASTS}

To reduce the LRG-BEASTS long-slit data, we used the same custom-made Python scripts as used in \cite{Kirk2017,Kirk2018,Kirk2019} and \cite{Alderson2020}, and we refer the reader to those papers for a more in depth description of the process.

In brief, the pipeline performs bias removal, tracing of the target and comparison spectra, standard aperture photometry using a linear polynomial interpolated across the stellar spectra (in the spatial direction) to estimate the sky background, cosmic-ray removal, and wavelength calibration. 

Master biases were created for each night and subtracted from the science data before extraction. For the first night 257 bias frames were median-combined to create a master bias, and 301 bias frames were combined for the second night. As with the ACCESS data, and our previous LRG-BEASTS studies, we chose not to apply a flat-field correction. In a test-run on LRG-BEASTS n1, we found that the use of a flat-field led to an identical transmission spectrum but with uncertainties $\sim10$\,\% larger than without the use of the flat-field.

We experimented with our choice of aperture width to extract the stellar spectra, finding an aperture width of 26 pixels to be optimal for night 1 and 40 pixels for night 2 (owing to the poorer seeing conditions). The optimal aperture width was determined as the aperture width that produced the minimum standard deviation following a fit of an analytic transit light curve with a cubic in time polynomial to the resulting white light curve. We note that the plate scale of ACAM is 0.25\,\arcsec \,pixel$^{-1}$.

The extracted, wavelength-calibrated LRG-BEASTS spectra of WASP-103 are shown in \autoref{fig:stellar_spectra}.

\subsection{Gemini/GMOS and VLT/FORS2 data}
\label{sec:dr_GMOS}

For the Gemini/GMOS and VLT/FORS2 transits, we used the stellar spectra as reduced by \cite{Lendl2017} and \cite{Wilson2020}. Importantly, we followed the approach of both \cite{Lendl2017} and \cite{Wilson2020} in using only one of the two comparison stars observed in those studies. This was due to the second comparison star leading to an increase in the noise in the data.

\subsection{Binning scheme and light curve creation}
\label{sec:binning_scheme}

\begin{figure}
\centering
\includegraphics[scale=0.47]{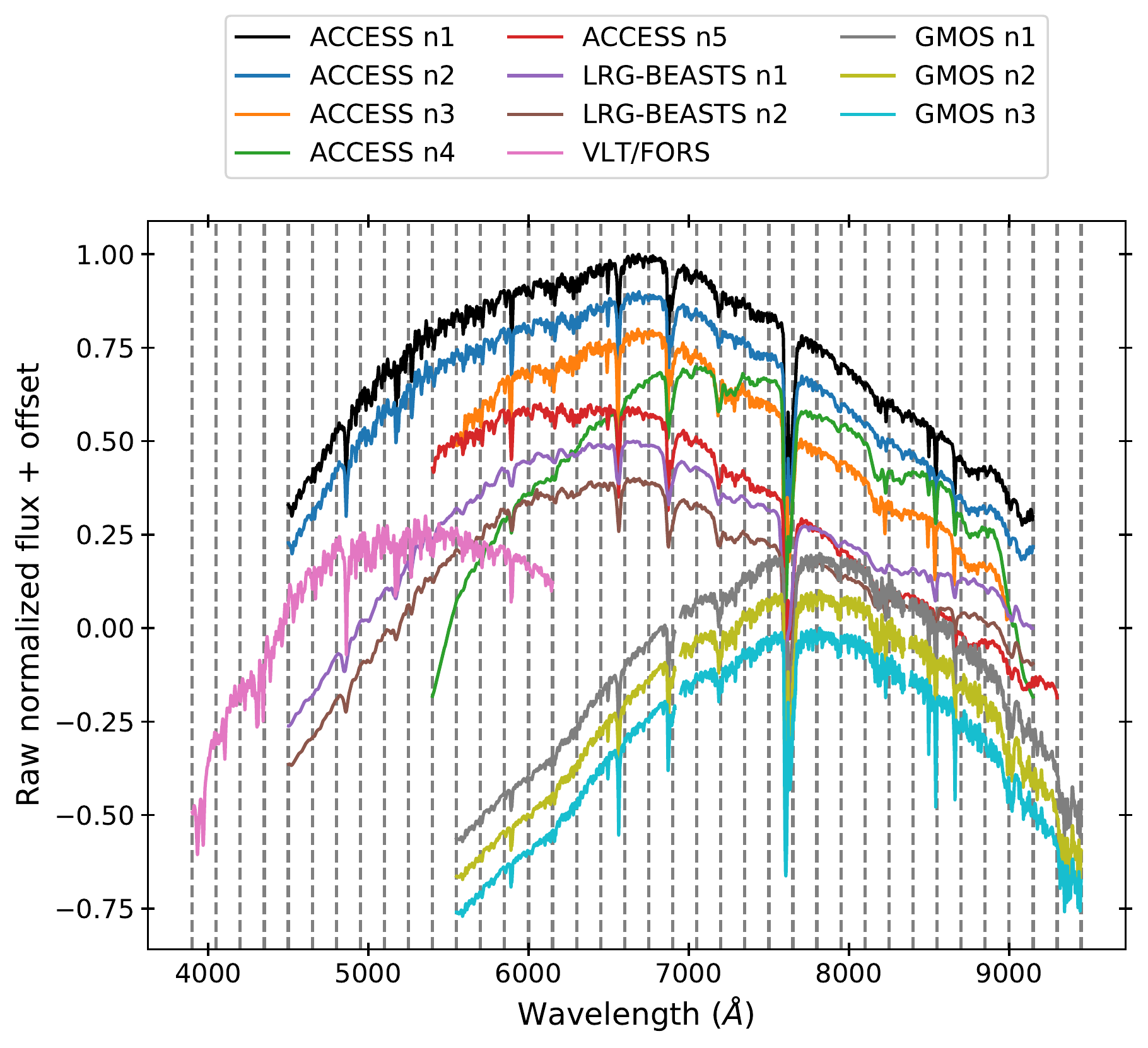}
\caption{WASP-103's nightly mean-averaged spectra as observed on each of the 11 nights used in our analysis, grouped by instrument and with an offset applied for clarity. The different spectral shapes are a result of the different instrument throughputs. The dashed vertical lines indicate the edges of 37 150-\AA-wide bins we used to generate the spectroscopic transit light curves.}
\label{fig:stellar_spectra}
\end{figure}

Given the number of different transits used in our analysis, and the subsequently different wavelength coverage, we opted for a straightforward binning scheme comprised of 150\,\AA-wide wavelength bins from 3900--9450\,\AA. The bins and nightly mean-averaged spectra of WASP-103 from all 11 transits are shown in \autoref{fig:stellar_spectra}. We note that we excluded the four bins of the Gemini/GMOS data that included chip gaps.

Given \cite{Lendl2017}'s finding of strong sodium and potassium in the atmosphere of WASP-103b in 20\,\AA-wide bins, we additionally constructed separate 20\,\AA-wide bins centered on the sodium doublet and each of the two absorption lines in the potassium doublet.

Using our 150\,\AA\ and 20\,\AA-wide wavelengths bins, we then created the spectroscopic light curves by integrating the spectra of the target and comparison stars over these wavelength ranges. The white light curves were created by integrating over the full wavelength range used for each night. 

For the LRG-BEASTS, GMOS, and FORS2 data, the target's light curves were divided by the comparisons' light curves at this point to correct for telluric absorption. However, the ACCESS light curves were treated differently due to the principal component analysis used in the ACCESS fitting routine (\autoref{sec:GPTransSpec}).

\begin{figure*}
    \centering
    \includegraphics[scale=0.6]{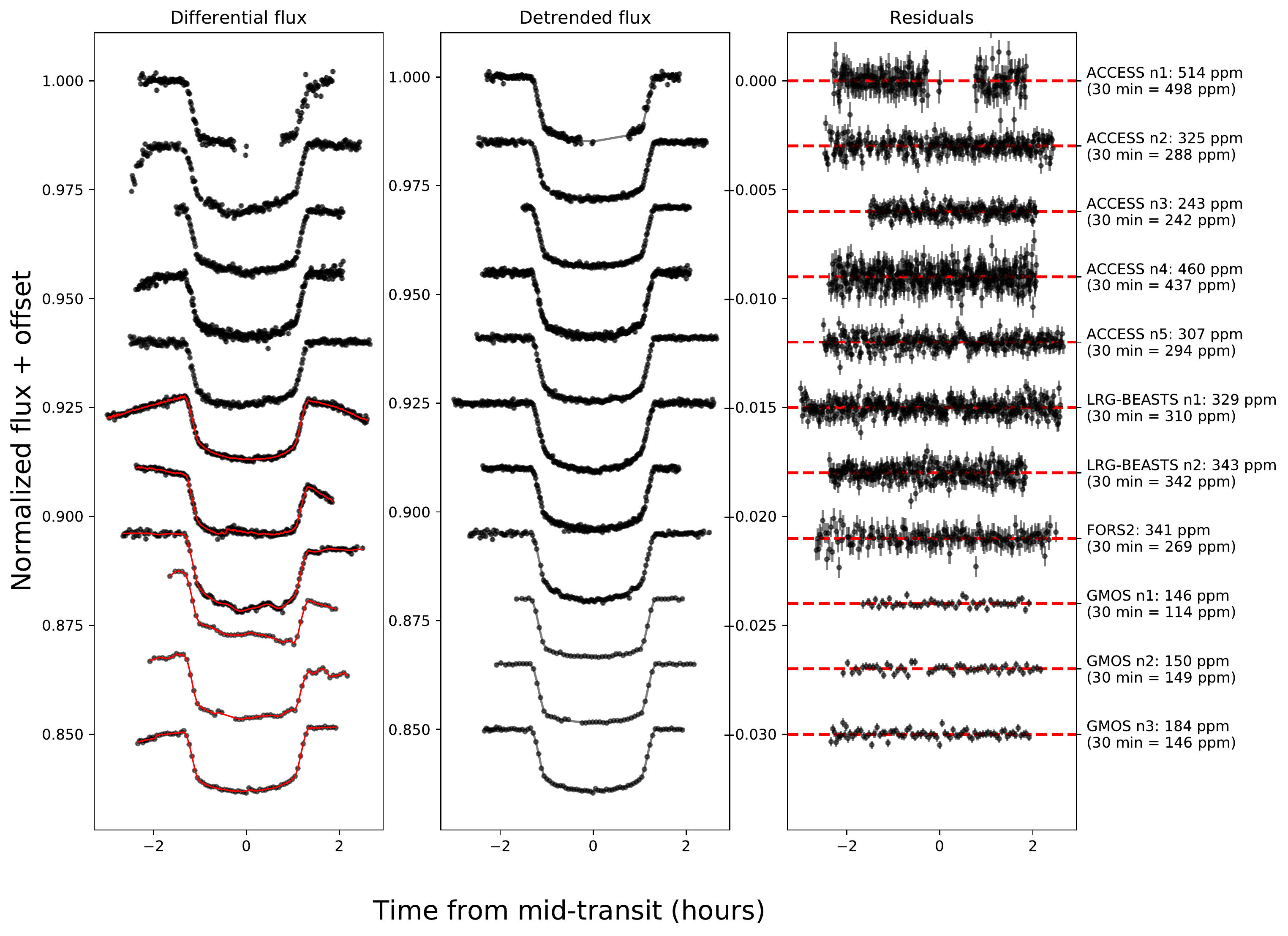}
    \caption{The white light curves from each of the 11 transits we analyzed. Left-hand panel: The white light curves after dividing by the comparison star and with each offset by -0.015 in flux. The red line shows the best-fitting transit + systematics models. We note that the ACCESS light curves are shown as WASP-103's light curve divided by the `best' comparison's light curve. However, these differential light curves are not used in the ACCESS fits, as explained in \autoref{sec:GPTransSpec}, and so do not include a best fit model. Middle panel: the detrended light curves following the removal of the systematics model. The gray lines show the best-fitting transit models. Right-hand panel: The residuals for each of the fits, this time with smaller offsets applied (-0.003) to allow for a zoomed-in view. The right-hand axis lists the RMS in ppm for each transit's residuals, along with the RMS in 30 minute bins.}
    \label{fig:wlc_fits}
\end{figure*}

The white light curves for all 11 transits are shown in \autoref{fig:wlc_fits}.

\section{Light curve fitting}
\label{sec:data_analysis}

Here we describe the different procedures we used to fit the ACCESS, LRG-BEASTS, and archival light curves. For both ACCESS and LRG-BEASTS, these surveys have custom pipelines (\texttt{GPTransSpec} for ACCESS, \gppmfit{} for LRG-BEASTS), written independently in Python, to perform the fitting. We refer the reader to the previous papers in the ACCESS and LRG-BEASTS series for more detail. In \autoref{sec:wb_comparing_GPTS_to_gppm_fit}, we compare the outputs of the two procedures on an identical data set and find excellent agreement between the two approaches.

We note that we treated the system parameters in two separate ways for all of our light curve fits with \texttt{GPTransSpec} and \gppmfit{}. Firstly we fixed the planet's orbital period ($P = 0.9255456$), scaled semi-major axis ($a/R_* = 2.999$) and inclination ($i = 87.3$) to the values of \cite{Southworth2015}. These were the system parameters used by \cite{Kreidberg2018} for their IR transmission spectrum of WASP-103b and our choice of a consistent set of system parameters allows for more accurate stitching of multi-epoch transmission spectra for our later retrieval analysis. We refer to this set of fits as our `fixed system parameters' fits. However, we also ran fits to our light curves with these system parameters as free parameters, which allows us to report updated system parameters. We refer to this set of fits as our `free system parameters' fits.

\subsection{Fitting the ACCESS light curves}
\label{sec:GPTransSpec}

\subsubsection{Model setup}

We extracted transmission spectra from the \Magellan{}/IMACS datasets with the procedure introduced by \citet{Espinoza2019} and detailed further in subsequent ACCESS studies \citep{Weaver2020,McGruder2020}.
We refer here to this procedure and the code implementing it as ``\texttt{GPTransSpec}''.

In short, we model the target light curve in magnitude space as a linear combination of the transit signal 
and `nuisance' signals introduced by instrumental systematics and observing conditions. We model the transit 
signal using the analytic expressions of \cite{Mandel2002} as implemented in \texttt{batman} \citep{batman}. The systematics, on the other hand, are modelled as a sum of two components. 
The first component is a set of basis signals obtained by performing principal 
component analysis (PCA) on a set of comparison stars \citep[see][for details]{Jordan2013, Espinoza2019}, and thus the raw light curves are modelled directly, not the differential light curves. The second component comes from a Gaussian process (GP, \citealp[e.g.,][]{Gibson2012}), where we use external parameters as regressors. The motivation for this two-component systematic modelling is that the
PCA signals inform our modelling of common systematics to both the target and the comparison stars 
(due to, e.g., atmospheric conditions), while the GP component allows us to incorporate 
systematics which are particular to our target lightcurve (e.g., position-dependent systematics, 
spectral profile fluctuations, etc.).

The number of basis signals $N$ obtained from our PCA analysis equals the number of 
comparison stars by the very definition of PCA. However, typically only a subset $n<N$ of 
those basis signals is actually informative for explaining/reconstructing the comparison (and target) 
star lightcurves, with basis signals having larger eigenvalues being the most important for 
this reconstruction. While in \cite{Jordan2013} the number of basis signals was decided via cross-validation, in \cite{Espinoza2019} Bayesian model averaging was used instead, where models 
using the $n = 1,2,...,N$ most important basis signals were fitted to the data, with the 
transit lightcurve parameters on each of these fits being later weighted by their log-evidences to 
report the final, ``model-averaged'' parameters. We use this same technique here in order to 
account for the uncertainty on the number $n$ of ``optimal'' basis signals.

We incorporate the GP in our modelling framework using 
the \texttt{george} Python package \citep{Foreman-Mackey2017}. 
In each of our fits, we use a multi-dimensional squared-exponential kernel, using as regressors 
the airmass, $x$ position of the trace, $y$ position of the trace, full-width at half maximum (FWHM), 
sky background and time as input variables. We assume a quadratic limb-darkening law for our 
transit lightcurve. Rather than fitting directly for the linear and quadratic coefficients ($u1$ and $u2$), \texttt{GPTransSpec} uses the change of variables introduced in \cite{Kipping2013} ($u1 \rightarrow q1$, $u2 \rightarrow q2$). In this way \texttt{GPTransSpec} treats both limb darkening coefficients as free parameters with uniform priors to prevent unphysical values.

In total, for each ACCESS white light curve fit, \texttt{GPTranSpec} fits for $n$ coefficients, 
which weigh the PCA-retrieved basis signals and 13 free parameters for both the transit lightcurve 
and the GP component. The latter parameters are the limb-darkening coefficients ($q1$ and $q2$), the six inverse length scales of our squared exponential GP kernels ($\tau_1$,...,$\tau_6$), the amplitude of the GP ($a$), a white noise term which accounts for an underestimation of the photometric error bars in the light curves ($\sigma^2$), time of mid-transit ($T_C$), the scaled planetary radius ($R_P/R_*$) and a normalization factor ($f$) to account for imperfect normalization of the light curves.

We used a wide uniform prior in log space for $a$ between 10$^{-10}$ and 0.01, and used a gamma prior on the GP length scales with shape parameter unity, following the approach of, e.g., \cite{Gibson2012} and \cite{Evans2015}. Finally, we also placed a wide uniform prior on $\sigma^2$, bounded by 10$^{-10}$ and 0.01 in milli-magnitudes$^2$ (mmag$^2$), which corresponds to $0.01-100$\, mmag bounds on the uncertainty in the photometric error bars. We used wide uniform priors on the transit parameters to prevent unphysical values.

For the `free system parameters' white light curve fits (where we wanted to refine the system parameters but which are not used in our final transmission spectra), we additionally fitted for the scaled semi-major axis ($a/R_*$) and the inclination of the planet's orbit ($i$).

To explore the parameter space, \texttt{GPTransSpec} uses nested sampling via the \texttt{PyMultiNest} package \citep{Buchner2014} with 1000 live points.

\subsubsection{Application of the model}

Prior to fitting the ACCESS light curves, we inspected the differential (target/comparison) light curves for each comparison star for all ACCESS nights. This enabled us to remove significant outliers from our photometry and to check whether all the comparison stars had similar noise characteristics to the target star and thus would be useful for inclusion in our \texttt{GPTransSpec} fitting. For all ACCESS nights we ended up using 5 of the 6 available comparison stars, with the same comparison stars used for nights that had a common slit mask (\autoref{tab:magellan_obs_log}).

We removed any data points from the light curves that were taken at airmasses greater than 2 and those that corresponded to cloudy conditions. These were defined as frames where WASP-103's flux dropped below 80\,\% of its nightly maximum. For ACCESS n1, a significant portion (30\,\%) of the transit was contaminated by clouds and clipped before fitting (\autoref{fig:wlc_fits}). For ACCESS n2, 10\,\% of the transit was cloudy. For the remaining three ACCESS nights, no data points were clipped. We note that we compared the results of fitting the light curves with and without this data clipping step and find the resulting transmission spectra to be consistent, but the data clipping leads to greater precision.

Following the fits to the ACCESS white light curves, we used the best-fitting systematics model from each night to remove the common mode systematics in the spectroscopic light curves prior to fitting. This reduces sources of noise that are common among the spectroscopic light curves and provides greater precision in the resulting transmission spectra. We then fitted the spectroscopic light curves with \texttt{GPTransSpec} but this time holding the time of mid-transit ($T_C$) fixed to the nightly best-fitting value.

\subsection{The LRG-BEASTS procedure}
\label{sec:gppm_fit}

\subsubsection{Model setup}

Like ACCESS's \texttt{GPTransSpec}, LRG-BEASTS's \gppmfit{} simultaneously fits quadratically limb-darkened transit light curves \citep{Mandel2002}, via the \texttt{batman} Python package \citep{batman}, together with a Gaussian process (also using the \texttt{george} Python package; \citealt{george}) to model correlated noise in the data. 

Unlike \texttt{GPTransSpec}, \gppmfit{} uses Markov chain Monte Carlo (MCMC) to perform the sampling of parameter space, via the \texttt{emcee} Python package \citep{emcee}. It also fits the differential light curves, unlike \texttt{GPTransSpec} which fits the raw stellar light curves in its PCA.

While \gppmfit{} also uses quadratic limb darkening, we instead hold the quadratic limb darkening coefficient ($u2$) fixed to a theoretical value calculated by the Limb Darkening Toolkit (\texttt{LDTk}, \citealt{ldtk}). To calculate these values we used the stellar parameters of \cite{Delrez2018}. We then only fitted for the linear coefficient ($u1$), for which we used a uniform prior to prevent unphysical values.

To fit the systematics in the light curves, we used a combination of squared exponential GP kernels, taking various ancillary data (airmass, FWHM, etc.) as input variables. We specify which variables were used for each night in the following subsections.  These kernels each had a separate length scale ($\tau$) but shared a common amplitude, $a$. We additionally included a white noise kernel, defined by the variance $\sigma^2$, to account for white noise not captured by the photometric error bars. We fitted for the natural logarithm of these values and fitted for the inverse length scale following previous studies \cite[e.g.,][]{Gibson2012,Gibson2017,Evans2017,Evans2018,Kirk2019,Alderson2020}.

Prior to fitting the data, the input variables were standardized by subtracting the mean and dividing by the standard deviation, following the procedure of, e.g., \cite{Evans2017,Evans2018,Kirk2019} and \cite{Alderson2020}. This gives each input variable a mean of zero and standard deviation of unity and helps the GP determine the inputs of importance for describing the noise characteristics. 

Similar to \cite{Gibson2017} and \cite{Alderson2020}, we placed truncated uniform priors in log space on the GP hyperparameters. The amplitude, $a$, was bounded between 0.01 and $100 \times$ the out-of-transit variance and the length scales were bounded between the minimum spacing and twice the maximum spacing of data points of the standardized variables. For the white noise term, we placed wide uniform priors in log space bounded by 10$^{-8}$ and $2.5 \times 10^{-7}$, which corresponds to 100--5000\,ppm bounds on the uncertainty in the photometric error bars. 

Prior to fitting the white and spectroscopic light curves with \gppmfit{}, we fit a transit model multiplied by a cubic polynomial using a Nelder-Mead algorithm \citep{nelder1965} to remove $>4 \sigma$ outliers from our light curves. This typically clipped at most 1--2 points per light curve.

To find the starting values for the GP hyperparameters, we optimized the GP model to the out-of-transit data prior to fitting each light curve. The MCMC chains were subsequently started with a small scatter around these values. Following the \texttt{george} documentation\footnote{\url{https://george.readthedocs.io/en/latest/tutorials/model/}}, \gppmfit{} runs two sets of chains for each light curve, with the second set of chains started with a smaller scatter around the best-fitting values from the first chain. 

For all our light curve fits with \gppmfit{}, the number of walkers equalled $6\times n_p$ (where $n_p$ is the number of free parameters) for 10000 steps each (split across the first and second runs as explained above).

\subsubsection{Application of the model}

Prior to fitting the LRG-BEASTS white and spectroscopic light curves, we binned the data from both nights to a time resolution of 60\,s. This reduced the number of data points to be fitted for all LRG-BEASTS light curves from 1400 to 337 for night 1 and from 1068 to 255 for night 2. We did this for computational reasons as GPs scale as $\mathcal{O}(n^3)$ for $n$ data points. 

For LRG-BEASTS night 1 we used five GP kernels, each taking one of airmass, FWHM, $x$ position of the trace, $y$ position of the trace and sky background as input variables. Due to the additional noise in LRG-BEASTS night 2 (\autoref{fig:wlc_fits}), we additionally included time with a sixth GP kernel for the analysis of these light curves. In total for the LRG-BEASTS white light curve fits, there were 10 free parameters ($T_C$, $R_P/R_*$, $u1$, $a$, $\tau_1$,...$\tau_5$, $\sigma^2$) for night 1 and 11 for night 2 (with the addition of the time kernel).

Similar to the ACCESS procedure, we then removed the common noise models from the spectroscopic light curves prior to fitting. We also again fixed $T_C$ to the best-fitting values from the white light curve fits.

\subsection{Fitting the GMOS and FORS2 light curves}

To fit the Gemini/GMOS and VLT/FORS2 light curves, we used the same \gppmfit{} code as we used to fit the LRG-BEASTS light curves.
For the GMOS data we again used 5 squared exponential GP kernels to fit the systematics in the data, with each taking one of airmass, FWHM, position angle, $y$ position of the trace, and sky background as input variables.
For the VLT data, we followed the approach of \cite{Wilson2020} in modelling the systematics with a single GP kernel, taking time as the input variable, combined with a linear in time polynomial.
The white light curve fits to the GMOS and FORS2 data are shown in \autoref{fig:wlc_fits}.

As with the ACCESS and LRG-BEASTS transits, we removed the common noise systematics model from the spectroscopic light curves prior to fitting. 
As we show in \autoref{sec:r_transmission_spectra}, our resulting GMOS and FORS2 transmission spectra are in good agreement with the initial analyses of \cite{Lendl2017} and \cite{Wilson2020}, giving us confidence in the repeatibility of our results.

\subsection{Comparing \texttt{GPTransSpec} with \gppmfit{}}

As an additional test of our light curve fitting routines, we generated a set of ACCESS differential light curves by dividing each night by the single `best' comparison star (as shown in \autoref{fig:wlc_fits}). The best comparison star was chosen as the comparison star that lead to the smallest median absolute deviation in the out-of-transit flux. We then ran these light curves through the same procedure as described in \autoref{sec:gppm_fit}. The results of this test are presented in \autoref{sec:wb_comparing_GPTS_to_gppm_fit} and show excellent agreement between \texttt{GPTransSpec} and \gppmfit{}.

\section{Correcting for third-light and nightside contamination}
\label{sec:3rd_light_corr}

Due to the presence of the blended contaminant within the WASP-103 spectra, we had to correct our resulting transmission spectra for this. To perform this correction, we followed a similar process to that used in \citet{Southworth2016,Cartier2017,Lendl2017,Turner2017} and \cite{Delrez2018}. We used PHOENIX \citep{Husser2013} stellar models for the target ($T_{\mathrm{eff}} = 6110$\,K, $\log g = 4.22$\,cgs, [Fe/H] = 0.06\,dex, \citealt{Delrez2018}) and contaminant ($T_{\mathrm{eff}} = 4400 \pm 200$\,K, \citealt{Cartier2017}), and scaled these to have the correct $J$-band magnitudes \citep{Cartier2017}. We then calculated the ratio of the contaminant's flux to the target's flux for each of our wavelength bins as defined in \autoref{sec:binning_scheme}.

\begin{figure}
    \centering
    \includegraphics[scale=0.55]{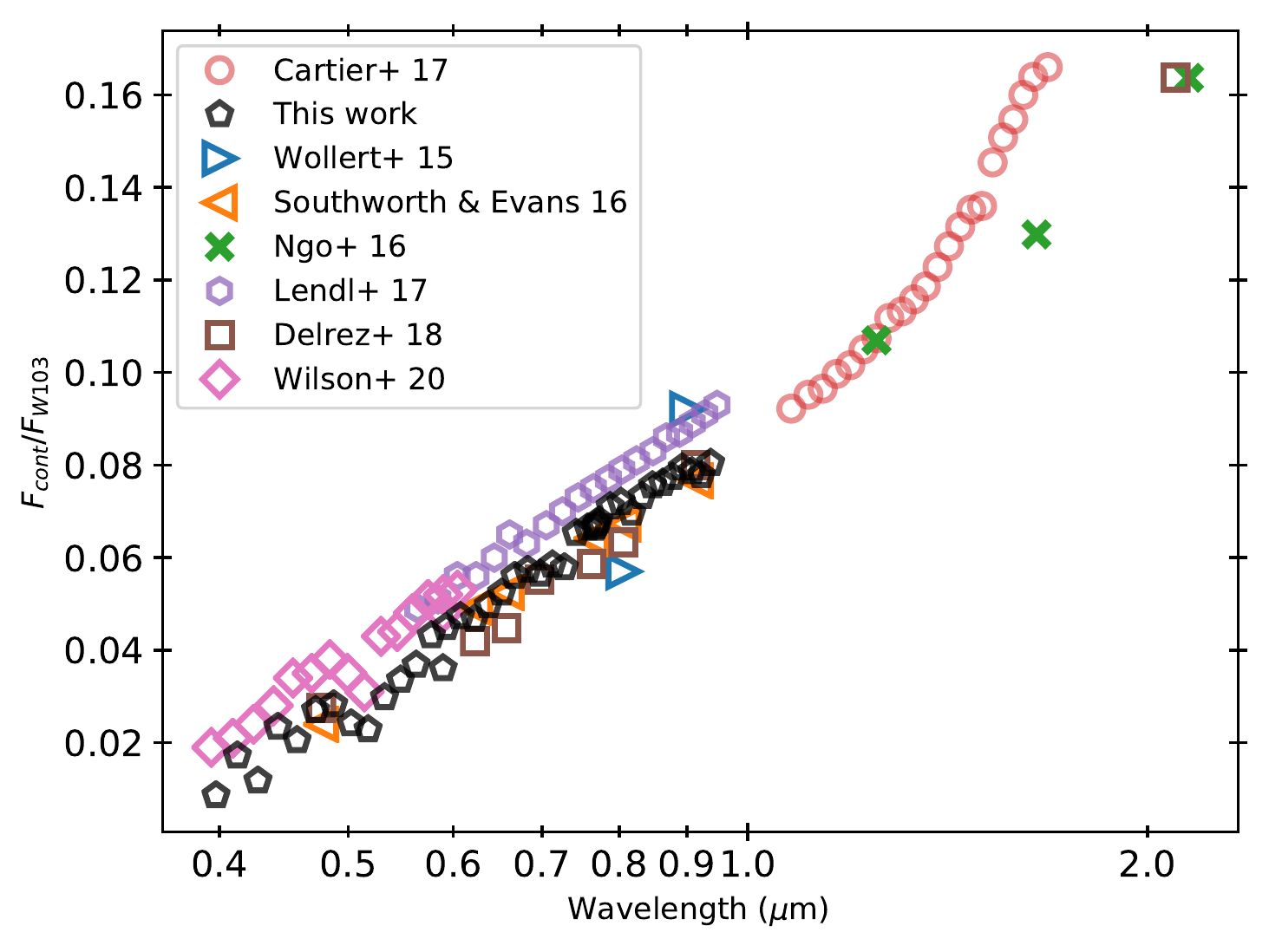}
    \caption{The ratio of the blended contaminant's flux ($F_\mathrm{cont}$) to WASP-103's flux ($F_\mathrm{W103}$) as a function of wavelength. The contamination ratios calculated for our wavelength bins are shown by the black pentagons. The associated values from several previous studies are also shown.}
    \label{fig:cont_ratios}
\end{figure}

\autoref{fig:cont_ratios} shows the contamination ratios we calculate as compared with those measured by \cite{Wollert2015,Ngo2016,Southworth2016,Cartier2017,Lendl2017,Delrez2018} and \cite{Wilson2020}. This figure shows the good agreement between our calculated flux ratios and these previous studies. The third-light correction factors for our wavelength bins are given in \autoref{tab:transmission_spectrum}.

To apply these flux corrections to our resulting transmission spectra, we used the following equation \cite[e.g.,][]{Kreidberg2018_review}
\begin{equation}
\label{eq:3rd}
    d_\mathrm{corr} = d_\mathrm{obs}(1 + F_\mathrm{cont}(\lambda)/F_\mathrm{W103}(\lambda)),
\end{equation}
where $d_\mathrm{obs}$ are the observed transit depths, $F_\mathrm{cont}(\lambda)$ is the flux of the contaminant, $F_\mathrm{W103}$ is WASP-103's flux and $d_\mathrm{corr}$ are the corrected transit depths. 

In addition to correcting for the contamination of the nearby star, we also considered dilution of the transit depth by the planet's nightside flux given the high equilibrium temperature of the planet \cite[e.g.,][]{Kipping2010}. While the effect is small in the optical, it can lead to a significant change in the infrared, as shown by \cite{Kreidberg2018} for WASP-103b. 

\cite{Kreidberg2018} used their \textit{Spitzer} phase curve observations to measure the nightside temperature of WASP-103b, finding it to be 1700\,K. We then used a PHOENIX model for the star and a blackbody at the nightside temperature to calculate the flux dilution in each of our wavelength bins from the nightside of the planet.

\begin{figure}
    \centering
    \includegraphics[scale=0.55]{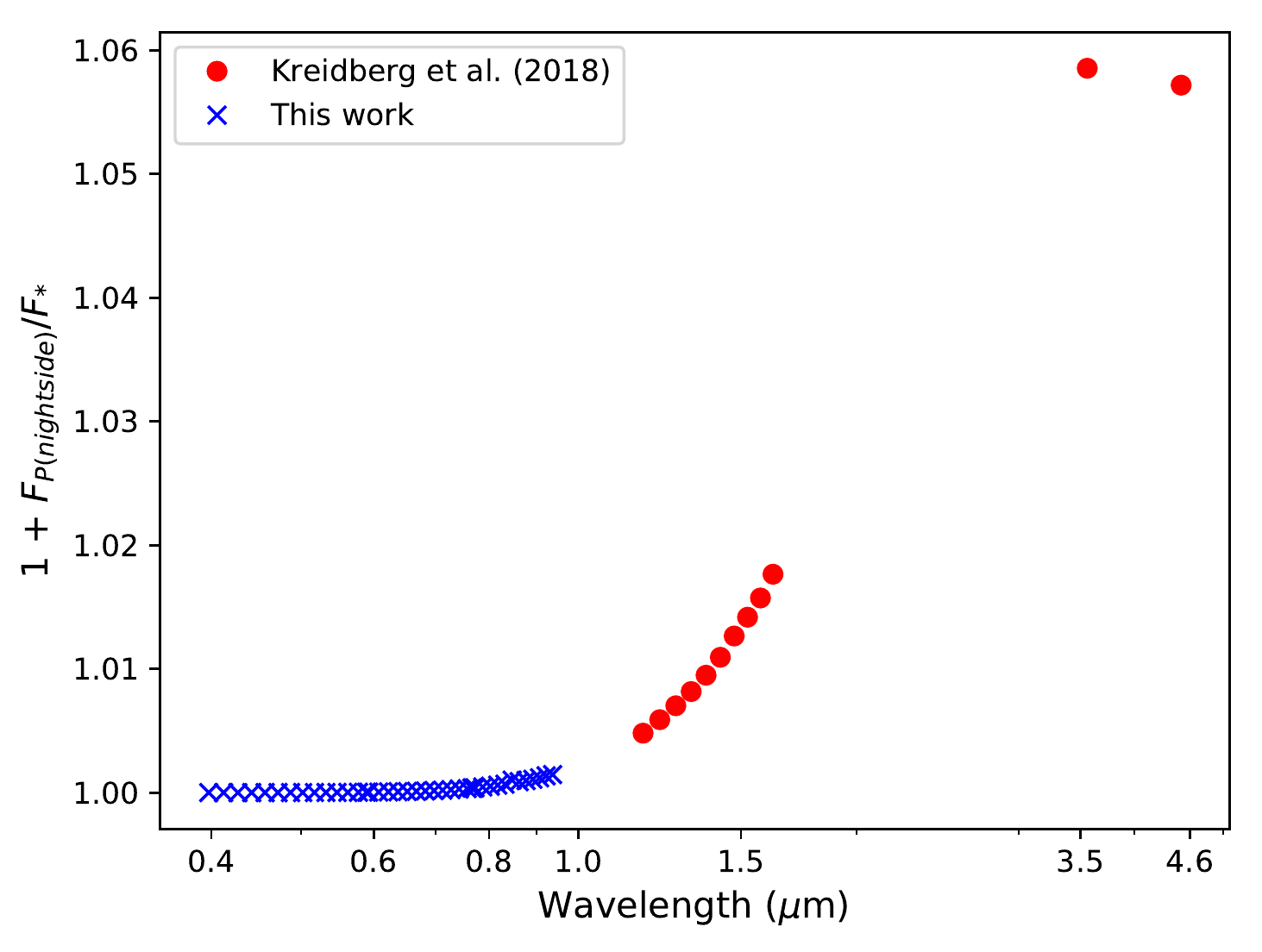}
    \caption{The correction factors due to the emission from WASP-103b's nightside (blue crosses). These were calculated using a 1700\,K blackbody for the planet's nightside. Also shown are the correction factors calculated by \protect\cite{Kreidberg2018} for their \textit{HST}/WFC3 and \textit{Spitzer}/IRAC wavelength bins (red circles).}
    \label{fig:nightside_corr}
\end{figure}

The nightside flux corrections were applied to our resulting transmission spectra using the following equation \cite[e.g.,][]{Kipping2010,Kreidberg2018_review}
\begin{equation}
\label{eq:ns}
    d_\mathrm{corr} = \frac{d_\mathrm{obs}}{1+F_{NS}(\lambda)/F_\mathrm{W103}(\lambda)},
\end{equation}
where $F_{NS}(\lambda)$ is the flux from the nightside of the planet. \autoref{fig:nightside_corr} shows our calculated correction factors (the denominator in \autoref{eq:ns}) along with those used by \cite{Kreidberg2018}. This figure demonstrates that the effect of the planet's nightside is significantly larger in the infrared as compared to the optical. In the wavelength range covered by our optical data, the maximum correction factor (corresponding to the reddest wavelength bin) leads to a 13\,ppm change in the transit depth. While we have applied this correction to our final transmission spectra, this correction is only 8\,\% of the typical error in the transit depths of our combined transmission spectrum and thus is negligible.

\section{Results}
\label{sec:results}

\subsection{White light fits}

The results of our fits to the white light curves are shown in \autoref{fig:wlc_fits}. We note that we also include the differential ACCESS light curves in \autoref{fig:wlc_fits}, which correspond to the target's light curve divided by the best comparisons' light curve, for ease of comparison with the other 6 transit light curves. However, we did not actually fit the differential ACCESS light curves but instead fit the raw stellar light curves as described in \autoref{sec:GPTransSpec}.

As the number of transits and wavelength bins resulted in over 322 spectroscopic light curves, we only include example fits to our ACCESS and LRG-BEASTS data in the Appendix (Figs.\ \ref{fig:wbfit_ACCESS_n2} and \ref{fig:wbfit_LRG-BEASTS_n1}). The rest of our light curve fits can be found online\footnote{\url{https://github.com/JamesKirk11/WASP103_lc_fits}}. 

\subsection{System parameters}
\label{sec:system_params}

\autoref{tab:r_system_parameters} lists the system parameters we derive from the `free system parameters' fits to our 11 white light curves. We note again that these results were not the ones used for our final transmission spectra, where we instead used the parameters of \cite{Southworth2015}, as used by \cite{Kreidberg2018}. 

\begin{figure}
    \centering
    \includegraphics[scale=0.55]{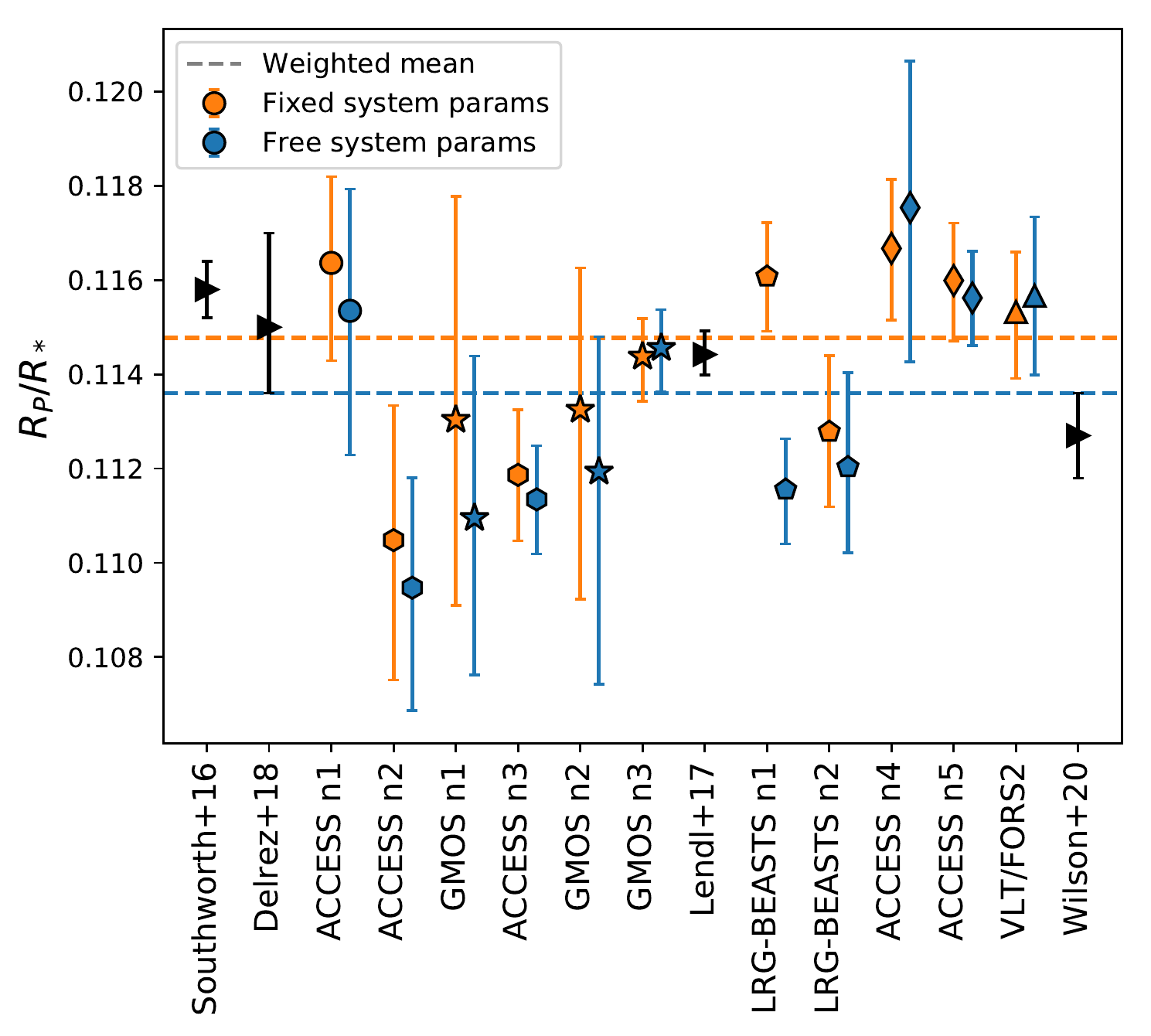}
    \caption{The reported white light $R_P/R_*$ values from literature studies (black triangles), along with our newly derived values when holding the remaining system parameters fixed to \protect\cite{Southworth2015} (orange points) and leaving these as free parameters (blue points). The plot symbols for our newly analyzed data are shared among transits that have a common wavelength range and comparison star. The dashed horizontal lines show the weighted means from our newly analyzed data, to which our transmission spectra (\autoref{fig:transmission_spectrum}) were normalized.}
    \label{fig:RpRs_variation}
\end{figure}

The $R_P/R_*$ values in \autoref{tab:r_system_parameters} are those after correcting for the third-light contamination and planetary nightside flux (\autoref{sec:3rd_light_corr}). This shows that while there is good overall agreement, there is scatter in each of these parameters. We plot the $R_P/R_*$ derived from each of our transits from both the `fixed' and `free' system parameters fits, along with the values of \cite{Southworth2016}, \cite{Lendl2017}, \cite{Delrez2018} and \cite{Wilson2020} in \autoref{fig:RpRs_variation}. This demonstrates that while there is some variation in $R_P/R_*$, each of our 11 transits are within 1--2$\sigma$ of the weighted mean. We note that ACCESS n5 and VLT/FORS2 are observations of the same transit epoch. 

\begin{table*}
\centering
\caption{The resulting system parameters from our white light curve fits with these as free parameters. In all cases we held the period fixed to 0.9255456\,d, as found by \protect\cite{Southworth2015} and used by \protect\cite{Kreidberg2018}. We note that these parameters were not used in the fitting of our spectroscopic light curves and hence our transmission spectrum. We also include the weighted mean of our results, labelled `Combined', and the results of \protect\cite{Southworth2015} and \protect\cite{Delrez2018} for comparison.}
\label{tab:r_system_parameters}
\begin{tabular}{lcccc} \hline
Data set & $T_C$ (BJD$_\mathrm{TDB}$) & $a/R_*$ & $i$ & $R_{P}/R_*$ \\ \hline\hline

ACCESS n1 &  $2457178.747991^{+0.000215}_{-0.000248}$ & $2.97^{+0.06}_{-0.06}$ & $85.14^{+1.8}_{-1.44}$ & $0.11535^{+0.00258}_{-0.00306}$
\\
ACCESS n2 & $2457179.673866^{+0.000123}_{-0.000128}$ & $2.97^{+0.04}_{-0.05}$ & $86.2^{+1.89}_{-1.5}$ & $0.10947^{+0.00234}_{-0.00261}$
 \\
ACCESS n3 & $2457205.588929^{+0.000119}_{-0.000108}$ & $3.03^{+0.01}_{-0.01}$ & $87.26^{+0.23}_{-0.23}$ & $0.11135^{+0.00114}_{-0.00116}$
 \\
ACCESS n4  & $2457848.843293^{+0.000191}_{-0.000164}$ & $2.96^{+0.04}_{-0.05}$ & $85.4^{+2.44}_{-1.61}$ & $0.11754^{+0.00311}_{-0.00327}$
 \\
ACCESS n5 & $2457874.758341^{+0.000087}_{-0.000089}$ & $3.00^{+0.01}_{-0.02}$ & $87.86^{+1.11}_{-1.32}$ & $0.11562^{+0.00099}_{-0.00101}$
 \\ 
LRG-BEASTS n1 & $2457541.562056^{+0.000136}_{-0.000129}$ & $3.02^{+0.03}_{-0.05}$ & $87.08^{+1.87}_{-1.83}$ & $0.11155^{+0.00108}_{-0.00115}$
 \\
LRG-BEASTS n2  & $2457566.551617^{+0.000117}_{-0.000106}$ & $3.01^{+0.02}_{-0.04}$ & $87.71^{+1.52}_{-1.82}$ & $0.11203^{+0.002}_{-0.00182}$
\\
FORS2 &  $2457874.758372^{+0.000310}_{-0.000343}$ & $2.97^{+0.05}_{-0.07}$ & $86.25^{+2.43}_{-2.2}$ & $0.11567^{+0.00167}_{-0.00168}$
\\
GMOS n1 & $2457200.961020^{+0.000325}_{-0.000372}$ & $2.93^{+0.07}_{-0.07}$ & $84.15^{+1.82}_{-1.39}$ & $0.11095^{+0.00344}_{-0.00333}$
\\
GMOS n2 & $2457213.918548^{+0.000348}_{-0.000351}$ & $2.82^{+0.06}_{-0.06}$ & $82.13^{+1.22}_{-1.08}$ & $0.11194^{+0.00287}_{-0.00451}$
\\
GMOS n3 & $2457511.944288^{+0.000085}_{-0.000091}$ & $3.00^{+0.01}_{-0.03}$ & $88.08^{+1.31}_{-1.51}$ & $0.11456^{+0.00081}_{-0.00094}$
\\ \hline
\textbf{Combined} & $2456836.296374 \pm 0.000040$ & $3.01 \pm 0.01$ & $87.0 \pm 0.21$ & $0.1136 \pm 0.00045$
\\
\protect\cite{Southworth2015} & $2456836.296445 \pm 0.000055$ & $2.999 \pm 0.031$  & $87.3 \pm 1.2$ & $0.1127 \pm 0.0009$\\
\protect\cite{Delrez2018} & $2456836.296427 \pm 0.000063$ & $3.010^{+0.008}_{-0.013}$ & $88.8^{+0.8}_{-1.1}$ &  $0.1150^{+0.0020}_{-0.0014}$ 

\\ \hline

\end{tabular}
\end{table*}

\subsection{Transmission spectrum}
\label{sec:r_transmission_spectra}

\begin{figure*}
\centering
\includegraphics[scale=0.45]{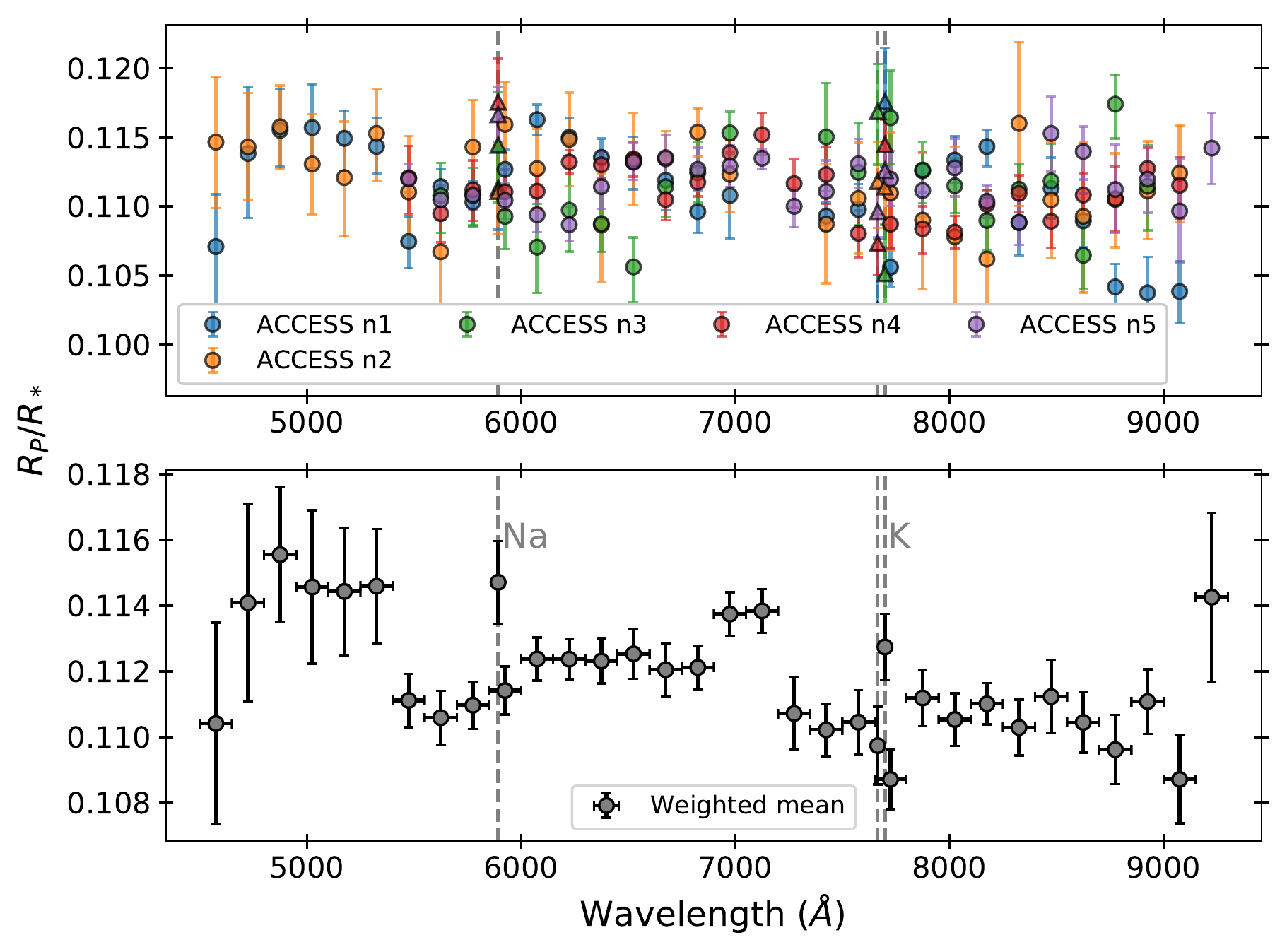}
\includegraphics[scale=0.45]{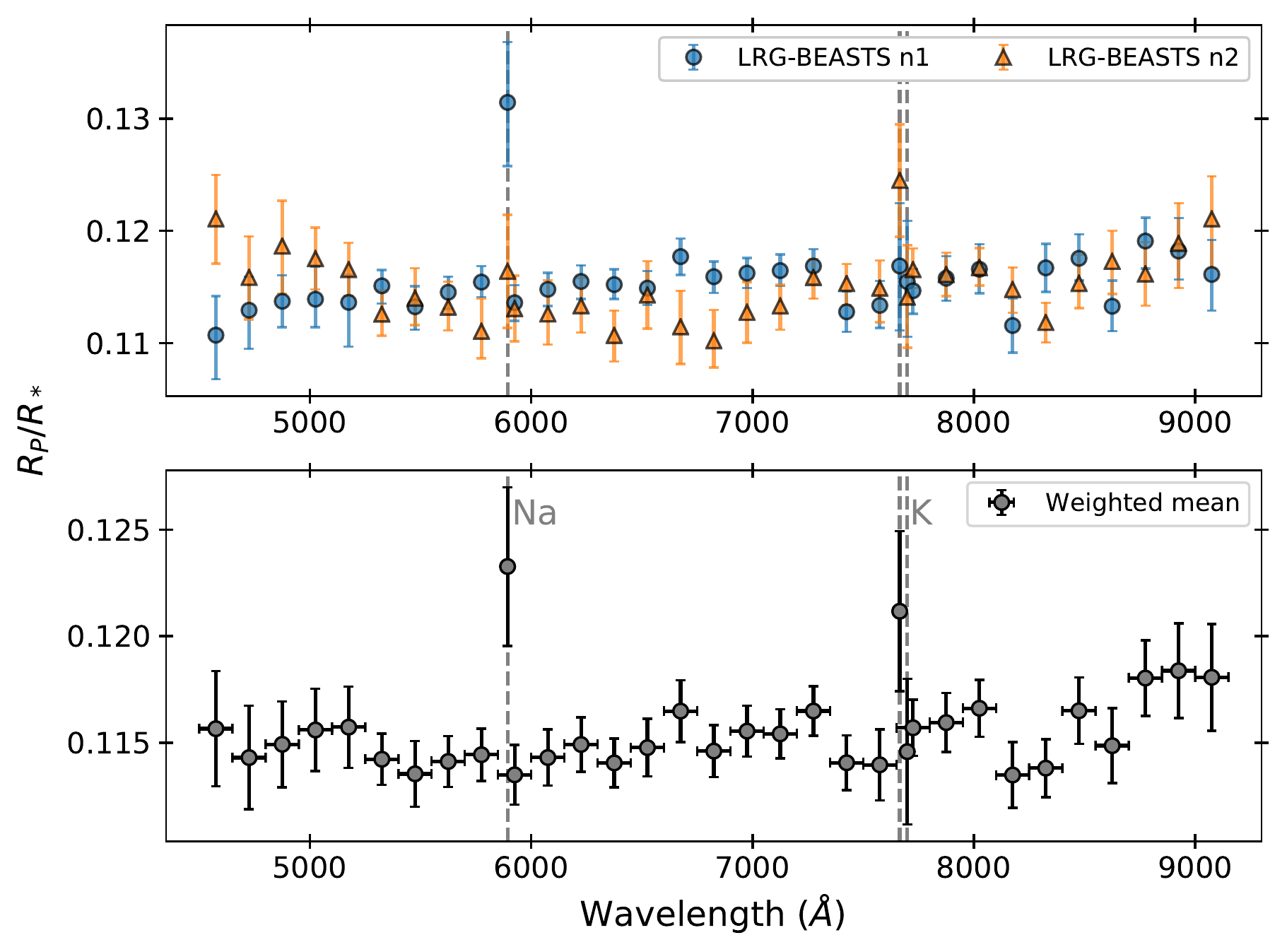}
\includegraphics[scale=0.45]{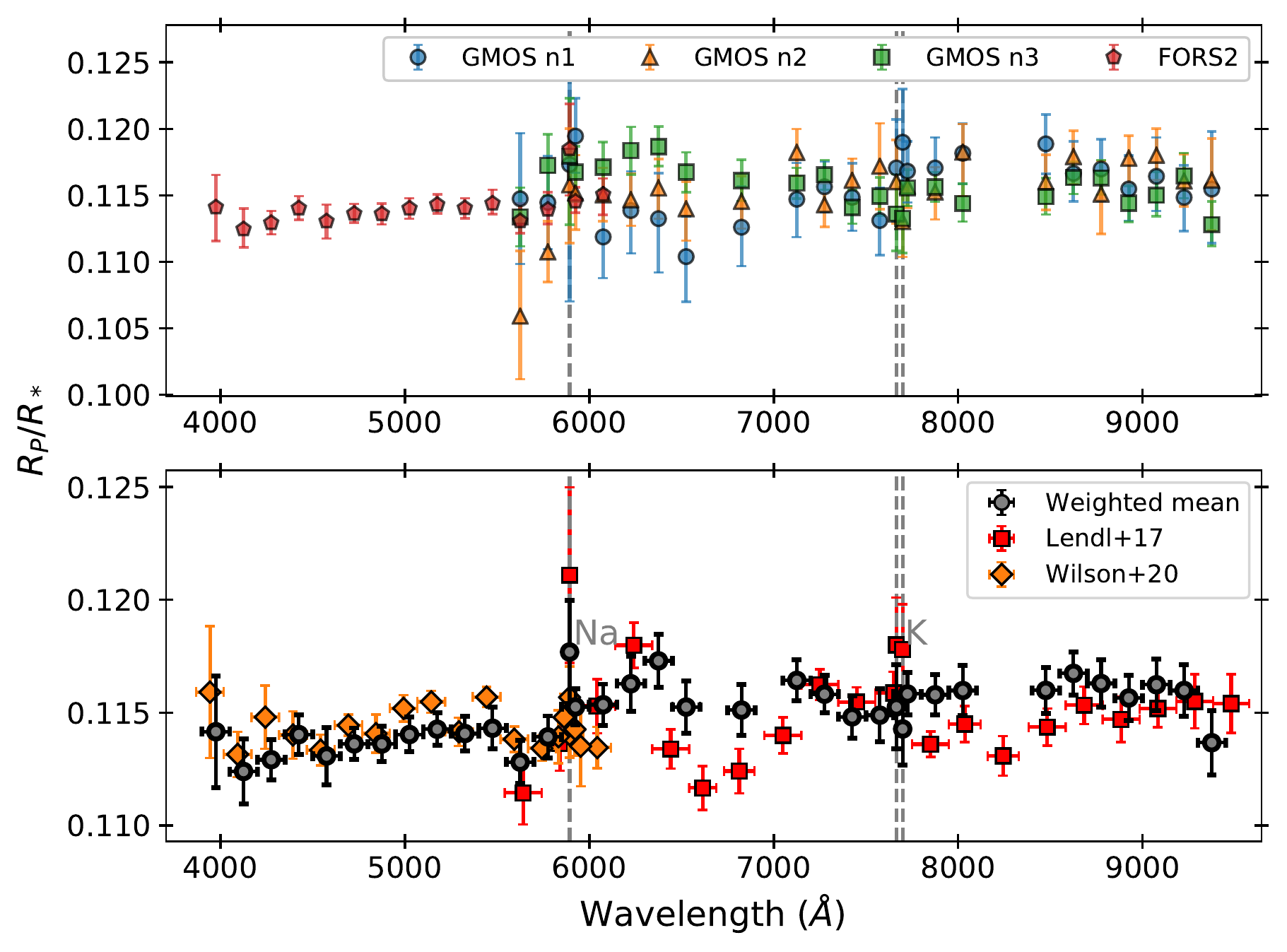}
\caption{The individual transmission spectra from all 11 transit observations are shown in the upper panels of all 3 figures. The lower panels show the weighted mean of the 5 ACCESS transits (upper left figure), 2 LRG-BEASTS transits (upper right figure), and the 1 FORS2 plus 3 GMOS transits (lower center figure). In the lower figure, we also include the Gemini/GMOS transmission spectrum as presented in \protect\cite{Lendl2017} and the VLT/FORS2 transmission spectrum as presented in \protect\cite{Wilson2020}.}
\label{fig:all_transmission_spectra}
\end{figure*}

In \autoref{fig:all_transmission_spectra}, we show the transmission spectra resulting from all 11 of our ground-based transits, following the correction for the third-light contamination and planetary nightside flux. The nightly transmission spectra in the top panels of \autoref{fig:all_transmission_spectra} have had offsets applied to ensure they have the same nightly mean $R_P/R_*$ as the weighted mean $R_P/R_*$ from all 11 nights, as discussed in \autoref{sec:system_params}. The nightly transmission spectra for all instruments are given in Tables \ref{tab:ts_ACCESS} and \ref{tab:ts_LRG-BEASTS} in the Appendix.

\autoref{fig:all_transmission_spectra} shows there is good overall agreement between each of the transits' transmission spectra. The most notable difference comes in the LRG-BEASTS transmission spectra, where the spectra diverge at blue wavelengths (although they are still consistent within $1\sigma$, other than in the bluest bin). 

We also note that the Na signal is much deeper in the first LRG-BEASTS night. As an additional check, we performed light curve fits to five 30\,\AA-wide bins centered on Na to see whether this was caused by systematics related to the 20\,\AA\ bin being too narrow. In this case we found the central Na bin to remain deep but a neighbouring bin to be anomalously shallow, giving us reason to question the authenticity of the deep LRG-BEASTS n1 bin. This behaviour did not occur for any of the 11 other transits we analysed. We are unsure of the causes of this deep Na bin, however we tried different treatment of the stellar limb darkening (fixed vs.\ free coefficients), GP inputs and GP kernels (Mat\'{e}rn 3/2 vs.\ Squared Exponential), and the use of a flat field, and found these made no difference to the depth of the bin. 

Notably, however, our final 11 transit-combined transmission spectrum changes by $\ll$1$\sigma$ regardless of whether the second LRG-BEASTS night or the LRG-BEASTS Na result is included. This is the advantage of combining 11 data sets. As a result, we do not exclude the LRG-BEASTS data from our final transmission spectrum. 

In \autoref{fig:all_transmission_spectra}, we also plot the Gemini/GMOS and VLT/FORS2 transmission spectra as derived by \cite{Lendl2017} and \cite{Wilson2020}. This shows the good agreement between our re-analysis of these data sets and the previously published analyses. We believe that the small differences between our analysis of the Gemini/GMOS data as compared to the analysis of \cite{Lendl2017} could be due to the different system parameters used and different systematics modelling approaches used.

\begin{figure*}
    \centering
    \includegraphics[scale=0.9]{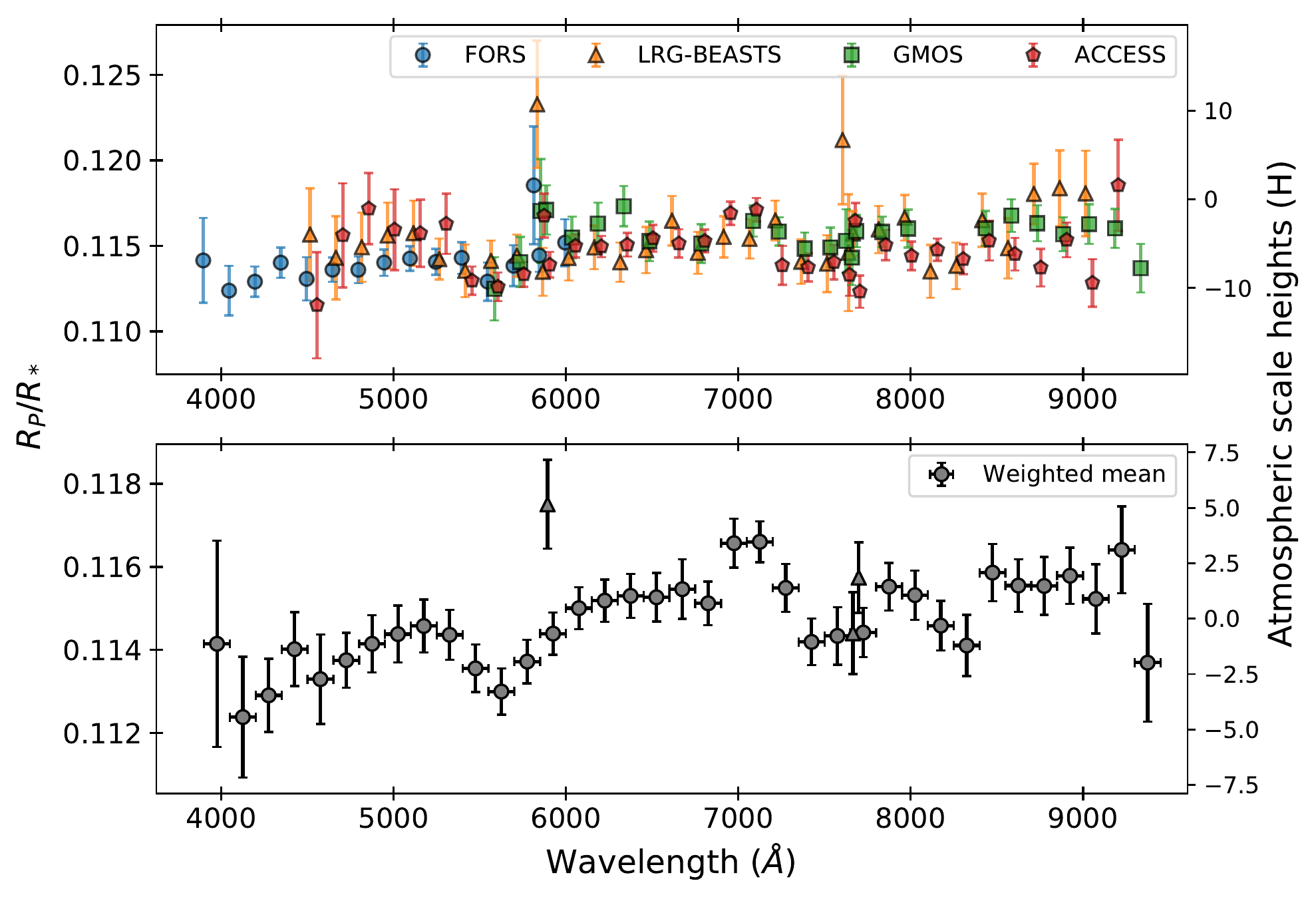}
    \caption{Top panel: the transmission spectra broken down into the component instruments. These are offset in wavelength by 20\,\AA\ for clarity. Bottom panel: the weighted mean of all 11 transmission spectra. The 20\,\AA-wide bins centered on Na and K are shown by the triangles.}
    \label{fig:transmission_spectrum}
\end{figure*}

In \autoref{fig:transmission_spectrum}, we show the weighted-mean transmission spectra for each instrument and the 11-transit weighted-mean combination. In this figure, all the spectra have been corrected for the third-light contamination as described in \autoref{sec:3rd_light_corr}. 

\autoref{fig:transmission_spectrum} shows an excellent agreement between all four instruments. This also highlights the precision we have achieved in our combined transmission spectrum with a median uncertainty on the transit depth of 148\,ppm. The transmission spectrum shows considerable structure, with an upwards slope from $\sim 4000$ to 5200\,\AA, before dropping towards 5600\,\AA\ and rising again towards 6200\,\AA. There is also a bump in the spectrum at around 7000\,\AA\ and a deeper transit centered on the Na doublet at 5893\,\AA, which we discuss in more detail in the following subsection.

\begin{table*}
\centering
\caption{The weighted mean transmission spectrum of WASP-103b from all 11 ground-based transit light curves. These results have been corrected for the third-light contamination and the planet's nightside flux. We note that we also include the third-light correction factors ($F_\mathrm{cont}/F_\mathrm{W103}$) in this table.}
\label{tab:transmission_spectrum}
\begin{tabular}{ccccc|ccccc}
\toprule
  Bin centre &  Bin width & $R_P/R_*$ & $\sigma(R_P/R_*)$ & $F_\mathrm{cont}/F_\mathrm{W103}$ & Bin centre &  Bin width & $R_P/R_*$ & $\sigma(R_P/R_*)$ & $F_\mathrm{cont}/F_\mathrm{W103}$ \\
  ($\AA$) & ($\AA$) & & & & ($\AA$) & ($\AA$) & & & \\ \hline
\midrule
 3975 &    150 &  0.11415 &  0.00248  & 0.0086  & 6825 &    150 &  0.11512 &  0.00052 & 0.0573  \\
 4125 &    150 &  0.11238 &  0.00145  & 0.0172 & 6975 &    150 &  0.11657 &  0.00059  & 0.0565 \\
 4275 &    150 &  0.11291 &  0.00088  & 0.0118 & 7125 &    150 &  0.11660 &  0.00050  & 0.0585 \\
 4425 &    150 &  0.11402 &  0.00089  & 0.0234 & 7275 &    150 &  0.11549 &  0.00058  & 0.0579 \\
 4575 &    150 &  0.11330 &  0.00108  & 0.0205 & 7425 &    150 &  0.11419 &  0.00056  & 0.0653 \\
 4725 &    150 &  0.11375 &  0.00066  & 0.0270 & 7575 &    150 &  0.11434 &  0.00069  & 0.0665 \\
 4875 &    150 &  0.11415 &  0.00069  & 0.0283 & 7665 &     20 &  0.11440 &  0.00098  & 0.0664 \\
 5025 &    150 &  0.11438 &  0.00068  & 0.0242 & 7699 &     20 &  0.11574 &  0.00085 & 0.0674 \\
 5175 &    150 &  0.11458 &  0.00063  & 0.0229 & 7725 &    150 &  0.11442 &  0.00059  & 0.0680 \\
 5325 &    150 &  0.11436 &  0.00060  & 0.0298 & 7875 &    150 &  0.11552 &  0.00057  & 0.0711 \\
 5475 &    150 &  0.11356 &  0.00057  & 0.0336 & 8025 &    150 &  0.11532 &  0.00059  & 0.0720 \\
 5625 &    150 &  0.11300 &  0.00056  & 0.0368 & 8175 &    150 &  0.11459 &  0.00060  & 0.0696 \\
 5775 &    150 &  0.11372 &  0.00052  & 0.0432 & 8325 &    150 &  0.11411 &  0.00074  & 0.0733 \\
 5893 &     20 &  0.11751 &  0.00107  & 0.0361 & 8475 &    150 &  0.11586 &  0.00069  & 0.0756 \\
 5925 &    150 &  0.11439 &  0.00051  & 0.0450 & 8625 &    150 &  0.11555 &  0.00063  & 0.0761 \\
 6075 &    150 &  0.11500 &  0.00050  & 0.0473 & 8775 &    150 &  0.11554 &  0.00070  & 0.0774 \\
 6225 &    150 &  0.11519 &  0.00051  & 0.0467 & 8925 &    150 &  0.11579 &  0.00068  & 0.0795 \\
 6375 &    150 &  0.11530 &  0.00053  & 0.0497 & 9075 &    150 &  0.11523 &  0.00084  & 0.0786 \\
 6525 &    150 &  0.11527 &  0.00058  & 0.0524 & 9225 &    150 &  0.11641 &  0.00105  & 0.0778 \\
 6675 &    150 &  0.11546 &  0.00072  & 0.0560 & 9375 &    150 &  0.11369 &  0.00142  & 0.0804 \\
 \hline
\bottomrule
\end{tabular}
\end{table*}

\subsection{Significance of the Na bin}
\label{sec:Na}

In order to estimate the significance of the sodium detection in our combined transmission spectrum (\autoref{fig:transmission_spectrum}), we fitted a straight line across the neighbouring bins between 5625 and 6075\,\AA\ where the continuum is rising. This is the same process used to estimate the significance of sodium as used in, e.g., \cite{Nikolov2016,Carter2020} and \cite{Alderson2020}.  Comparing the difference between our Na bin and the fitted continuum, we find the Na bin is higher at $3.1\sigma$ confidence.

At face value this appears significant. However, when we exclude the LRG-BEASTS Na bin from our combined transmission spectrum, which we believe could be affected by systematics (\autoref{fig:all_transmission_spectra}), the significance of Na by this method drops to $2.8\sigma$. While the inclusion of the LRG-BEASTS Na bin changes the transmission spectrum by $< 1\sigma$, owing to the fact we are combining 11 data sets, we favour our conservative $2.8\sigma$ evidence for Na, considering to the possible systematics affecting the LRG-BEASTS Na bin. \cite{Wilson2020} detect Na at $2.0\sigma$ confidence while \cite{Lendl2017} find signs of strong Na absorption. Based on our analysis, we cannot claim a significant detection of Na in WASP-103b's atmosphere.

\subsection{Comparing the ACCESS \& LRG-BEASTS fitting pipelines}
\label{sec:wb_comparing_GPTS_to_gppm_fit}

\begin{figure}
    \centering
    \includegraphics[scale=0.47]{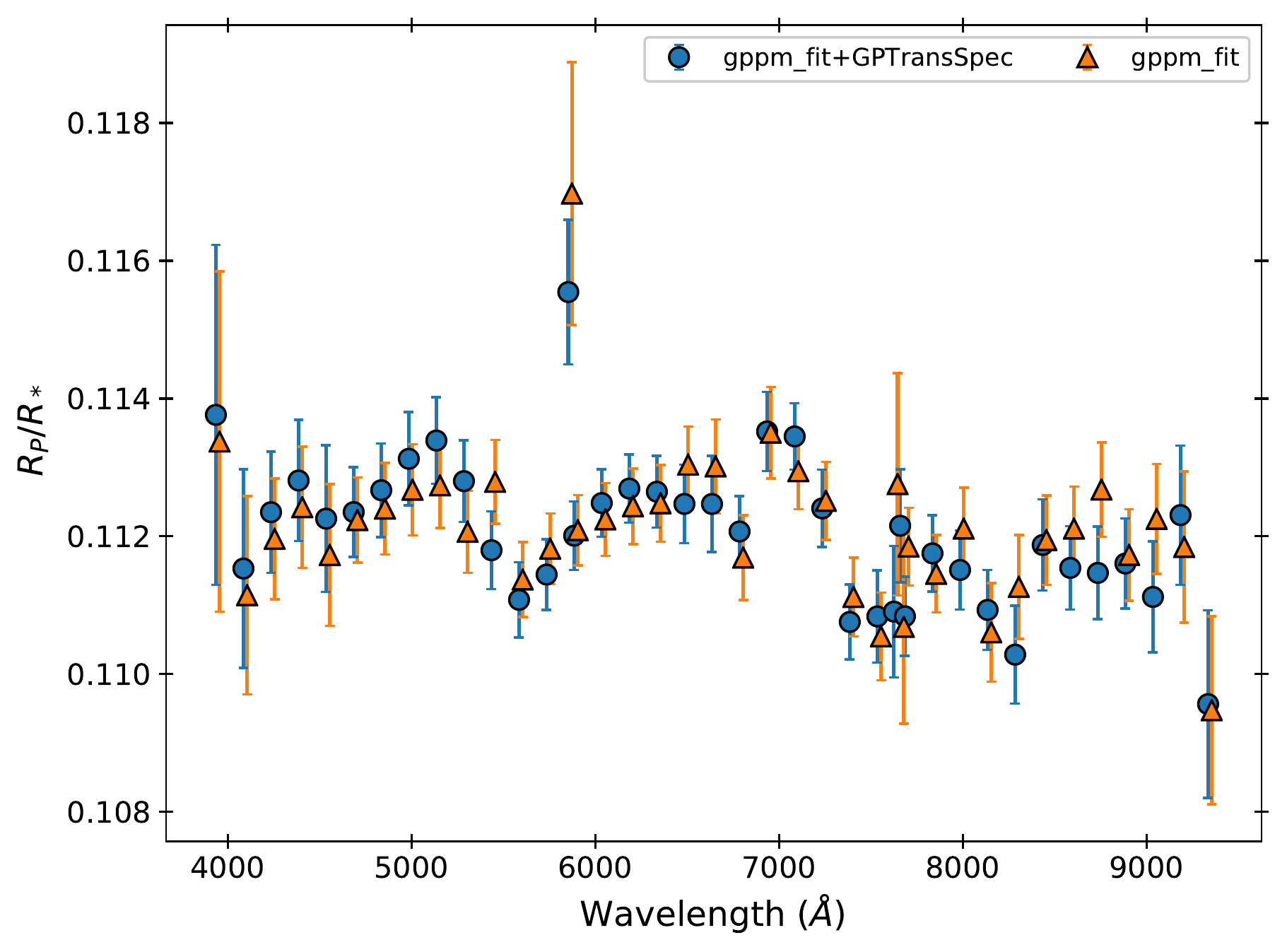}
    \caption{A comparison of the transmission spectra resulting from the use of \gppmfit{} only and \gppmfit{} + \texttt{GPTransSpec}. In orange are the results from fitting all 11 transits with \gppmfit{}. In blue are the results from fitting the 5 ACCESS transits with \texttt{GPTransSpec} and the remaining 6 transits with \gppmfit{}, as was used for our final analysis.}
    \label{fig:GPTS_gppm_fit}
\end{figure}

As an additional check of our results, we also fitted the ACCESS light curves with the LRG-BEASTS fitting code (\gppmfit{}, \autoref{sec:gppm_fit}). \autoref{fig:GPTS_gppm_fit} shows the comparison between our final transmission spectra from both methods, prior to applying the third-light correction. This shows the excellent agreement between the two fitting routines despite the different approaches to the stellar limb darkening, systematics models, sampling algorithm, and the priors on the transit and systematic models' parameters. \autoref{fig:GPTS_gppm_fit} demonstrates our results and conclusions are not dependent on the different fitting methods used. 

\subsection{Impact of third-light correction}

\begin{figure}
    \centering
    \includegraphics[scale=0.55]{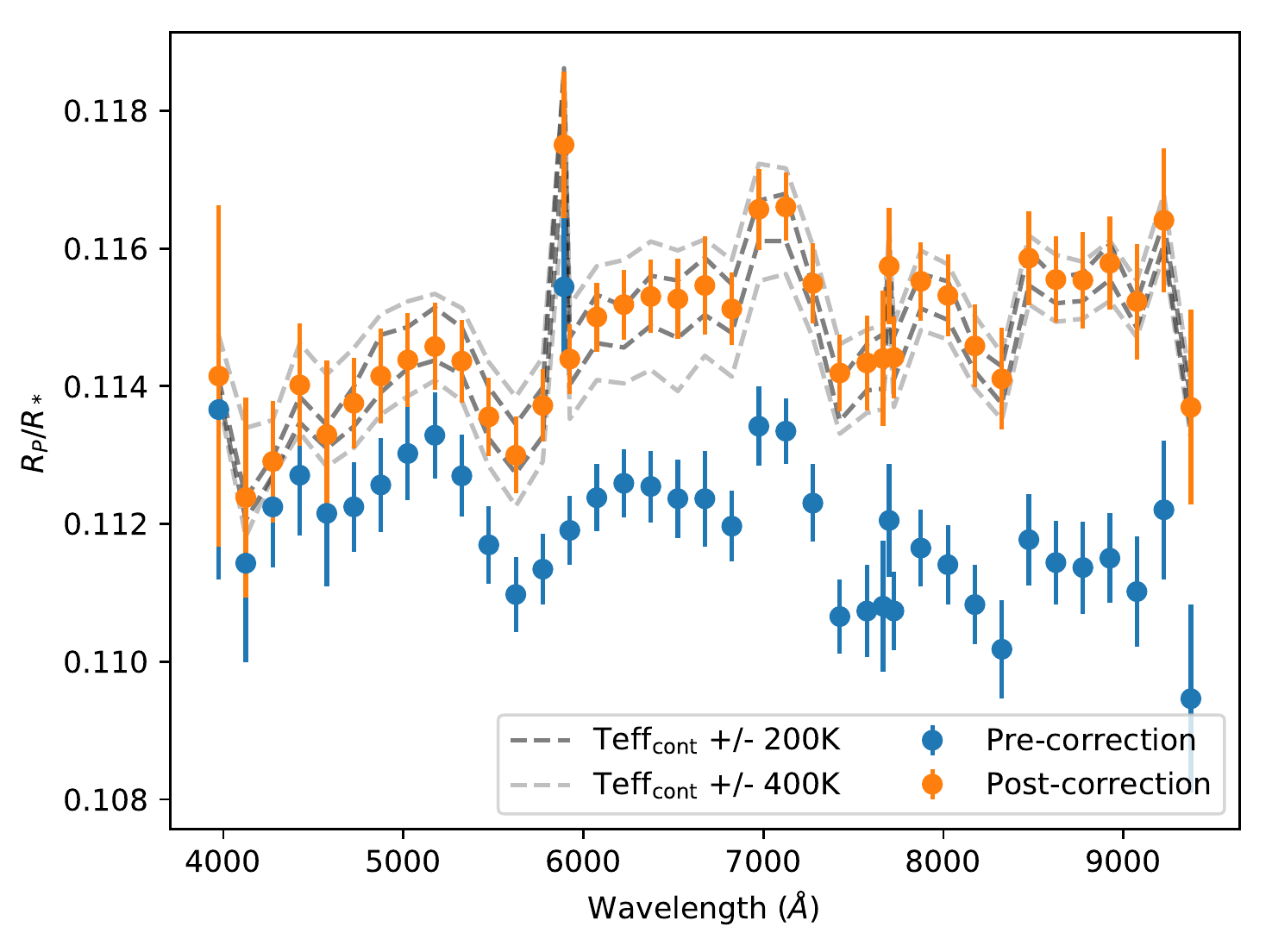}
    \caption{The effect of applying the third-light correction to our transmission spectrum, assuming a K5 contaminant with an effective temperature of $4400 \pm 200$\,K \protect\citep{Cartier2017}. The blue points correspond to the pre-correction data and the orange points to the post-correction data. The grey dashed lines show the 1 and $2\sigma$ uncertainties on the corrected spectrum taking the reported uncertainty in the contaminant's effective temperature (200\,K, \protect\citealt{Cartier2017}).}
    \label{fig:3rd_light_corr}
\end{figure}

As described in \autoref{sec:3rd_light_corr}, we had to correct our transmission spectrum for the flux of the blended contaminant ($4400 \pm 200$\,K, \citealt{Cartier2017}). In \autoref{fig:3rd_light_corr} we plot the pre- and post-third-light-corrected transmission spectra. We additionally include (in dotted lines), the impact of the contamination assuming the effective temperature of the contaminant $\pm$ the 1 and 2$\sigma$ uncertainties reported by \cite{Cartier2017}, which shows that the uncertainty in the contaminant does not result in a significant change in the transmission spectrum. This was also found by \cite{Lendl2017}.

\autoref{fig:3rd_light_corr} demonstrates that while the third-light correction has a significant impact over a broad wavelength range, changing the transmission spectrum from gently sloping down towards red wavelengths to sloping upwards, it has a minor impact on the more localised features, such as the dip at ${\sim}5600$\,\AA, the Na bin, and the bump at ${\sim}7000$\,\AA. We explore the impact this correction has on our retrievals in \autoref{sec:POSEIDON}.

\section{Forward models \& retrievals with petitRADTRANS}
\label{sec:pRT}

To interpret WASP-103b's transmission spectrum, we used both forward models and retrieval analyses using two seperate codes, firstly using petitRADTRANS \citep{Molliere2019}.

Given the significant structure seen in WASP-103b's transmission spectrum, its hot equilibrium temperature, and the evidence for a thermal inversion in its atmosphere, we generated forward models using petitRADTRANS that included only TiO, VO, and FeH as optical opacity sources. We assumed a cloud-free isothermal atmosphere at the equilibrium temperature of the planet (2489\,K), and took the planet's surface gravity and stellar radius from \cite{Delrez2018}. For these exploratory forward models, we assumed atmospheric mass fractions for TiO, VO, and FeH of 0.1\,\%. Importantly, as we discuss later, these initial TiO and VO forward models use the \cite{Plez1998,Plez1999} line lists as is the default for petitRADTRANS.

\begin{figure*}
    \centering
    \includegraphics[scale=0.8]{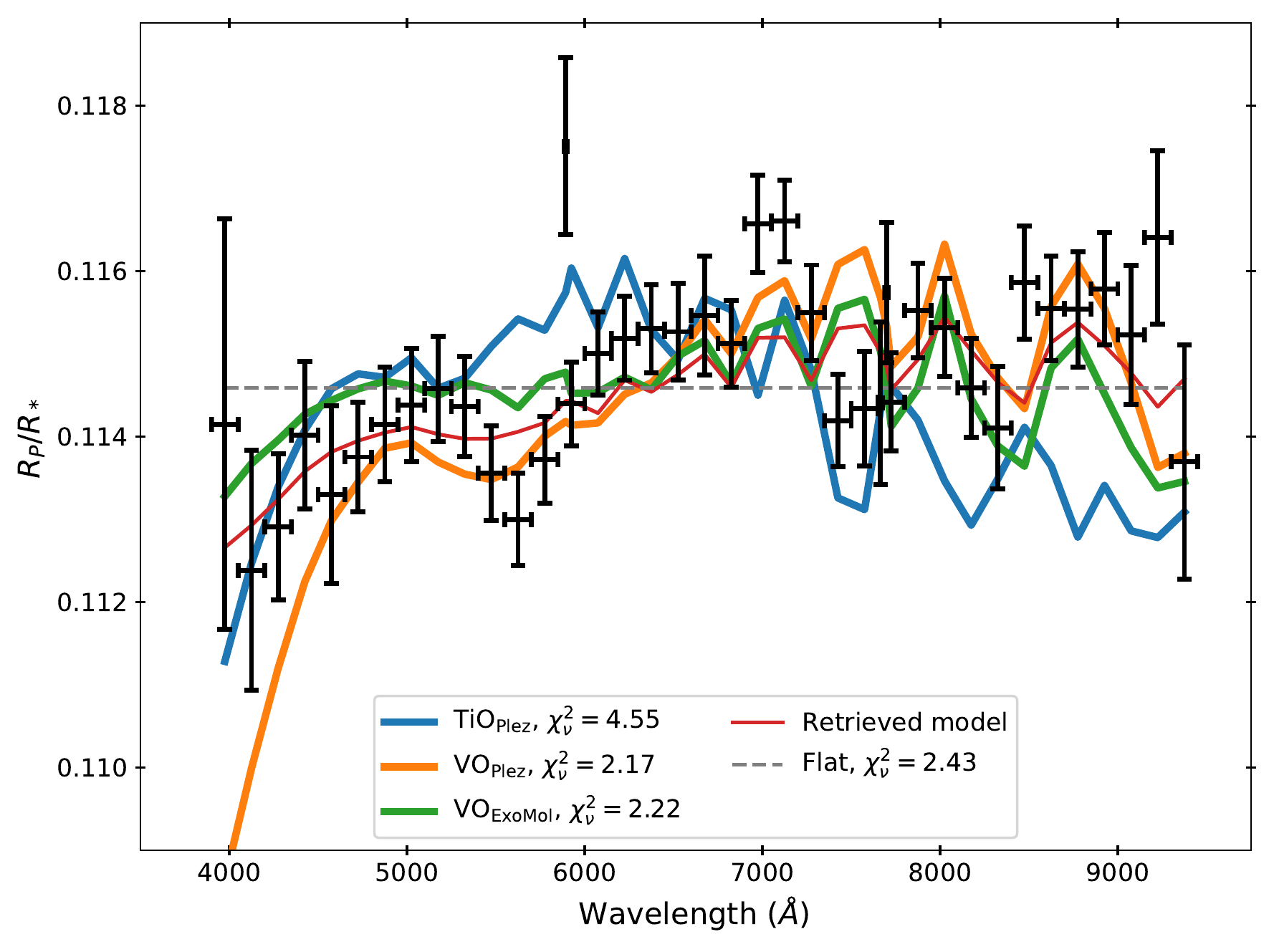}
    \caption{The forward models generated using petitRADTRANS, along with the reduced $\chi^{2}$ for each with 39 degrees of freedom. We include two VO-only forward models, one using the \protect\cite{Plez1999} line list (orange) and one with the ExoMol line list \protect\citep{McKemmish2016} (green). The VO and TiO-only models have atmospheric mass fractions of 0.1\,\%. The retrieved model is shown in red and uses the \protect\cite{Plez1999} line list. This finds a VO abundance of $-5.65^{+0.46}_{-0.64}$\,dex ($\sim10^4 \times$ solar) and no evidence for TiO.}
    \label{fig:pRT_models}
\end{figure*}

The forward models of TiO and VO are shown in \autoref{fig:pRT_models} along with a flat line. We do not include the FeH model, owing to the poor fit to the shape of the spectrum. These exploratory models show that, while none of these models provide an acceptable $\chi^2$, both TiO and VO can fit the downturn in the transmission spectrum we observe in the blue, while only VO is able to replicate the downturn seen at ${\sim}5600$\,\AA. 

In order to test this hypothesis further, we also ran a retrieval using petitRADTRANS. We used an MCMC, through \texttt{emcee} \citep{emcee}, to sample the parameter space with 240 walkers, each with 4200 steps. We fitted for the abundances of CO, H$_2$O, CH$_4$, NH$_3$, CO$_2$, H$_2$S, Na, K, TiO, VO, and FeH. We also fitted for the surface gravity and reference pressure. Following the petitRADTRANS retrieval example\footnote{\url{https://petitradtrans.readthedocs.io}}, we also fitted for 6 parameters that defined the temperature-pressure profile, although we note that these were poorly constrained. The abundances of IR species were also poorly constrained. However, the retrievals did constrain the volume mixing ratio (VMR) of VO to be $-5.65^{+0.46}_{-0.64}$\,dex, which is about $10^{4} \times$ the solar abundance \citep{Woitke2018}. The TiO abundance was unconstrained, instead bumping up against the lower boundary of the uniform-in-log-space prior ($-10$ dex). This agreed with our forward model analysis that the transmission spectrum is better matched with VO than TiO, when using the Plez line lists. The retrieved model using the Plez line lists is also shown in \autoref{fig:pRT_models}.

Motivated by our retrievals run with POSEIDON (\autoref{sec:POSEIDON}), we also generated a VO forward model with petitRADTRANS but using the updated VO line list from the ExoMol group \citep{McKemmish2016}. This model is shown in green in \autoref{fig:pRT_models}. The difference between the orange and green models in \autoref{fig:pRT_models} is just the choice of VO line list, all other parameters were equal. This demonstrates how the choice of VO line list used makes a significant difference to the gradient of the slope in the blue and, to a lesser extent, the shallower transit depths seen around 5600\,\AA. However, we note that the goodness of fit is unaffected by the choice of line list.

\section{Retrievals with POSEIDON}
\label{sec:POSEIDON}

We additionally conducted a retrieval analysis of WASP-103b's transmission spectrum with the POSEIDON atmospheric retrieval code \citep{MacDonald2017}. While our petitRADTRANS retrieval (\autoref{sec:pRT}) focused on our optical transmission spectrum, our POSEIDON analysis further considers the addition of the near-infrared \textit{HST}/WFC3 and \textit{Spitzer}/IRAC observations from \citet{Kreidberg2018}. Our POSEIDON retrievals also include a prescription for unocculted stellar heterogeneity, which is a potential source of bias in exoplanet transmission spectra \citep[e.g.,][]{McCullough2014,Oshagh2014,Rackham2017,Rackham2018,Rackham2019}.

Our POSEIDON retrievals consider a wide range of atmospheric and stellar properties. We include the following chemical species with both prominent absorption features over our wavelength range and an expectation of occurrence in hot giant planet atmospheres \citep[e.g.][]{Sharp2007,Madhusudhan2016}: Na, K, H${-}$, TiO, VO, AlO, CrH, FeH, H$_2$O, CO, CO$_2$, and HCN. In our initial exploratory retrievals, we also included opacity from Li, Fe, Fe${+}$, Ti, Ti${+}$, H$_3{+}$, SiO, TiH, SiH, CH$_4$, NH$_3$, N$_2$O, NO$_2$, and NO but found these to be unconstrained and so did not include them in our final analysis.

Besides the line absorption from these species, POSEIDON also includes collision-induced absorption from H$_2$-H$_2$ and H$_2$-He pairs and Rayleigh scattering. We allow for inhomogenous clouds and hazes following the prescription in \citet{MacDonald2017}. We assume an isothermal temperature profile, given the quality of the present observations \citep{Rocchetto2016}. Our treatment of unocculted starspots or faculae is based on the parametrization of \citet{Pinhas2018}. In short, the heterogeneity `contamination factor' \citep{Rackham2017} is computed by interpolating stellar models from the Castelli-Kurucz 2004 atlas \citep{Castelli2003} using the pysynphot package \citep{pysynphot}. 

In total, our POSEIDON retrievals have a maximum of 22 free parameters. The planetary atmosphere is defined by the volume mixing ratios ($X_i$) of the 12 chemical species listed above, the terminator temperature ($T$), a reference radius at 10\,bar ($R_{\rm{p, \, 10\,bar}}$), and the four-parameter cloud and hazes prescription from \cite{MacDonald2017}. The stellar heterogeneity contribution is described by the covering fraction of the active regions on the stellar disc ($f_{\rm{het}}$) and the temperatures of the active regions ($T_{*, \, \rm{het}}$) and stellar photosphere ($T_{*, \, \rm{phot}}$), respectively. For retrievals including the data from \citet{Kreidberg2018}, we opted to fit for a relative offset, $\delta_{\rm rel}$, between our optical transmission spectrum and the IR transmission spectrum (noting that instrumental offsets in the optical data were taken care of in \autoref{sec:system_params}). We ascribe uniform-in-the-logarithm priors on the mixing ratios ($10^{-16}$ to $10^{-1}$) and uniform priors on $T$ (400 to 3000\,K), $R_{\rm{p, \, 10 \, bar}}$ ($0.85 - 1.15\,R_p$), $f_{\rm{het}}$ (0 to 0.5), $T_{*, \, \rm{het}}$ (0.65 to 1.2 $T_{*, \, \rm{phot}, \rm{a priori}}$), and $\delta_{\rm rel}$ ($\pm$ 1000\,ppm). Since the stellar photosphere temperature is well-known \textit{a priori}, we place an informative Gaussian prior on $T_{*, \, \rm{phot}}$ (mean: 6110\,K; standard deviation: 160\,K - \citealp{Gillon2014}). The cloud and haze priors are as in \citet{MacDonald2017}.

We ran POSEIDON on four combinations of our new optical dataset and the IR transmission spectrum of \cite{Kreidberg2018}: the optical data alone following the application of the third-light correction (`Ground Corrected'); the optical data alone \textit{prior} to the application of the third-light correction (`Ground Uncorrected'); the HST and Spitzer data without the optical data; and the third-light-corrected optical data combined with the HST and Spitzer data. For each dataset, we conducted Bayesian parameter estimation and model comparisons via the nested sampling algorithm MultiNest \citep{Feroz2008,Feroz2009,Feroz2019,Buchner2014}, with 4,000 live points per retrieval. We report the results from our retrievals in \autoref{tab:POSEIDON_retrieval_results}. The main result is that the optical data reveals tentative evidence for TiO ($\sim 2\sigma$), while also showing strong evidence of contamination from unocculted stellar heterogeneities. We discuss these results in more detail in the following subsections.

\subsection{On the optical data alone}

Our retrieved spectra for both the third-light corrected and uncorrected optical data are shown in \autoref{fig:POSEIDON_retrieved_spectra} (top panels). Prior to the third-light correction (\autoref{sec:3rd_light_corr}), there is a gentle upwards slope towards blue wavelengths which is best fit by unocculted \textit{cool} active regions. Following the third-light correction, the spectral morphology changes into a steep downwards slope best explained by unocculted \textit{hot} active regions, which we interpret as faculae. Quantitatively, the evidence for stellar activity increases from 1.5 to $4.0\sigma$ following the third-light correction (see \autoref{tab:POSEIDON_retrieval_results}). Clearly the third-light correction makes a significant impact on the conclusions about stellar activity. 

We additionally infer tentative evidence for atmospheric TiO absorption from our optical data. This TiO inference persists for both the uncorrected and third-light corrected data, changing from 2.3 to $1.7\sigma$ following the third-light correction. The TiO mixing ratio, however, is poorly constrained from the optical data alone. No other chemical species are favored by our Bayesian model comparisons with POSEIDON, with $2\,\sigma$ upper limits on their mixing ratios reported in \autoref{tab:POSEIDON_retrieval_results}. The non-detection of Na, despite the 20-\AA bin centered on the Na D-lines being ${\sim}3\sigma$ deeper than the surrounding continuum (\autoref{sec:Na}), is likely due to the lack of pressure-broadened Na wings in the surrounding data points \citep[e.g.,][]{Nikolov2018,Alam2021}.

The tentative evidence for TiO from POSEIDON is an interesting contrast with our earlier petitRADTRANS retrieval (\autoref{sec:pRT}), which instead inferred VO without the need for TiO. To better understand this discrepancy, we ran additional retrievals with both codes varying properties such as the prior ranges on the VMRs and the top of atmosphere pressure. Ultimately this made little difference. The most notable difference we identified is that POSEIDON uses updated ExoMol line lists for VO and TiO, resulting in considerably different optical band shapes compared to the petitRADTRANS VO and TiO cross sections (see \autoref{fig:pRT_models}). However, as we discuss in \autoref{sec:d_optical_trans_spec}, we cannot be certain that this is the cause the disagreement. Indeed, our $4\sigma$ preference for unocculted faculae with POSEIDON (for the corrected data) automatically considered VO as an alternative explanation via the Bayesian evidence computations. Therefore, we believe that the inclusion of stellar heterogeneity in our POSEIDON retrievals is the more likely cause of the disagreement.

\subsection{On the IR data alone}

Our IR-only retrievals on the HST/WFC3 and Spitzer/IRAC data from \citet{Kreidberg2018} weakly favor unocculted faculae and H$_2$O. We find a lower detection significance for faculae than that from the optical-only corrected spectrum ($2.4\sigma$ vs. $4.0\sigma$). This is expected, since the influence of unocculted stellar heterogeneities is largest at short wavelengths \citep[e.g.,][]{Rackham2018, Rackham2019}. The retrieved heterogeneity parameters from the IR data are consistent with those from our `Ground Corrected' retrieval (\autoref{tab:POSEIDON_retrieval_results}), giving us added confidence in this interpretation. Our inference of H$_2$O is marginal at best ($1.7\sigma$) and notably less significant than that found by \citet{Wilson2020} ($4.0\sigma$). We believe this discrepancy is due to our consideration of stellar heterogeneities, which were not included in the retrieval analyses of \citet{Wilson2020}. Finally, we note that TiO is not inferred from our IR-only retrievals, though the upper limit on its abundance is consistent with the `Ground Corrected' retrieval. In all, these consistent findings between each portion of the transmission spectrum strengthen the conclusions derived from the combined optical and IR data set.

\subsection{On the combined optical and IR data}

\begin{figure*}[ht!]
    \centering
    \includegraphics[width=0.492\textwidth]{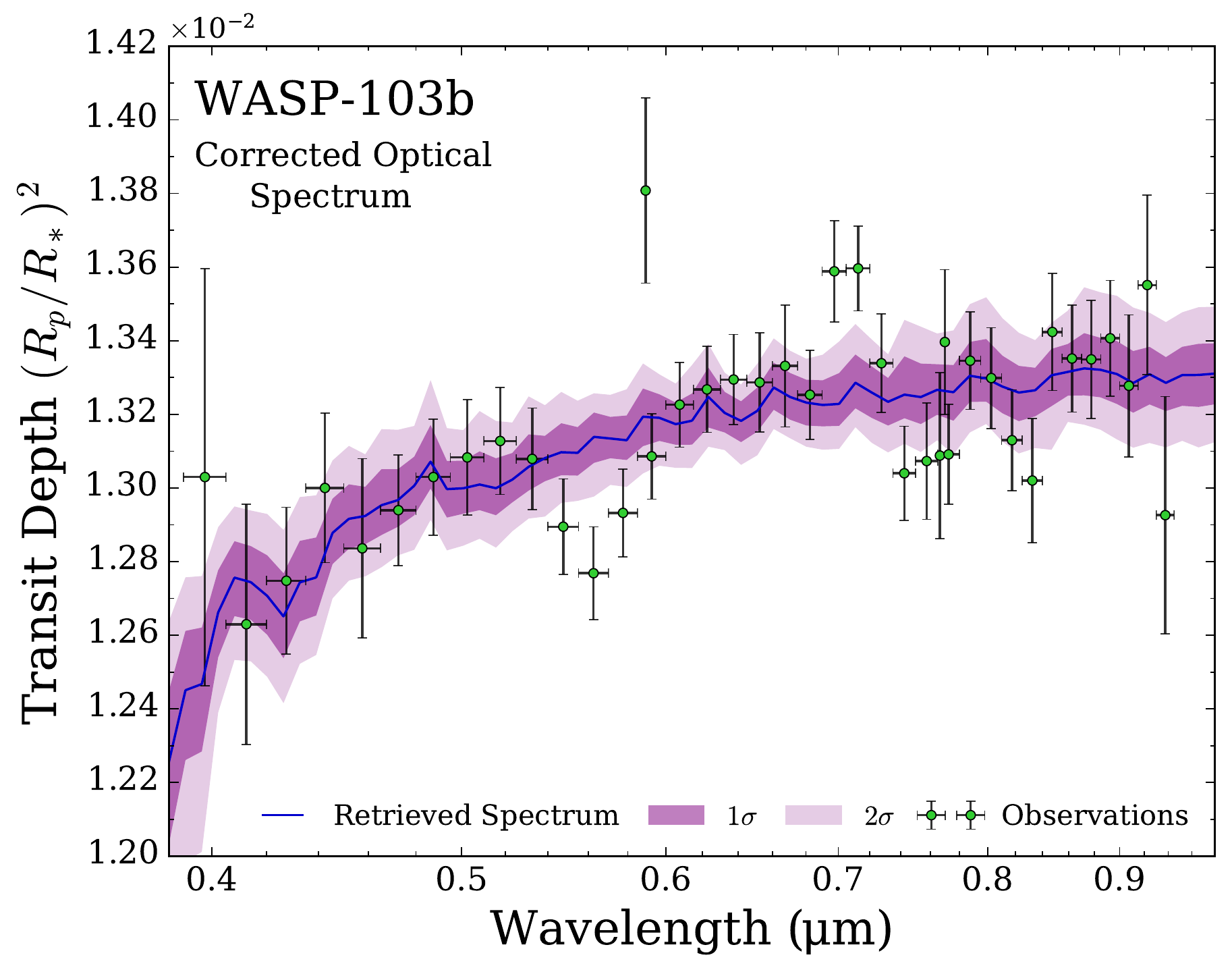}
    \includegraphics[width=0.492\textwidth, trim={0.0cm 0.0cm 0.0cm 0.0cm}]{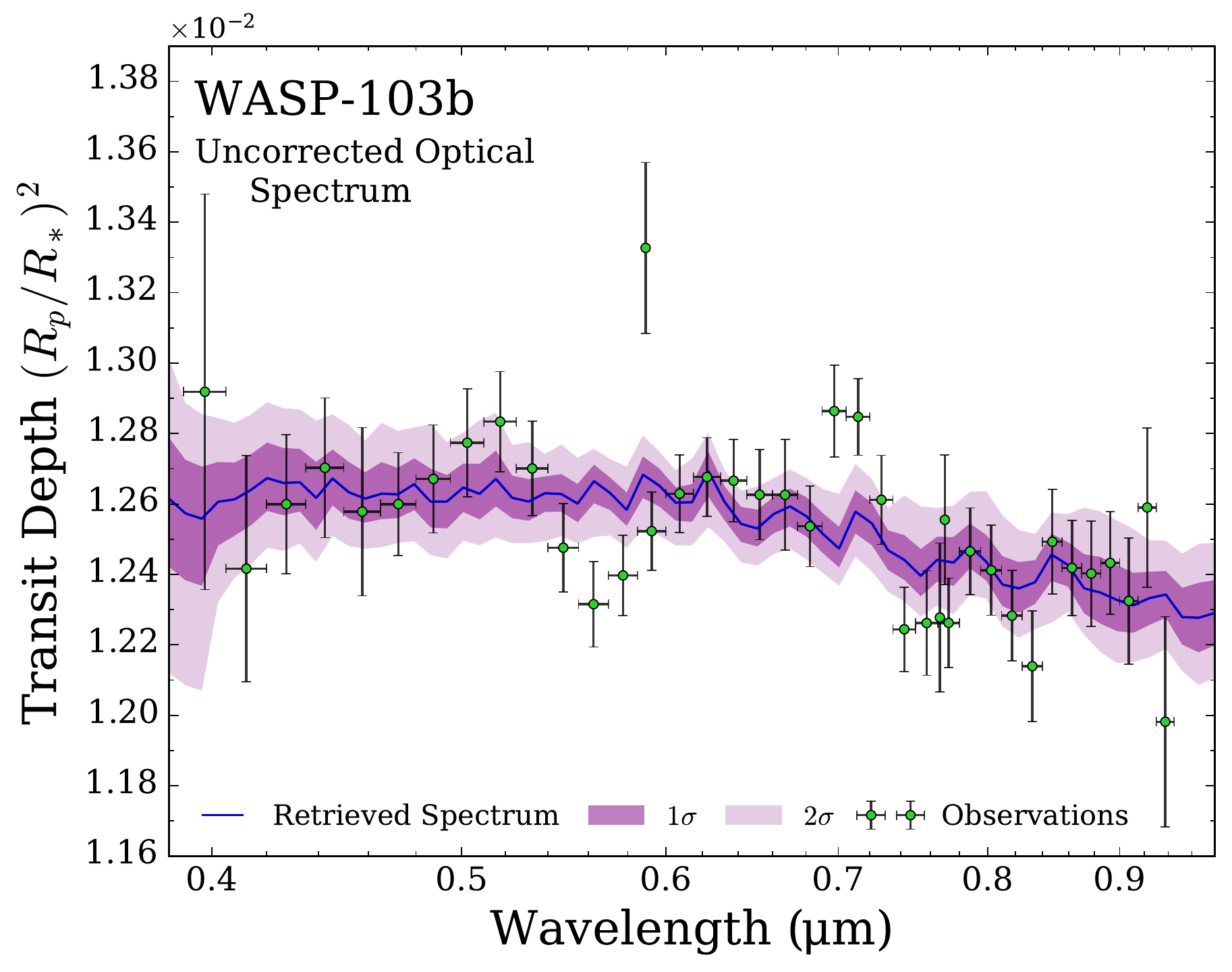}
    \includegraphics[width=\textwidth]{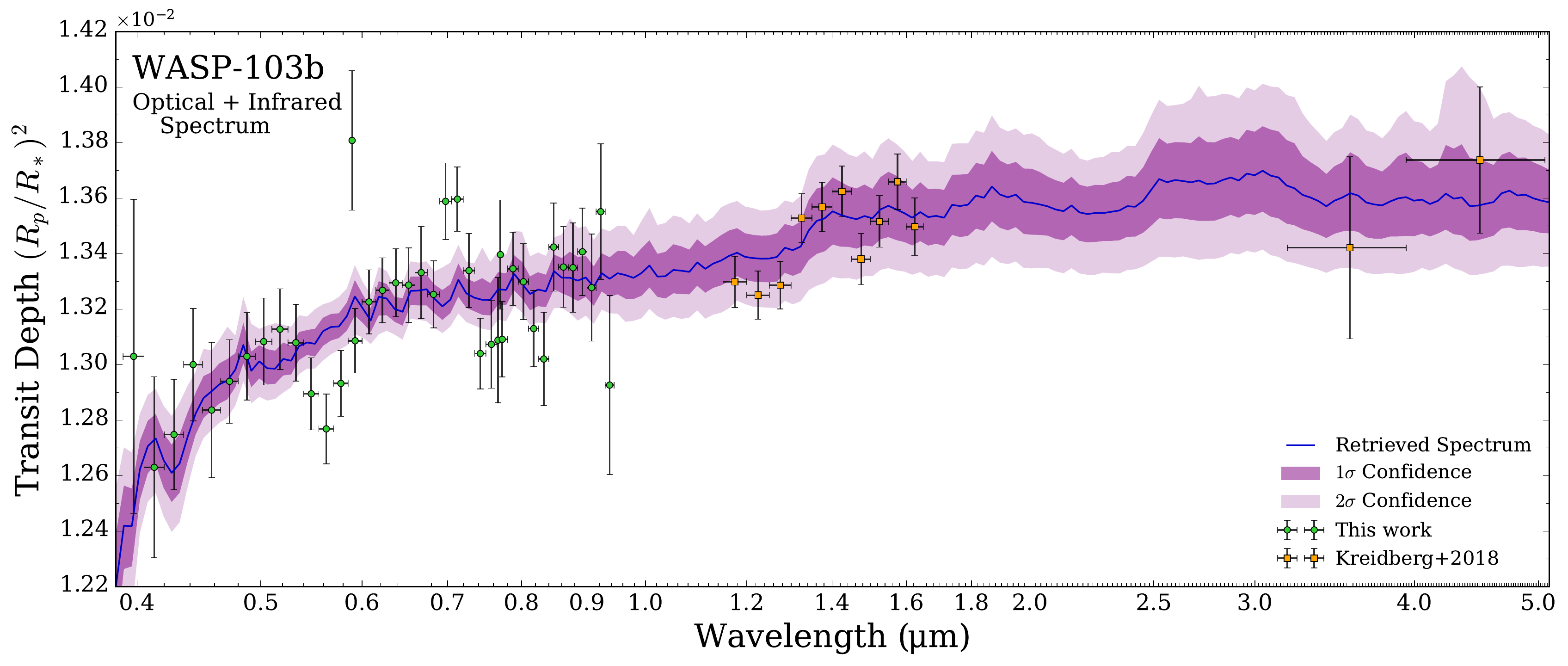}
    \caption{Retrieved transmission spectra of WASP-103b by POSEIDON. Each panel shows the median retrieved model (blue line) and associated $1\,\sigma$ and $2\,\sigma$ confidence regions (purple contours) from three separate retrievals of different datasets: our third-light and nightside corrected optical spectrum (top left); our uncorrected spectrum (top right); and the combination of our corrected optical spectrum with the WFC3 and Spitzer data from \citet{Kreidberg2018} (bottom). A 210\,ppm offset has been applied to the infrared data to represent the median retrieved offset between the ground and space-based observations. The corrected optical spectrum is best fit by unocculted stellar faculae with residual hints of atmospheric TiO. The uncorrected optical spectrum also contains hints of TiO, but with unocculted stellar spots rather than faculae. The full, corrected, optical + infrared spectrum of WASP-103b is best fit by unocculted faculae and hints of TiO, H$_2$O, and HCN.}
    \label{fig:POSEIDON_retrieved_spectra}
\end{figure*}

Our retrieval of the combined optical+IR data, shown in \autoref{fig:POSEIDON_retrieved_spectra} (bottom panel), identifies strong evidence of unocculted faculae ($4.3\sigma$) and tentative evidence for TiO ($2.1\sigma$), H$_2$O ($1.9\sigma$), and HCN ($1.7\sigma$). The inferred faculae are $\sim 350$\,K hotter than the stellar photosphere with a coverage fraction of $22^{+12}_{-9}$\,\%. The fitted offset between the optical and IR data is $210^{+83}_{-85}$\,ppm, consistent with the $1\sigma$ uncertainties in our optical transmission spectrum. Due to the weak evidence for individual molecules, we also derived the combined significance of TiO + H$_2$O + HCN from a Bayesian model comparison with all three species excluded. We find that the significance of \emph{at least one} of TiO, H$_2$O, or HCN, is $2.8\sigma$. Therefore, we conclude there is moderate evidence for molecular opacity from WASP-103b alongside the stellar faculae contamination. 

\begin{figure*}[ht!]
    \centering
    \includegraphics[width=\textwidth]{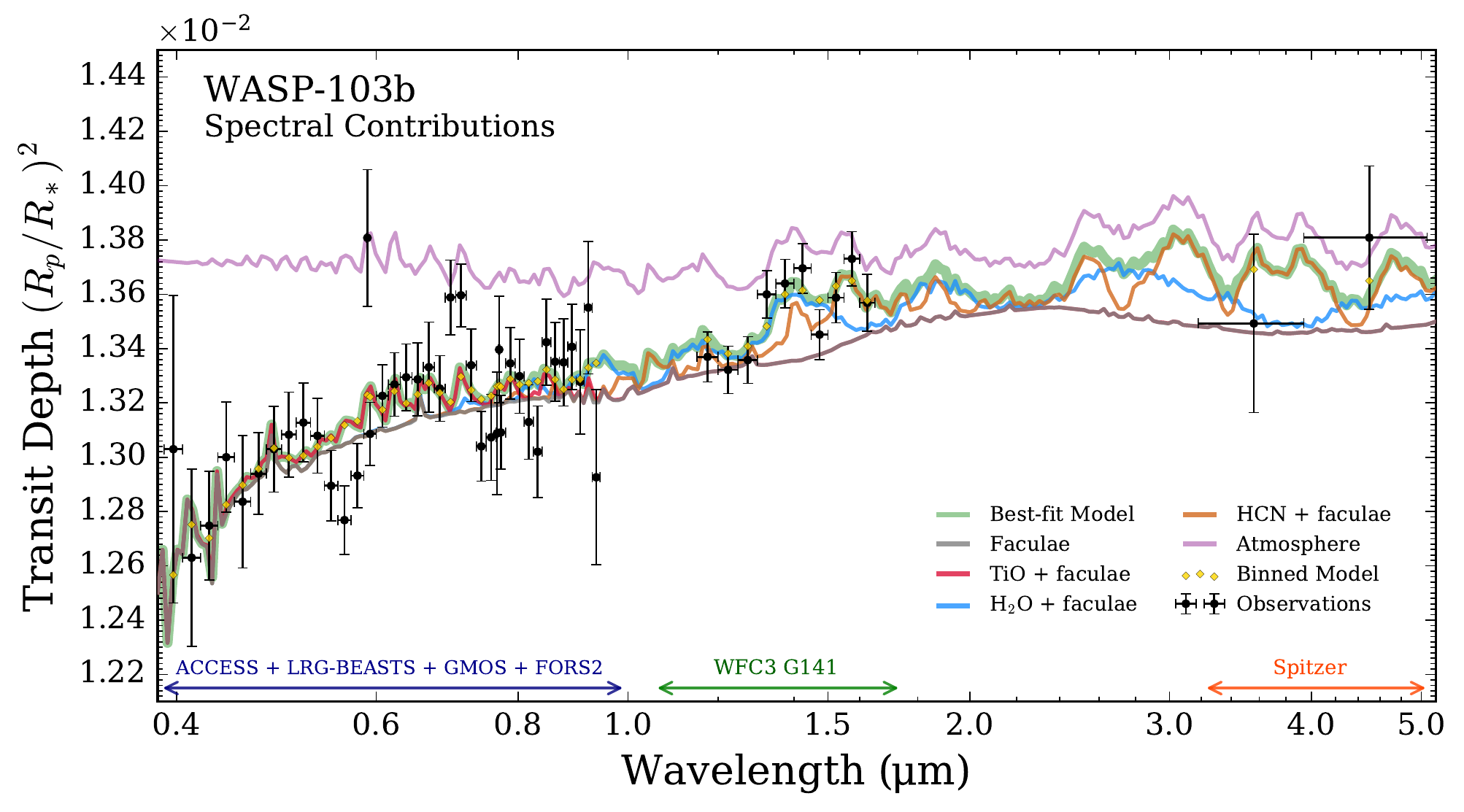}
    \caption{Spectral contributions to the best-fitting POSEIDON model of the transmission spectrum of WASP-103b. The maximum likelihood retrieved spectrum (green shading) is composed of contributions from unocculted faculae (grey), TiO (red), H$_2$O (blue), and HCN (orange). The equivalent model without faculae (purple) is shown for comparison. All models include background atmosphere H$_2$-H$_2$ collision-induced absorption (CIA) -- producing the broad continuum feature centered at 2.2\,$\micron$ visible in the faculae-only model. The instrument modes corresponding to each observed dataset are highlighted at the bottom. The best-fitting model, binned to the resolution of the data, is overlaid (gold diamonds). This `minimal' model achieves a reduced chi square of $\chi_{\nu}^2$ = 1.36 (for 43 degrees of freedom).}
    \label{fig:POSEIDON_best_fit_model}
\end{figure*}

\begin{figure*}[ht!]
    \centering
    \includegraphics[width=\textwidth]{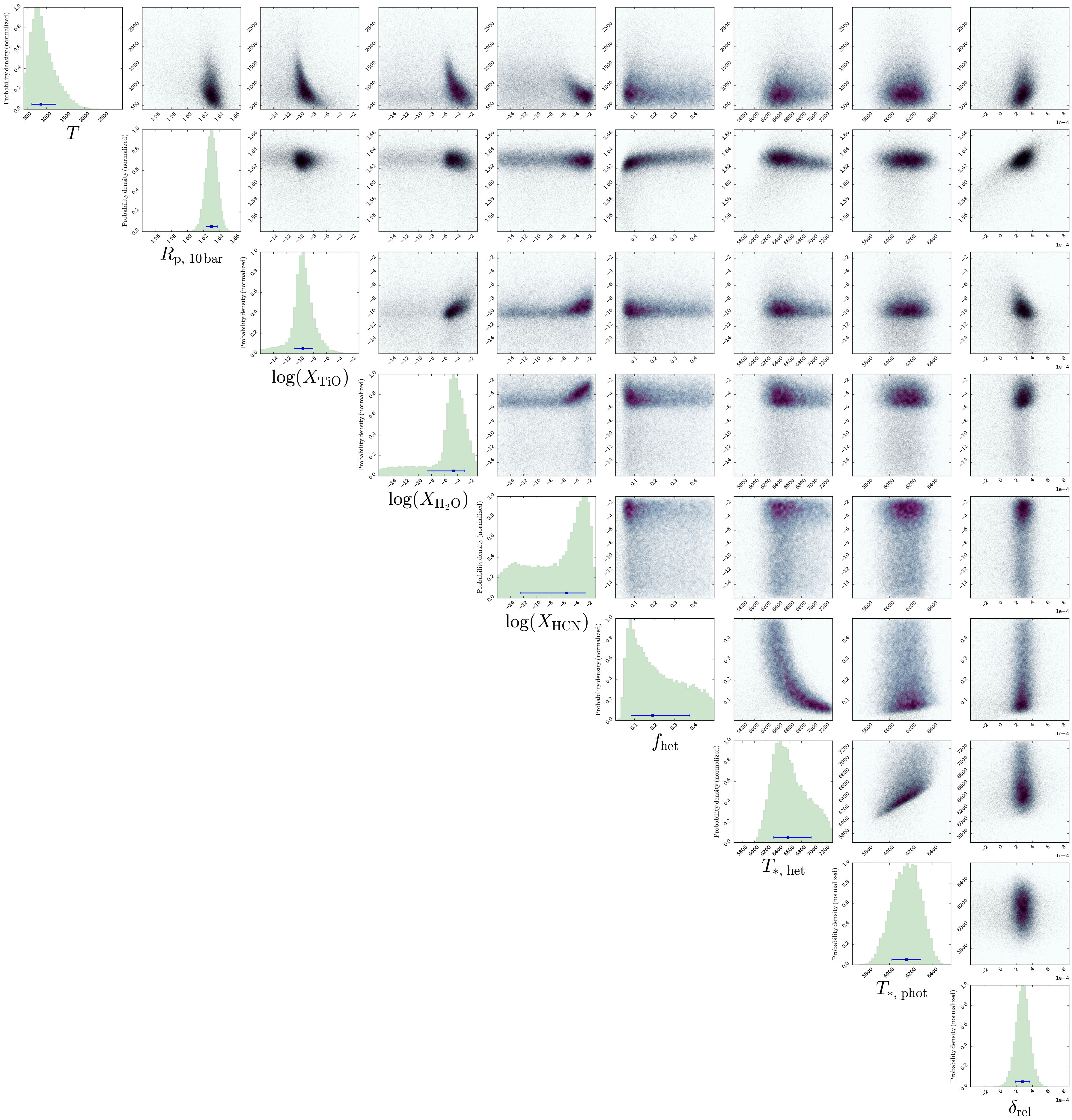}
    \caption{Posterior distribution from the best-fitting POSEIDON retrieval of WASP-103b's transmission spectrum. This posterior corresponds to the fit in Figure~\ref{fig:POSEIDON_best_fit_model} and the final column in Table~\ref{tab:POSEIDON_retrieval_results}. The marginalized histograms for each parameter are overlaid with the median retrieved value and $\pm 1\sigma$ confidence levels (blue error bars). The retrieved parameters indicate strong evidence of unocculted faculae ($T_{*, \, \rm{het}} > T_{*, \, \rm{phot}}$), a bounded constraint on WASP-103b's TiO abundance, suggestive (but weak) constraints on the H$_2$O and HCN abundances, and a $\sim$200\,ppm relative offset between the optical and infrared datasets.}
    \label{fig:POSEIDON_corner_plot}
\end{figure*}

We illustrate the contributions of each chemical species and the unocculted faculae to our best-fitting POSEIDON model in \autoref{fig:POSEIDON_best_fit_model}. This figure demonstrates that unocculted faculae still induce a slight downward slope between the red and blue ends of the HST/WFC3/IR/G141 bandpass, explaining why our IR-only retrievals also favor unocculted faculae. Unlike our retrievals with petitRADTRANS, we see that POSEIDON does not fit the dip in the transmission spectrum around $0.55\,\micron$, which may be due to the ExoMol line lists used (\autoref{fig:pRT_models}) or the inclusion of faculae -- we discuss this more in \autoref{sec:d_optical_trans_spec}. We note that the `minimal' model shown in this figure uses only those parameters required to fit the combined optical + IR data (i.e.\ without unconstrained chemical species or cloud parameters). This 9-parameter model achieves a minimum reduced chi-square of $\chi^2_{\nu} = 1.36$, which we consider an excellent fit to our transmission spectrum of WASP-103b. The full posterior corresponding to this best-fitting model is shown in \autoref{fig:POSEIDON_corner_plot}, with the parameter constraints also given in \autoref{tab:POSEIDON_retrieval_results}. We caution, however, that our recommended atmospheric and stellar properties are those from the `full' optical + IR retrieval (see \autoref{tab:POSEIDON_retrieval_results}), since the full retrieval accounts for marginalization over the entire parameter space. 

Our inferred atmospheric properties for WASP-103b are summarized in \autoref{tab:POSEIDON_retrieval_results}. We find a TiO mixing ratio of $-7.31^{+2.14}_{-2.06}$\,dex, broadly consistent with expectations for a solar abundance atmosphere at WASP-103b's equilibrium temperature \citep{Woitke2018}. The retrieved H$_2$O mixing ratio, $-3.28^{+1.46}_{-4.08}$\,dex, spans a wide range from super-solar to sub-solar abundances. The latter is consistent with the suggestion by \citet{Kreidberg2018} of H$_2$O dissociation in WASP-103b's atmosphere. We suggest that \citet{Wilson2020}'s tighter constraint on the H$_2$O abundance of ${\sim}40\times$ solar, without sub-solar solutions, may arise from the absence of marginalization over unocculted faculae in their retrievals. A hint of HCN near $1.6\,\micron$ (see \autoref{fig:POSEIDON_best_fit_model}) emerges only from our combined optical + IR retrievals, though the abundance is poorly constrained with the present data. Future observations of WASP-103b can serve to confirm the presence of TiO, H$_2$O, and/or HCN, while strengthening their abundance determinations.

\begin{deluxetable*}{lcccccc}[ht!]
    \tabletypesize{\small}
    \tablecaption{Atmospheric Retrieval Analysis Summary (POSEIDON) \label{tab:POSEIDON_retrieval_results}}
    \tablehead{
    \textbf{Dataset} & \phm{-} & Ground & Ground & WFC3 + & \multicolumn{2}{c}{\bfseries Ground Corrected +} \\
    & \phm{-} & Corrected & Uncorrected & Spitzer & \multicolumn{2}{c}{\bfseries WFC3 + Spitzer} \\
    \cmidrule{1-1} \cmidrule{3-7}
    Retrieval & & Full & Full & Full & \textbf{Full} & Minimal
    }
    \startdata \\[-8pt]
    \textbf{Planetary Atmosphere} \\ 
    \hspace{0.5em} $T$ (K) & & $ \hspace{0.7em} 861^{+469}_{-303}$ & $ \hspace{0.7em} 968^{+519}_{-348}$ & $ \hspace{0.7em} 924^{+462}_{-321}$ & $ \hspace{0.7em} 782^{+283}_{-231}$ & $ \hspace{0.7em} 850^{+392}_{-245}$  \\
    \hspace{0.5em} $R_{\rm{p, \, ref}}$ ($R_J$) & & $ \hspace{0.7em} 1.60^{+0.02}_{-0.03}$ & $ \hspace{0.7em} 1.48^{+0.02}_{-0.03}$ & $ \hspace{0.7em} 1.62^{+0.02}_{-0.02}$ & $ \hspace{0.7em} 1.62^{+0.01}_{-0.01}$ & $ \hspace{0.7em} 1.63^{+0.01}_{-0.01}$ \\
    \hspace{0.5em} log($X_{\rm{Na}}$) & & $ \hspace{-3.2em} < -1.86$ & $ \hspace{-3.2em} < -2.07$ & $ \hspace{-3.2em} < -1.82$ & $ \hspace{-3.2em} < -2.36$ & --- \\
    \hspace{0.5em} log($X_{\rm{K}}$) & & $ \hspace{-3.2em} < -2.44$ & $ \hspace{-3.2em} < -2.60$ & $ \hspace{-3.2em} < -1.90$ & $ \hspace{-3.2em} < -4.30$ & --- \\
    \hspace{0.5em} log($X_{\rm{H^{-}}}$) & & $ \hspace{-3.2em} < -2.54$ & $ \hspace{-3.2em} < -4.40$ & $ \hspace{-3.2em} < -4.40$ & $ \hspace{-3.2em} < -8.30$ & --- \\
    \hspace{0.5em} log($X_{\rm{TiO}}$) & & $ -6.36^{+2.24}_{-3.81}$ & $ -4.70^{+1.65}_{-1.88}$ & $ \hspace{-3.2em} < -2.26$ & $ -7.31^{+2.14}_{-2.06}$ & $ -9.56^{+1.53}_{-1.32}$ \\ 
    \hspace{0.5em} log($X_{\rm{VO}}$) & & $ \hspace{-3.2em} < -1.84$ & $ \hspace{-3.2em} < -2.44$ & $ \hspace{-3.2em} < -2.09$ & $ \hspace{-3.2em} < -5.40$ & --- \\ 
    \hspace{0.5em} log($X_{\rm{AlO}}$) & & $ \hspace{-3.2em} < -2.07$ & $ \hspace{-3.2em} < -2.70$ & $ \hspace{-3.2em} < -1.99$ & $ \hspace{-3.2em} < -3.50$ & --- \\ 
    \hspace{0.5em} log($X_{\rm{CrH}}$) & & $ \hspace{-3.2em} < -2.12$ & $ \hspace{-3.2em} < -2.39$ & $ \hspace{-3.2em} < -2.29$ & $ \hspace{-3.2em} < -3.16$ & --- \\ 
    \hspace{0.5em} log($X_{\rm{FeH}}$) & & $ \hspace{-3.2em} < -1.84$ & $ \hspace{-3.2em} < -2.25$ & $ \hspace{-3.2em} < -2.90$ & $ \hspace{-3.2em} < -4.30$ & --- \\ 
    \hspace{0.5em} log($X_{\rm{H_2 O}}$) & & $ \hspace{-3.2em} < -1.72$ & $ \hspace{-3.2em} < -1.85$ & $ -4.15^{+2.28}_{-6.99}$ & $ -3.28^{+1.46}_{-4.08}$ & $ -4.68^{+1.68}_{-3.98}$ \\
    \hspace{0.5em} log($X_{\rm{CO}}$) & & $ \hspace{-3.2em} < -1.84$ & $ \hspace{-3.2em} < -1.98$ & $ \hspace{-3.2em} < -1.70$ & $ \hspace{-3.2em} < -1.69$ & --- \\
    \hspace{0.5em} log($X_{\rm{CO_2}}$) & & $ \hspace{-3.2em} < -1.96$ & $ \hspace{-3.2em} < -2.22$ & $ \hspace{-3.2em} < -1.85$ & $ \hspace{-3.2em} < -1.91$ & --- \\ 
    \hspace{0.5em} log($X_{\rm{HCN}}$) & & $ \hspace{-3.2em} < -1.86$ & $ \hspace{-3.2em} < -1.90$ & $ \hspace{-3.2em} < -1.66$ & $ -4.07^{+2.08}_{-6.32}$ & $ -5.45^{+2.93}_{-7.00}$ \\[3pt]
    \midrule
    \textbf{Stellar Properties} \\ 
    \hspace{0.5em} $f_{\rm{het}}$ & & $ \hspace{0.4em} 0.18^{+0.12}_{-0.08}$ & $ \hspace{0.4em} 0.09^{+0.03}_{-0.03}$ & $ \hspace{0.4em} 0.25^{+0.12}_{-0.10}$ & $ \hspace{0.4em} 0.22^{+0.12}_{-0.09}$ & $ \hspace{0.4em} 0.19^{+0.18}_{-0.11}$ \\[3pt]
    \hspace{0.5em} $T_{*, \, \rm{het}}$ (K) & & $ 6501^{+243}_{-191}$ & $ \hspace{-3.0em} < 5972$ & $ 6728^{+282}_{-261}$ & $ 6499^{+218}_{-184}$ & $ 6570^{+399}_{-246}$ \\
    \hspace{0.5em} $T_{*, \, \rm{phot}}$ (K) & & $ 6150^{+121}_{-117}$ & $ 6093^{+119}_{-126}$ & $ 6103^{+124}_{-121}$ & $ 6158^{+124}_{-125}$ & $ 6158^{+130}_{-140}$ \\
    \midrule
    \textbf{Optical - Infrared Offset} \\ 
    \hspace{0.5em} $\delta_{\mathrm{rel}}$ (ppm) & & --- & --- & --- & $ \hspace{0.2em} 210^{+83}_{-85}$ & $ \hspace{0.2em} 273^{+88}_{-88}$ \\[3pt]
    \midrule
    \textbf{Detection Significances} \\ 
    \hspace{0.5em} TiO & & $1.7\,\sigma$ & $2.3\,\sigma$ & --- & $2.1\,\sigma$ & $2.2\,\sigma$ \\
    \hspace{0.5em} H$_2$O & & --- & --- & $1.7\,\sigma$ & $1.9\,\sigma$ & $2.1\,\sigma$ \\
    \hspace{0.5em} HCN & & --- & --- & --- & $1.7\,\sigma$ & $1.6\,\sigma$ \\
    \hspace{0.5em} TiO + H$_2$O + HCN & & $1.7\,\sigma$ & $2.3\,\sigma$ & $1.9\,\sigma$ & $2.8\,\sigma$ & $3.0\,\sigma$ \\
    \hspace{0.5em} Stellar heterogeneity & & $4.0\,\sigma$ & $1.5\,\sigma$ & $2.4\,\sigma$ & $4.3\,\sigma$ & $4.1\,\sigma$ \\[3pt]
    \midrule
    \textbf{Model Statistics} \\ 
    \hspace{0.5em} ln(Evidence) & & $276.6$ & $277.6$ & $84.7$ & $363.9$ & $365.7$ \\
    \hspace{0.5em} $\chi^2_{\nu, \, \rm{min}}$ & & $2.62$ & $2.53$ & N/A & $1.96$ & $1.36$ \\
    \hspace{0.5em} $N_{\rm param}$ & & $21$ & $21$ & $21$ & $22$ & $9$ \\
    \hspace{0.5em} d.o.f. & & $19$ & $19$ & N/A & $30$ & $43$ \\[3pt]
    \enddata 
    \tablecomments{Parameters without clear lower and upper bounds (e.g. abundances of undetected chemical species) are represented by 2$\sigma$ upper bounds. Cloud and/or haze parameters were completely unconstrained in all our retrievals. The detection significances were established by nested Bayesian model comparisons with respect to the model in each column. The combined significance `TiO + H$_2$O + HCN' represents the probability of \emph{at least one} of these molecules influencing WASP-103b's transmission spectrum. The reduced chi-square and degrees of freedom (d.o.f) are undefined for the WFC3 + Spitzer retrieval ($N_{\rm param} > N_{\rm data}$); we include this retrieval for comparison with the optical-only and optical + infrared retrievals. While the `minimal' model provides the best-fit (via the evidence or reduced chi-square), the `full' models include marginalization over a wider range of parameters---accounting for overlapping absorption features and other degeneracies---so our recommended atmospheric and stellar parameters are the `full' optical + infrared parameters (highlighted in bold / 5$^{\rm th}$ column).}
\end{deluxetable*}

\section{Discussion}
\label{sec:discussion}

\subsection{The optical transmission spectrum of WASP-103b}
\label{sec:d_optical_trans_spec}

As shown in \autoref{sec:pRT}, our petitRADTRANS retrievals that do not include stellar activity favor VO, albeit with an implausibly high volume mixing ratio (\autoref{fig:pRT_models}). However, retrievals that do include the impacts of stellar activity significantly favor stellar-activity contamination with tentative evidence for TiO (\autoref{sec:POSEIDON}). This may be due to the different opacity at bluer wavelengths; the \cite{Plez1999} VO line list used by petitRADTRANS leads to a steeper drop off than the ExoMol line list used by POSEIDON (\autoref{fig:pRT_models}). As a result, the POSEIDON retrievals with the ExoMol line list necessitate unocculted faculae to fit the downwards trend we observe at blue wavelengths (\autoref{fig:POSEIDON_best_fit_model}).

In order to place WASP-103b into context, we compare our transmission spectrum with that of another ultrahot Jupiter, WASP-121b \citep{Delrez2016}. WASP-121b's equilibrium temperature of $2358 \pm 52$\,K is 130\,K cooler than WASP-103b's, and it orbits an F6 dwarf, compared to WASP-103, which is an F8 dwarf. 

\begin{figure}
    \centering
    \includegraphics[scale=0.6]{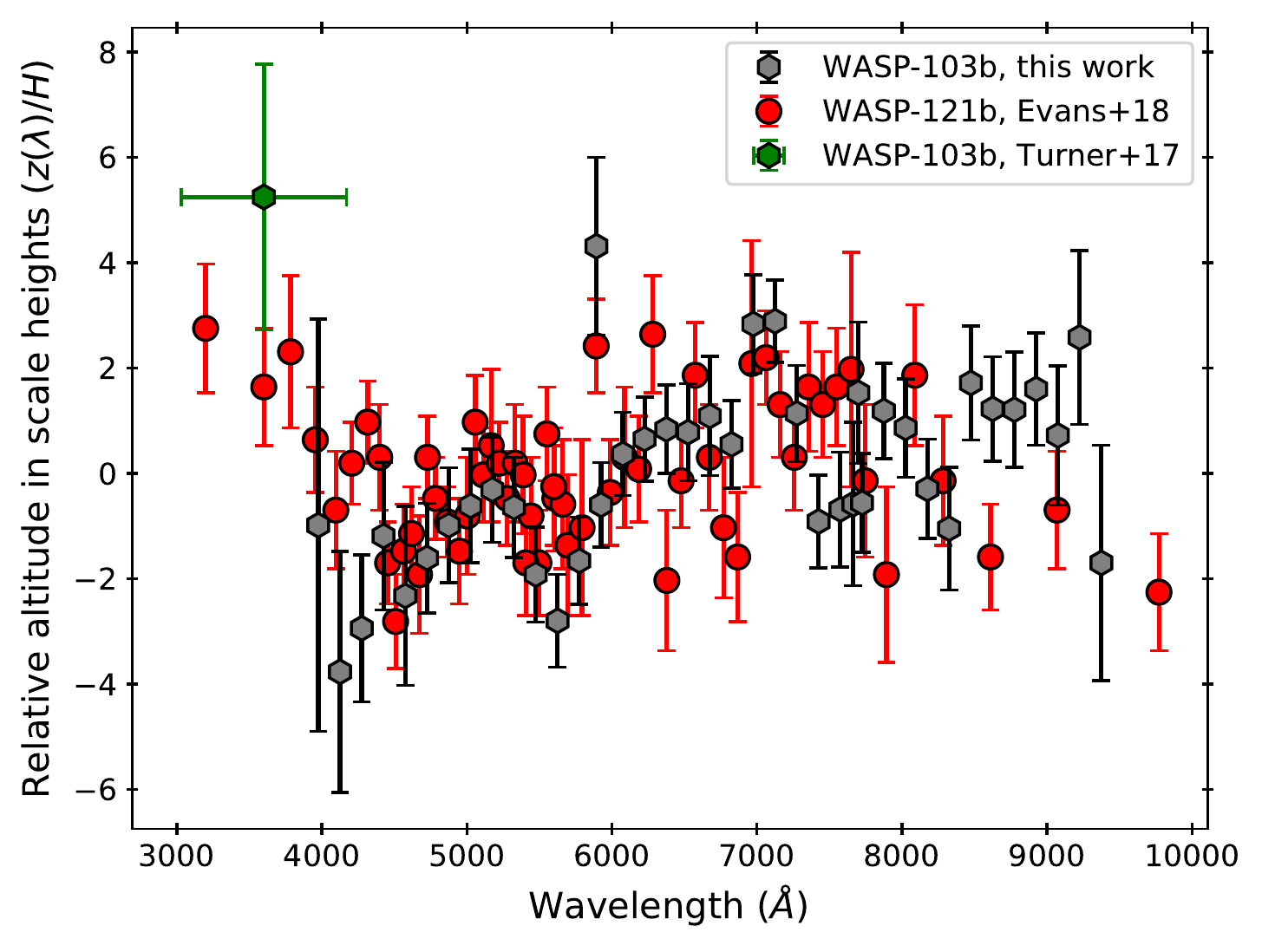}
    \caption{Our transmission spectrum of WASP-103b (grey hexagons) compared to the HST/STIS transmission spectrum of WASP-121b from \protect\cite{Evans2018} (red circles). The transmission spectra have both been scaled to their respective scale heights: 152\,ppm for WASP-103b and 239\,ppm for WASP-121b (derived from \protect\citealt{Delrez2016}). We also include the near-UV photometric data point from \protect\cite{Turner2017} (green hexagon).}
    \label{fig:w121_comparison}
\end{figure}

WASP-121b's low-resolution HST/STIS transmission spectrum \citep{Evans2018} is shown in \autoref{fig:w121_comparison}. There is an excellent agreement between the transmission spectra of WASP-121b and WASP-103b, particularly between ${\sim}4000$ and $6500$\,\AA. \autoref{fig:w121_comparison} additionally demonstrates our ability to achieve HST-quality transmission spectra from the ground.

Figure \ref{fig:w121_comparison} also includes the near-UV photometric data point from \cite{Turner2017}. We did not analyze that data set in the current work nor include it in our retrieval analysis due to the different system parameters that \cite{Turner2017} used. However, at face value this point also seems to follow the subsequent rise in the transmission spectrum towards the UV, as seen in WASP-121b. 

\subsubsection{On the potential for VO \& TiO}
\label{sec:d_VO}

In the case of WASP-121b, its low-resolution transmission spectral features are taken as strong evidence for VO \citep{Evans2016,Evans2018}. In secondary eclipse, WASP-121b shows water in emission, which is direct evidence for a temperature inversion in the planet's atmosphere \citep{Evans2017,Evans2019,Evans2020,Daylan2019,Bourrier2020}. It has been suggested that VO is responsible for this temperature inversion \citep{Evans2017,Evans2018,Evans2020,Daylan2019,Bourrier2020} although the presence of this molecule is debated at high-resolution \citep{Ben-Yami2020,Cabot2020,Hoeijmakers2020,Merritt2020,Borsa2021}, leading to suggestions that neutral Fe may instead explain WASP-121b's temperature inversion \citep{Gibson2020}.

Given the similarity between the low-resolution transmission spectra of WASP-103b and WASP-121b, VO seems a plausible explanation. If present, VO could also be responsible for WASP-103b's temperature inversion \citep{Kreidberg2018}. 

The comparison to WASP-121b is strengthened when taking our transmission spectrum with the near-UV data point of \cite{Turner2017}, as this resembles the same rise in the transmission spectrum towards the UV as seen in WASP-121b (\autoref{fig:w121_comparison}). If this rise does indeed hold, then this would be evidence of a planetary atmospheric contribution and not unocculted faculae, which would continue trending downwards towards the UV. However, we caution that due to the small variations in $R_P/R_*$ seen in our own analysis (\autoref{fig:RpRs_variation}), any rise in WASP-103b's transmission spectrum towards the UV should be confirmed with future HST observations that cover a broad wavelength range. 

In \autoref{sec:pRT} we suggested that the different line lists used by petitRADTRANS and POSEIDON could be responsible for the different interpretations of WASP-103b's transmission spectrum. However, in \cite{Evans2018}, the authors used the ATMO retrieval code which uses the ExoMol line list for VO \citep{Goyal2018}, like POSEIDON. Despite this, \cite{Evans2018} still find strong evidence for VO. This may be because of their greater spectral resolution (median wavelength bin size of 97\,\AA), their coverage of the UV, and the fact that ATMO does not fit for stellar activity. Taken together, it is unclear to what extent the choice of line list impacts our conclusions.

While the comparison between WASP-121b and WASP-103b is evidence for their similar atmospheric chemistry (in the absence of stellar contamination), we caution that the very high VO abundances derived from our petitRADTRANS retrievals are implausible (\autoref{sec:pRT}). 

However, the retrievals run with POSEIDON (\autoref{sec:POSEIDON}) favor TiO, although with only marginal significance. Given the updated line lists that POSEIDON uses and the plausible $\sim$ solar TiO abundances it finds for both the optical and IR data, it is perhaps more likely that TiO is present in WASP-103b's atmosphere. If confirmed by follow-up studies, this would make WASP-103b more similar to WASP-33b and WASP-76b for which TiO and temperature inversions have been observed \citep{Haynes2015,Nugroho2017,Fu2020}.

Further high-resolution observations are needed of WASP-103b to search for TiO and VO via cross-correlation techniques. In any case, the tentative evidence for these species in our transmission spectrum is suggestive that they could be responsible for WASP-103b's temperature inversion as postulated by \cite{Kreidberg2018}.

\subsubsection{On stellar-activity contamination}

If instead stellar contamination is the cause of some structures in WASP-103b's transmission spectrum, then we can also compare with WASP-121b. WASP-103 is an F8 dwarf \citep{Gillon2014} while WASP-121 is an active F6 dwarf \citep{Delrez2016}. 

In the case of WASP-121, it shows no photometric modulation above the 1~mmag threshold of the WASP photometry, no emission in the Ca\,{\footnotesize II} H and K line cores, and no spot-crossing events during transit \citep{Delrez2016}. However, it does show significant RV jitter. \cite{Delrez2016} conclude that this scenario for WASP-121 can be explained if the star is dominated by plage (the chromospheric counterpart to photospheric active regions), with a covering fraction of 8\,\%. This is significantly larger than the typical facular coverage of F6 dwarfs estimated from the variability of Kepler stars \citep{Rackham2019}. However, under this assumption, WASP-121b's transmission spectrum would likely also be contaminated. Ultimately though, the rise in WASP-121b's transmission spectrum at wavelengths shorter than 4000\,\AA\ cannot be caused by unocculted faculae, which would lead to a continued decrease in transit depth towards bluer wavelengths \citep[e.g.,][]{Rackham2017,Rackham2019}.

In the case of WASP-103, it shows a photometric modulation of $5 \pm 1$\,mmag in C14 automated imaging telescope data from Fairborn Observatory \citep{Kreidberg2018}. \cite{Gillon2014} do not report any activity in WASP-103's RV data but \cite{Staab2017} measure WASP-103's $\log R^{'}_{\mathrm{HK}}$ as $-4.57$, consistent with other active exoplanet host stars, such as WASP-19 \citep{Sing2016} and WASP-52 \citep{Hebrard2013}. Our retrievals that include activity suggest unocculted faculae $\sim350$\,K hotter than the surrounding photosphere with large covering fractions of $18^{+12}_{-8}$\,\% for the optical data and $22^{+12}_{-9}$\,\% for the combined optical and IR data (\autoref{tab:POSEIDON_retrieval_results}). Our retrieved faculae temperatures are consistent with solar faculae \citep{Topka1997}, while the relatively large uncertainties in the covering fractions would only make WASP-103 a $1.5\sigma$ outlier as compared to the Kepler sample \citep{Rackham2019}. However, as \cite{Delrez2016} and \cite{Rackham2019} discuss, estimates based from photometric modulation do not necessarily reveal the underlying activity coverage, particularly in the case of bright regions. Therefore we do not rule out the large covering fractions of faculae our retrievals with POSEIDON derive, particularly given WASP-103's high $\log R^{'}_{\mathrm{HK}}$ \citep{Staab2017}. UV spectra are needed to determine whether WASP-103b is indeed contaminated by unocculted faculae or instead follows the same trend as WASP-121b.

\subsection{Comparison with previous optical transmission spectra}

\begin{figure}
    \centering
    \includegraphics[scale=0.45]{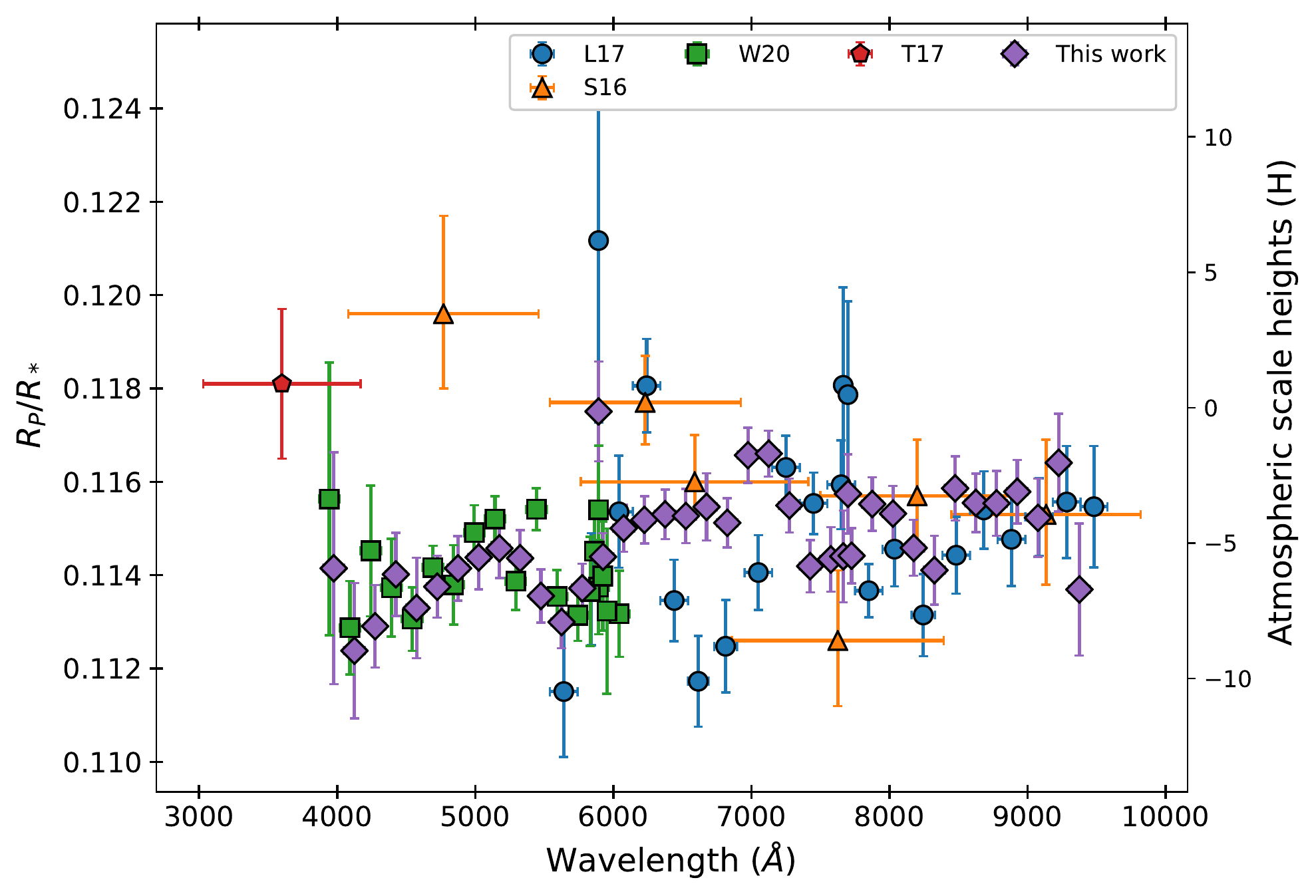}
    \caption{A comparison between our new transmission spectrum (purple diamonds) and the previously published transmission spectra of \protect\cite{Southworth2016} (orange triangles), \protect\cite{Turner2017} (red pentagon), \protect\cite{Lendl2017} (blue circles), and \protect\cite{Wilson2020} (green squares).}
    \label{fig:literature_comparison}
\end{figure}

Figure \ref{fig:literature_comparison} shows the comparison between our newly derived transmission spectrum and the previously published results of \protect\cite{Southworth2016}, \protect\cite{Turner2017}, \protect\cite{Lendl2017}, and \protect\cite{Wilson2020}. In all cases the transmission spectra have been corrected for the third-light contamination. We note that in this figure we applied an offset in $R_P/R_*$ of 0.00065 and 0.0021 to the transmission spectra of \cite{Lendl2017} and \cite{Wilson2020}, respectively, so that their median $R_P/R_*$ is equal to the median of our newly derived transmission spectrum. We believe the larger offset for the \cite{Wilson2020} data is because those authors used a Gaussian prior on $R_P/R_*$ and took the mean and standard deviation from \cite{Southworth2015}, who did not account for the third light contamination (\autoref{sec:3rd_light_corr}). We have not applied an offset to the transmission spectrum of \cite{Southworth2016}, where a steep slope is found, nor to the single wavelength bin of \cite{Turner2017}.

This figure demonstrates the improved precision that we achieved by combining our 11 transits. It also shows the slight disagreement between our new transmission spectrum and some of the Gemini/GMOS and VLT/FORS2 bins of \cite{Lendl2017} and \cite{Wilson2020}, which highlights the benefits of combining many transits to form our transmission spectrum. 

Additionally, we do not see the blueward rise as observed by \cite{Southworth2016}. However, other than \cite{Southworth2016}'s bluest photometric data point, our results are roughly consistent to $1\sigma$. Additionally, \cite{Turner2017}'s near-UV photometric band indicates a rise towards wavelengths bluer than our spectroscopic coverage. However, we caution against reading too much into this due to the different system parameters used and the offsets in $R_P/R_*$ needed to bring the spectroscopic data in agreement.

\section{Conclusions}
\label{sec:conclusions}

We have presented a new ground-based optical transmission spectrum of the ultrahot Jupiter WASP-103b. We have combined 5 new transits from the ACCESS survey and 2 new transits from the LRG-BEASTS survey with a reanalysis of 3 published Gemini/GMOS transits and 1 VLT/FORS2 transit. Our 11-transit combined transmission spectrum has a median uncertainty in the transit depth of 148\,ppm ($<1 H$), which is of HST quality and is one of the most precise ground-based transmission spectra to date.

Our optical transmission spectrum shows evidence for sodium in a 20\,\AA-wide bin at $2.8\sigma$ and significant structure with a downwards slope towards blue wavelengths. Our transmission spectrum is in excellent agreement with WASP-121b, another ultrahot Jupiter of comparable temperature for which the presence of VO has been inferred. In our analysis of WASP-103b's optical transmission spectrum using the petitRADTRANS retrieval software, we find that VO can provide a good fit to the planet's spectral features but only if we ignore stellar activity and use implausibly high VO abundances.

In our separate optical-only retrievals with POSEIDON, which include unocculted stellar heterogeneity contamination and updated VO and TiO line lists, we find strong evidence for unocculted faculae ($4.0\sigma$), an absence of VO, and tentative evidence for TiO ($1.7\sigma$) at roughly solar abundances. Given WASP-103's high $\log R^{'}_{\mathrm{HK}}$, this is our favored conclusion.

When we combine our optical transmission spectrum with the previously published HST/WFC3/IR/G141 and Spitzer/IRAC transmission spectrum, we find a slightly higher evidence for unocculted faculae ($4.3\sigma$) and TiO ($2.1\sigma$), and weak evidence for H$_2$O ($1.9\sigma$) and HCN ($1.7\sigma$). If TiO is confirmed by future high-resolution observations in the optical and HST observations in the UV, this would make WASP-103b the fourth planet for which evidence of TiO and a temperature inversion have been observed in the same exoplanet.

Our result highlights the precision that ground-based transmission spectroscopy can reach, in addition to the need for a careful treatment of stellar activity contamination for exoplanets orbiting F-type stars.


\acknowledgments

The William Herschel Telescope is operated on the island of La Palma by the Isaac Newton Group in the Spanish Observatorio del Roque de los Muchachos of the Instituto de Astrof\'{i}sica de Canarias. This paper includes data gathered with the 6.5 meter Magellan Telescopes located at Las Campanas Observatory, Chile. We thank the observing personnel for providing the facilities and guidance necessary for making the collection of the ACCESS datasets possible. B.V.R.\ thanks the Heising-Simons Foundation for support. A.J.\ acknowledges support from FONDECYT project 1210718, and ANID - Millennium Science Initiative - ICN12$\_$009. The results reported herein benefited from collaborations and/or information exchange within NASA's Nexus for Exoplanet System Science (NExSS) research coordination network sponsored by NASA's Science Mission Directorate, and funding through the NExSS Earths in Other Solar System (PI: Apai) and ACCESS (PI: L\'opez-Morales) teams. ICW and MLM thank the Brinson Foundation for their support. This paper includes data gathered with the 6.5 meter Magellan Telescopes located at Las Campanas Observatory (LCO), Chile. PJW acknowledges support from STFC under consolidated grants ST/P000495/1 and ST/T000406/1.

%

\vspace{5mm}
\facilities{Magellan/IMACS, WHT/ACAM, Gemini/GMOS, VLT/FORS2}


\software{Astropy \citep{astropy:2013,astropy:2018}, Batman \citep{batman}, emcee \citep{emcee}, George \citep{george,george2}, LDTk \citep{ldtk}, Matplotlib \citep{matplotlib}, MultiNest \citep{Feroz2008,Feroz2009,Feroz2019}, Numpy \citep{numpy}, petitRADTRANS \citep{Molliere2019}, PLATON \citep{Zhang2019,Zhang2020}, POSEIDON \citep{MacDonald2017}, PyMultiNest \citep{Buchner2014}, Scipy \citep{scipy}}



\bibliography{wasp103_bib}
\bibliographystyle{aasjournal}

\appendix
\section{Appendix: nightly transmission spectra}

The nightly transmission spectra for the ACCESS data are given in \autoref{tab:ts_ACCESS}, and those for the LRG-BEASTS, VLT/FORS2 and Gemini/GMOS data are given in \autoref{tab:ts_LRG-BEASTS}.

\begin{table*}[ht!]
\caption{The nightly transmission spectra for the ACCESS data. Note these are following the application of the $R_P/R_*$ offset (\autoref{fig:RpRs_variation}) and the correction for the third-light contamination and planet's nightside temperature (\autoref{sec:3rd_light_corr}).}
\label{tab:ts_ACCESS}
\centering
\begin{tabular}{ccccccc} \hline
$\lambda$ &     $\Delta(\lambda)$     & $R_P/R_*$ & $R_P/R_*$ & $R_P/R_*$ & $R_P/R_*$ & $R_P/R_*$ \\ 
(\AA) & (\AA) & ACCESS n1 & ACCESS n2 & ACCESS n3 & ACCESS n4 & ACCESS n5 \\ \hline
3975     &      150 &                              -- &                              -- &                              -- &                              -- &                              -- \\
4125     &      150 &                              -- &                              -- &                              -- &                              -- &                              -- \\
4275     &      150 &                              -- &                              -- &                              -- &                              -- &                              -- \\
4425     &      150 &                              -- &                              -- &                              -- &                              -- &                              -- \\
4575     &      150 &   $0.1082^{+0.00382}_{-0.00432}$ &  $0.11579^{+0.00473}_{-0.00484}$ &                              -- &                              -- &                              -- \\
4725     &      150 &  $0.11538^{+0.00488}_{-0.00473}$ &  $0.11578^{+0.00395}_{-0.00393}$ &                              -- &                              -- &                              -- \\
4875     &      150 &  $0.11716^{+0.00305}_{-0.00263}$ &   $0.11733^{+0.00304}_{-0.0031}$ &                              -- &                              -- &                              -- \\
5025     &      150 &  $0.11712^{+0.00319}_{-0.00298}$ &  $0.11438^{+0.00366}_{-0.00367}$ &                              -- &                              -- &                              -- \\
5175     &      150 &  $0.11625^{+0.00205}_{-0.00238}$ &   $0.11332^{+0.0041}_{-0.00432}$ &                              -- &                              -- &                              -- \\
5325     &      150 &  $0.11606^{+0.00212}_{-0.00201}$ &   $0.11693^{+0.00325}_{-0.0035}$ &                              -- &                              -- &                              -- \\
5475     &      150 &  $0.10928^{+0.00185}_{-0.00197}$ &  $0.11282^{+0.00412}_{-0.00395}$ &                              -- &  $0.11397^{+0.00236}_{-0.00264}$ &  $0.11388^{+0.00106}_{-0.00097}$ \\
5625     &      150 &  $0.11349^{+0.00134}_{-0.00144}$ &   $0.10859^{+0.0039}_{-0.00427}$ &  $0.11275^{+0.00241}_{-0.00274}$ &  $0.11152^{+0.00203}_{-0.00213}$ &   $0.11251^{+0.00142}_{-0.0015}$ \\
5775     &      150 &  $0.11271^{+0.00157}_{-0.00178}$ &  $0.11665^{+0.00349}_{-0.00329}$ &  $0.11318^{+0.00196}_{-0.00229}$ &   $0.11362^{+0.00216}_{-0.0023}$ &  $0.11319^{+0.00098}_{-0.00104}$ \\
5893     &       20 &  $0.11335^{+0.00282}_{-0.00306}$ &    $0.113^{+0.00337}_{-0.00313}$ &  $0.11643^{+0.00391}_{-0.00424}$ &   $0.11969^{+0.0032}_{-0.00325}$ &  $0.11876^{+0.00204}_{-0.00203}$ \\
5925     &      150 &  $0.11523^{+0.00145}_{-0.00134}$ &  $0.11843^{+0.00315}_{-0.00302}$ &  $0.11164^{+0.00235}_{-0.00242}$ &   $0.11358^{+0.00165}_{-0.0014}$ &  $0.11292^{+0.00132}_{-0.00136}$ \\
6075     &      150 &  $0.11904^{+0.00112}_{-0.00117}$ &  $0.11527^{+0.00296}_{-0.00302}$ &    $0.10948^{+0.0032}_{-0.0034}$ &  $0.11374^{+0.00125}_{-0.00124}$ &   $0.11198^{+0.00132}_{-0.0013}$ \\
6225     &      150 &   $0.1177^{+0.00142}_{-0.00143}$ &   $0.1174^{+0.00348}_{-0.00346}$ &    $0.1122^{+0.0032}_{-0.00308}$ &  $0.11587^{+0.00088}_{-0.00089}$ &  $0.11121^{+0.00129}_{-0.00123}$ \\
6375     &      150 &    $0.11637^{+0.0014}_{-0.0016}$ &  $0.11121^{+0.00382}_{-0.00418}$ &  $0.11135^{+0.00241}_{-0.00209}$ &  $0.11583^{+0.00101}_{-0.00101}$ &  $0.11419^{+0.00162}_{-0.00164}$ \\
6525     &      150 &  $0.11617^{+0.00189}_{-0.00203}$ &  $0.11619^{+0.00345}_{-0.00332}$ &  $0.10827^{+0.00218}_{-0.00262}$ &  $0.11642^{+0.00128}_{-0.00131}$ &   $0.11619^{+0.0014}_{-0.00134}$ \\
6675     &      150 &  $0.11503^{+0.00191}_{-0.00223}$ &  $0.11646^{+0.00206}_{-0.00207}$ &  $0.11442^{+0.00228}_{-0.00229}$ &  $0.11359^{+0.00139}_{-0.00152}$ &   $0.11669^{+0.0017}_{-0.00175}$ \\
6825     &      150 &  $0.11276^{+0.00162}_{-0.00157}$ &   $0.11851^{+0.0018}_{-0.00179}$ &  $0.11553^{+0.00225}_{-0.00222}$ &  $0.11497^{+0.00106}_{-0.00106}$ &   $0.11589^{+0.0016}_{-0.00163}$ \\
6975     &      150 &  $0.11392^{+0.00326}_{-0.00324}$ &  $0.11532^{+0.00256}_{-0.00278}$ &  $0.11845^{+0.00159}_{-0.00157}$ &   $0.1171^{+0.00092}_{-0.00097}$ &   $0.1161^{+0.00155}_{-0.00155}$ \\
7125     &      150 &                              -- &                              -- &                              -- &  $0.11856^{+0.00162}_{-0.00154}$ &  $0.11678^{+0.00069}_{-0.00082}$ \\
7275     &      150 &                              -- &                              -- &                              -- &  $0.11488^{+0.00179}_{-0.00179}$ &  $0.11317^{+0.00139}_{-0.00157}$ \\
7425     &      150 &  $0.11285^{+0.00099}_{-0.00106}$ &  $0.11206^{+0.00451}_{-0.00442}$ &  $0.11861^{+0.00404}_{-0.00404}$ &  $0.11596^{+0.00207}_{-0.00219}$ &  $0.11468^{+0.00232}_{-0.00249}$ \\
7575     &      150 &   $0.1134^{+0.00188}_{-0.00217}$ &  $0.11404^{+0.00417}_{-0.00413}$ &  $0.11604^{+0.00369}_{-0.00395}$ &   $0.11163^{+0.0017}_{-0.00179}$ &  $0.11681^{+0.00187}_{-0.00184}$ \\
7665     &       20 &   $0.10592^{+0.00526}_{-0.0052}$ &   $0.1153^{+0.00297}_{-0.00348}$ &  $0.12057^{+0.00355}_{-0.00396}$ &  $0.11087^{+0.00219}_{-0.00239}$ &  $0.11324^{+0.00188}_{-0.00193}$ \\
7699     &       20 &  $0.12152^{+0.00401}_{-0.00485}$ &  $0.11494^{+0.00269}_{-0.00276}$ &  $0.10851^{+0.00363}_{-0.00396}$ &  $0.11831^{+0.00188}_{-0.00201}$ &  $0.11636^{+0.00156}_{-0.00163}$ \\
7725     &      150 &  $0.10917^{+0.00142}_{-0.00145}$ &  $0.11452^{+0.00451}_{-0.00433}$ &   $0.1202^{+0.00352}_{-0.00348}$ &   $0.1124^{+0.00192}_{-0.00185}$ &  $0.11574^{+0.00211}_{-0.00199}$ \\
7875     &      150 &  $0.11654^{+0.00141}_{-0.00159}$ &  $0.11264^{+0.00505}_{-0.00518}$ &  $0.11644^{+0.00206}_{-0.00247}$ &  $0.11221^{+0.00167}_{-0.00187}$ &  $0.11506^{+0.00192}_{-0.00195}$ \\
8025     &      150 &   $0.1174^{+0.00177}_{-0.00174}$ &  $0.11143^{+0.00672}_{-0.00688}$ &  $0.11531^{+0.00204}_{-0.00205}$ &    $0.112^{+0.00122}_{-0.00125}$ &   $0.1168^{+0.00216}_{-0.00216}$ \\
8175     &      150 &  $0.11827^{+0.00125}_{-0.00146}$ &  $0.10965^{+0.00461}_{-0.00491}$ &  $0.11258^{+0.00223}_{-0.00222}$ &  $0.11395^{+0.00108}_{-0.00102}$ &   $0.11415^{+0.0012}_{-0.00121}$ \\
8325     &      150 &  $0.11277^{+0.00261}_{-0.00243}$ &   $0.11999^{+0.00611}_{-0.0062}$ &   $0.1151^{+0.00192}_{-0.00235}$ &  $0.11497^{+0.00136}_{-0.00137}$ &  $0.11278^{+0.00164}_{-0.00169}$ \\
8475     &      150 &   $0.11544^{+0.0023}_{-0.00224}$ &  $0.11435^{+0.00441}_{-0.00435}$ &   $0.11583^{+0.0028}_{-0.00278}$ &  $0.11298^{+0.00201}_{-0.00203}$ &  $0.11955^{+0.00277}_{-0.00288}$ \\
8625     &      150 &  $0.11304^{+0.00204}_{-0.00199}$ &  $0.11316^{+0.00553}_{-0.00571}$ &  $0.11028^{+0.00244}_{-0.00252}$ &   $0.11499^{+0.00164}_{-0.0015}$ &   $0.11823^{+0.00188}_{-0.0021}$ \\
8775     &      150 &  $0.10814^{+0.00171}_{-0.00181}$ &  $0.11449^{+0.00347}_{-0.00359}$ &    $0.1217^{+0.00222}_{-0.0026}$ &  $0.11478^{+0.00246}_{-0.00246}$ &  $0.11544^{+0.00336}_{-0.00322}$ \\
8925     &      150 &   $0.1078^{+0.00269}_{-0.00279}$ &  $0.11522^{+0.00374}_{-0.00362}$ &  $0.11564^{+0.00304}_{-0.00332}$ &  $0.11716^{+0.00155}_{-0.00129}$ &  $0.11632^{+0.00244}_{-0.00252}$ \\
9075     &      150 &  $0.10784^{+0.00225}_{-0.00236}$ &   $0.11652^{+0.0036}_{-0.00369}$ &                              -- &  $0.11584^{+0.00211}_{-0.00249}$ &  $0.11388^{+0.00388}_{-0.00391}$ \\
9225     &      150 &                              -- &                              -- &                              -- &                              -- &  $0.11855^{+0.00262}_{-0.00272}$ \\
9375     &      150 &                              -- &                              -- &                              -- &                              -- &                              -- \\ \hline
\end{tabular}
\end{table*}

\begin{table*}[ht!]
\caption{The nightly transmission spectra for the LRG-BEASTS, VLT/FORS2 and Gemini/GMOS data. Note these are following the application of the $R_P/R_*$ offset (\autoref{fig:RpRs_variation}) and the correction for the third-light contamination and planet's nightside temperature (\autoref{sec:3rd_light_corr}).}
\label{tab:ts_LRG-BEASTS}
\centering
\begin{tabular}{cccccccc} \hline
$\lambda$ &     $\Delta(\lambda)$     & $R_P/R_*$ & $R_P/R_*$ & $R_P/R_*$ & $R_P/R_*$ & $R_P/R_*$  & $R_P/R_*$ \\ 
(\AA) & (\AA) & LRG-BEASTS n1 & LRG-BEASTS n2 & VLT/FORS2 & GMOS n1 & GMOS n2 & GMOS n3 \\ \hline
3975     &      150 &                              -- &                              -- &  $0.11411^{+0.00243}_{-0.00253}$ &                              -- &                              -- &                              -- \\
4125     &      150 &                              -- &                              -- &  $0.11246^{+0.00154}_{-0.00136}$ &                              -- &                              -- &                              -- \\
4275     &      150 &                              -- &                              -- &  $0.11293^{+0.00091}_{-0.00086}$ &                              -- &                              -- &                              -- \\
4425     &      150 &                              -- &                              -- &   $0.11404^{+0.0009}_{-0.00088}$ &                              -- &                              -- &                              -- \\
4575     &      150 &  $0.11071^{+0.00346}_{-0.00392}$ &  $0.12107^{+0.00392}_{-0.00398}$ &  $0.11306^{+0.00124}_{-0.00129}$ &                              -- &                              -- &                              -- \\
4725     &      150 &  $0.11293^{+0.00299}_{-0.00344}$ &  $0.11585^{+0.00367}_{-0.00375}$ &  $0.11363^{+0.00072}_{-0.00069}$ &                              -- &                              -- &                              -- \\
4875     &      150 &  $0.11373^{+0.00231}_{-0.00232}$ &  $0.11865^{+0.00403}_{-0.00423}$ &   $0.1136^{+0.00078}_{-0.00079}$ &                              -- &                              -- &                              -- \\
5025     &      150 &   $0.11391^{+0.00292}_{-0.0025}$ &  $0.11752^{+0.00278}_{-0.00272}$ &  $0.11402^{+0.00076}_{-0.00078}$ &                              -- &                              -- &                              -- \\
5175     &      150 &  $0.11364^{+0.00245}_{-0.00394}$ &  $0.11655^{+0.00238}_{-0.00236}$ &   $0.1143^{+0.00077}_{-0.00066}$ &                              -- &                              -- &                              -- \\
5325     &      150 &   $0.1151^{+0.00141}_{-0.00157}$ &   $0.11259^{+0.0021}_{-0.00191}$ &  $0.11405^{+0.00073}_{-0.00079}$ &                              -- &                              -- &                              -- \\
5475     &      150 &  $0.11324^{+0.00183}_{-0.00206}$ &    $0.114^{+0.00266}_{-0.00238}$ &   $0.1144^{+0.00103}_{-0.00081}$ &                              -- &                              -- &                              -- \\
5625     &      150 &   $0.11454^{+0.0014}_{-0.00147}$ &   $0.1132^{+0.00229}_{-0.00206}$ &   $0.11307^{+0.0013}_{-0.00093}$ &  $0.11474^{+0.00494}_{-0.00491}$ &  $0.10592^{+0.00492}_{-0.00474}$ &  $0.11339^{+0.00219}_{-0.00222}$ \\
5775     &      150 &  $0.11544^{+0.00142}_{-0.00136}$ &    $0.11106^{+0.0029}_{-0.0024}$ &   $0.11394^{+0.0013}_{-0.00109}$ &  $0.11447^{+0.00348}_{-0.00352}$ &  $0.11074^{+0.00231}_{-0.00226}$ &  $0.11724^{+0.00235}_{-0.00228}$ \\
5893     &       20 &  $0.13144^{+0.00539}_{-0.00568}$ &   $0.1164^{+0.00503}_{-0.00504}$ &   $0.1185^{+0.00338}_{-0.00351}$ &  $0.11732^{+0.00922}_{-0.01028}$ &  $0.11579^{+0.00423}_{-0.00437}$ &  $0.11796^{+0.00434}_{-0.00517}$ \\
5925     &      150 &   $0.1136^{+0.00156}_{-0.00162}$ &   $0.11307^{+0.00294}_{-0.0029}$ &  $0.11454^{+0.00106}_{-0.00084}$ &  $0.11945^{+0.00285}_{-0.00275}$ &  $0.11516^{+0.00287}_{-0.00273}$ &  $0.11675^{+0.00194}_{-0.00224}$ \\
6075     &      150 &  $0.11479^{+0.00149}_{-0.00147}$ &  $0.11259^{+0.00303}_{-0.00272}$ &  $0.11507^{+0.00119}_{-0.00154}$ &  $0.11188^{+0.00294}_{-0.00309}$ &  $0.11504^{+0.00202}_{-0.00199}$ &   $0.11712^{+0.00188}_{-0.0017}$ \\
6225     &      150 &  $0.11549^{+0.00144}_{-0.00154}$ &  $0.11332^{+0.00249}_{-0.00238}$ &                              -- &  $0.11386^{+0.00295}_{-0.00323}$ &  $0.11467^{+0.00197}_{-0.00197}$ &   $0.11837^{+0.00178}_{-0.0018}$ \\
6375     &      150 &  $0.11522^{+0.00136}_{-0.00129}$ &  $0.11069^{+0.00222}_{-0.00232}$ &                              -- &  $0.11325^{+0.00342}_{-0.00406}$ &   $0.11555^{+0.00218}_{-0.0023}$ &  $0.11867^{+0.00151}_{-0.00147}$ \\
6525     &      150 &   $0.1149^{+0.00152}_{-0.00151}$ &  $0.11429^{+0.00301}_{-0.00298}$ &                              -- &  $0.11039^{+0.00304}_{-0.00339}$ &  $0.11398^{+0.00206}_{-0.00238}$ &  $0.11676^{+0.00147}_{-0.00152}$ \\
6675     &      150 &  $0.11771^{+0.00161}_{-0.00163}$ &   $0.11146^{+0.00322}_{-0.0033}$ &                              -- &                              -- &                              -- &                              -- \\
6825     &      150 &  $0.11592^{+0.00135}_{-0.00144}$ &  $0.11022^{+0.00274}_{-0.00237}$ &                              -- &  $0.11261^{+0.00309}_{-0.00292}$ &  $0.11454^{+0.00191}_{-0.00205}$ &  $0.11612^{+0.00156}_{-0.00156}$ \\
6975     &      150 &  $0.11624^{+0.00135}_{-0.00133}$ &   $0.11274^{+0.0027}_{-0.00271}$ &                              -- &                              -- &                              -- &                              -- \\
7125     &      150 &  $0.11646^{+0.00145}_{-0.00137}$ &    $0.1133^{+0.00188}_{-0.0021}$ &                              -- &  $0.11471^{+0.00274}_{-0.00284}$ &   $0.11823^{+0.00176}_{-0.0018}$ &  $0.11593^{+0.00113}_{-0.00118}$ \\
7275     &      150 &   $0.11687^{+0.0015}_{-0.00158}$ &  $0.11584^{+0.00161}_{-0.00189}$ &                              -- &    $0.11564^{+0.0019}_{-0.0019}$ &   $0.11429^{+0.0016}_{-0.00167}$ &  $0.11655^{+0.00109}_{-0.00112}$ \\
7425     &      150 &    $0.1128^{+0.00183}_{-0.0018}$ &   $0.1153^{+0.00175}_{-0.00193}$ &                              -- &  $0.11486^{+0.00256}_{-0.00253}$ &  $0.11613^{+0.00164}_{-0.00176}$ &  $0.11409^{+0.00131}_{-0.00121}$ \\
7575     &      150 &  $0.11336^{+0.00218}_{-0.00201}$ &   $0.1149^{+0.00247}_{-0.00305}$ &                              -- &  $0.11311^{+0.00243}_{-0.00261}$ &  $0.11719^{+0.00323}_{-0.00322}$ &  $0.11495^{+0.00139}_{-0.00147}$ \\
7665     &       20 &   $0.11687^{+0.0056}_{-0.00572}$ &    $0.12449^{+0.005}_{-0.00502}$ &                              -- &  $0.11706^{+0.00365}_{-0.00403}$ &  $0.11602^{+0.00316}_{-0.00313}$ &     $0.1136^{+0.003}_{-0.00277}$ \\
7699     &       20 &   $0.11548^{+0.00542}_{-0.0049}$ &  $0.11409^{+0.00462}_{-0.00449}$ &                              -- &    $0.11899^{+0.004}_{-0.00343}$ &  $0.11303^{+0.00246}_{-0.00265}$ &  $0.11327^{+0.00235}_{-0.00261}$ \\
7725     &      150 &  $0.11467^{+0.00195}_{-0.00206}$ &  $0.11655^{+0.00187}_{-0.00164}$ &                              -- &  $0.11682^{+0.00224}_{-0.00235}$ &  $0.11568^{+0.00152}_{-0.00162}$ &  $0.11555^{+0.00138}_{-0.00136}$ \\
7875     &      150 &  $0.11578^{+0.00196}_{-0.00201}$ &   $0.1161^{+0.00196}_{-0.00191}$ &                              -- &   $0.11706^{+0.0023}_{-0.00211}$ &  $0.11526^{+0.00186}_{-0.00206}$ &  $0.11563^{+0.00112}_{-0.00112}$ \\
8025     &      150 &  $0.11658^{+0.00224}_{-0.00214}$ &   $0.11672^{+0.00176}_{-0.0016}$ &                              -- &  $0.11817^{+0.00222}_{-0.00229}$ &   $0.11825^{+0.0021}_{-0.00331}$ &   $0.1144^{+0.00147}_{-0.00137}$ \\
8175     &      150 &  $0.11157^{+0.00239}_{-0.00243}$ &  $0.11477^{+0.00196}_{-0.00206}$ &                              -- &                              -- &                              -- &                              -- \\
8325     &      150 &   $0.1167^{+0.00214}_{-0.00214}$ &  $0.11184^{+0.00175}_{-0.00177}$ &                              -- &                              -- &                              -- &                              -- \\
8475     &      150 &  $0.11756^{+0.00214}_{-0.00197}$ &   $0.11531^{+0.00257}_{-0.0022}$ &                              -- &  $0.11887^{+0.00222}_{-0.00224}$ &    $0.116^{+0.00205}_{-0.00211}$ &    $0.1149^{+0.0014}_{-0.00134}$ \\
8625     &      150 &  $0.11329^{+0.00228}_{-0.00223}$ &  $0.11728^{+0.00272}_{-0.00291}$ &                              -- &   $0.11665^{+0.0024}_{-0.00211}$ &  $0.11791^{+0.00196}_{-0.00207}$ &  $0.11634^{+0.00121}_{-0.00124}$ \\
8775     &      150 &  $0.11909^{+0.00207}_{-0.00246}$ &  $0.11616^{+0.00285}_{-0.00281}$ &                              -- &  $0.11698^{+0.00224}_{-0.00238}$ &  $0.11509^{+0.00249}_{-0.00299}$ &  $0.11631^{+0.00134}_{-0.00126}$ \\
8925     &      150 &   $0.1182^{+0.00295}_{-0.00253}$ &   $0.11888^{+0.0036}_{-0.00397}$ &                              -- &   $0.11548^{+0.00251}_{-0.0024}$ &   $0.11779^{+0.0017}_{-0.00191}$ &  $0.11441^{+0.00131}_{-0.00142}$ \\
9075     &      150 &  $0.11612^{+0.00308}_{-0.00321}$ &   $0.12106^{+0.0038}_{-0.00444}$ &                              -- &   $0.11643^{+0.00294}_{-0.0026}$ &  $0.11801^{+0.00201}_{-0.00188}$ &  $0.11501^{+0.00177}_{-0.00158}$ \\
9225     &      150 &                              -- &                              -- &                              -- &   $0.11485^{+0.0024}_{-0.00253}$ &   $0.11608^{+0.00203}_{-0.0019}$ &  $0.11648^{+0.00172}_{-0.00168}$ \\
9375     &      150 &                              -- &                              -- &                              -- &  $0.11547^{+0.00432}_{-0.00407}$ &  $0.11614^{+0.00314}_{-0.00378}$ &  $0.11281^{+0.00173}_{-0.00161}$ \\ \hline
\end{tabular}
\end{table*}



\section{Appendix: spectroscopic light curve fits}

\begin{figure}[ht!]
    \centering
    \includegraphics[scale=0.65]{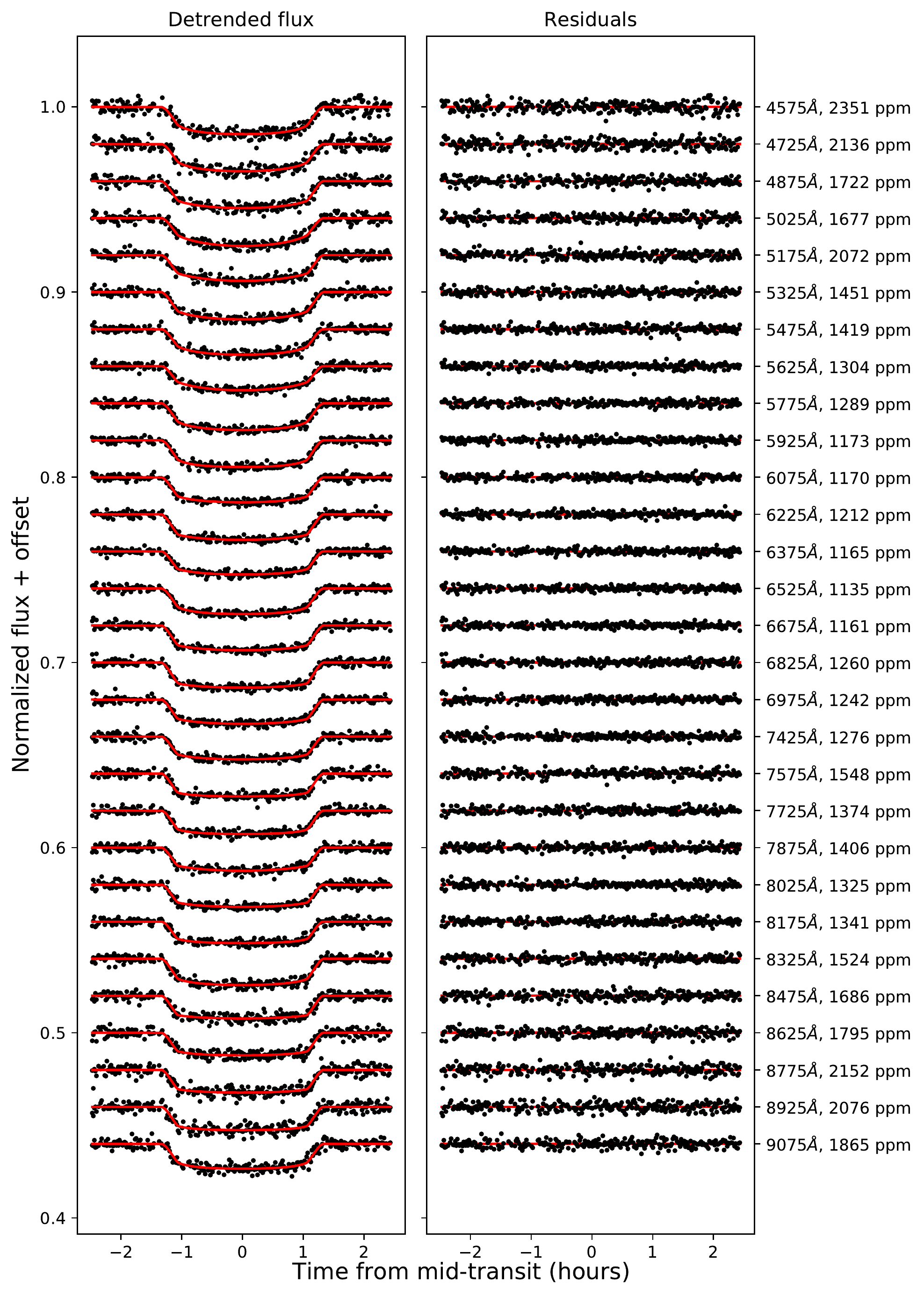}
    \caption{The spectroscopic light curve fits to the 150\,\AA-wide bins of ACCESS n2, using \texttt{GPTransSpec}. Left-hand panel: the black points show the detrended data (after removing the PCA and GP components) and the red line shows the best fitting transit model. Right-hand panel: the residuals to the fits in the left-hand panel. The RMS of these is given in ppm on the right-hand axis.}
    \label{fig:wbfit_ACCESS_n2}
\end{figure}

\begin{figure}[ht!]
    \centering
    \includegraphics[scale=0.8]{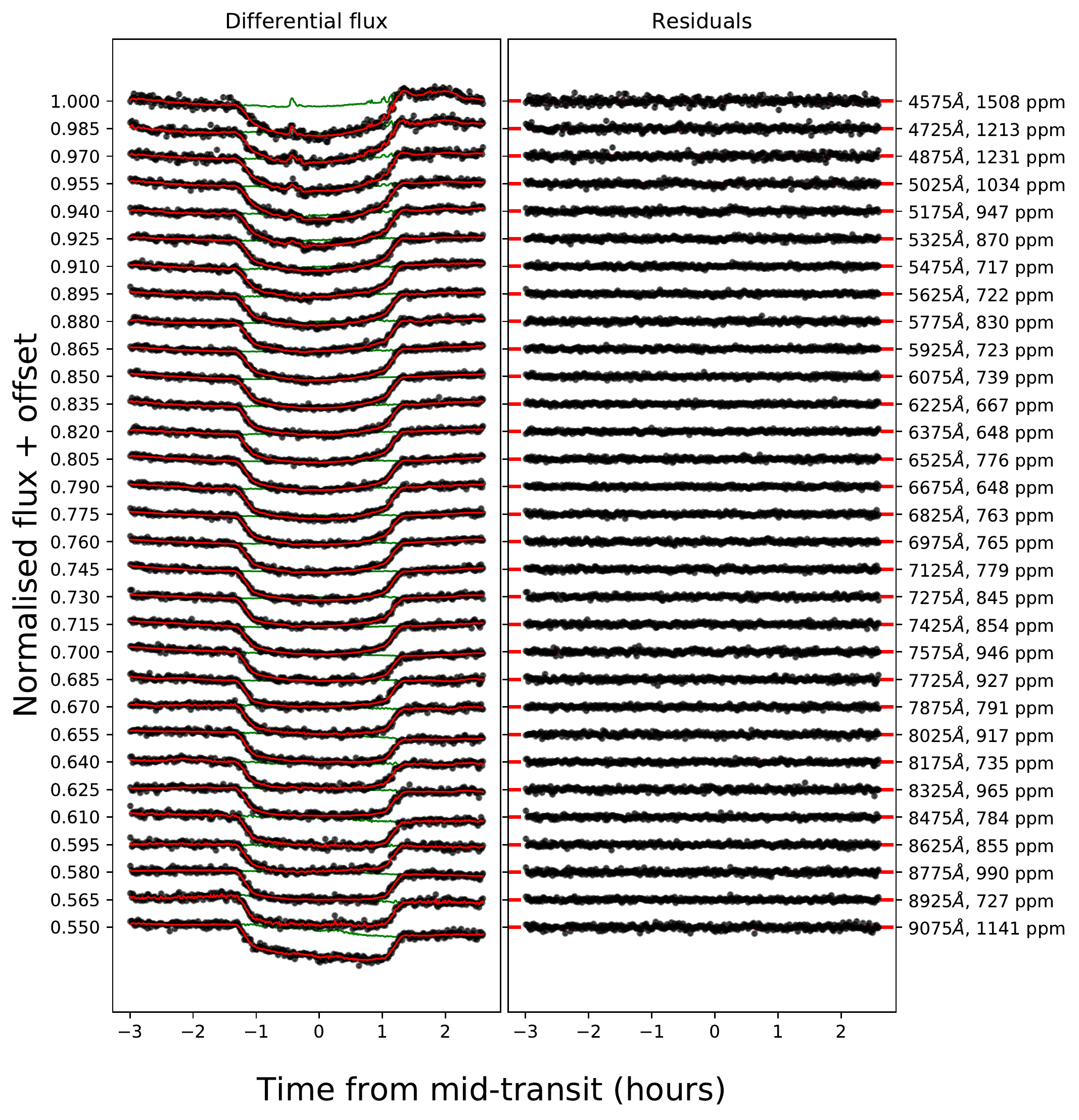}
    \caption{The spectroscopic light curve fits to the 150\,\AA-wide bins of LRG-BEASTS n1, using \gppmfit{}. Left-hand panel: the black points show the data, the red line shows the best fitting systematics + transit model and the green line shows the best fitting systematics model only. Right-hand panel: the residuals to the fits in the left-hand panel. The RMS of these is given in ppm on the right-hand axis.}
    \label{fig:wbfit_LRG-BEASTS_n1}
\end{figure}

\autoref{fig:wbfit_ACCESS_n2} shows spectroscopic light curve fits for an example ACCESS dataset (ACCESS n2), and \autoref{fig:wbfit_LRG-BEASTS_n1} does the same for an example LRG-BEASTS dataset (LRG-BEASTS n1).

\end{document}